\documentclass[12pt]{article}
\usepackage{amsmath, amssymb}
\usepackage{epsfig}
\usepackage{color}
\usepackage{curves}
\usepackage{eepic}
\usepackage{epic}
\usepackage{theorem}
\usepackage{amscd}
\usepackage{multirow}
{\theorembodyfont{\normalfont}

}

\begin{document}
\newcommand{\1}{\mbox{1}\hspace{-0.25em}\mbox{l}}

\title{Band rearrangement  through the 3D-Dirac equation with  
boundary conditions, and the corresponding topological change }

\maketitle
\begin{center}
{Toshihiro Iwai${}^1$ and Boris Zhilinskii${}^2$} 
\end{center}

\begin{center}
\begin{list}{}{}
\item[${}^1$]
Kyoto University, 606-8501 Kyoto, Japan
\item[${}^2$]
Universit{\'e} du Littoral C{\^o}te d'Opale, Dunkerque, France
\item
E-mail: iwai.toshihiro.63u@st.kyoto-u.ac.jp,  zhilin@univ-littoral.fr 
\end{list}
\end{center}

\begin{abstract}
\noindent 
Rearrangement of energy bands against a parameter is studied through the 3D-Dirac equation on a ball in 
$\mathbb{R}^3$ under the APS and the chiral bag boundary conditions on the boundary two-sphere, 
where APS is an abbreviation of Atiyah-Patodi-Singer.
The notion of spectral flow and its extension is introduced to characterize the energy eigenvalue 
redistribution against the parameter, which reflects an analytical property of edge eigenstates.  
It is shown that though band rearrangement takes place the net spectral flow is zero 
($1+(-1)=0$) for both boundary conditions. 
The corresponding semi-quantum Hamiltonian defined on the 3D momentum space 
is studied in parallel and it is shown that a change against the parameter is observed 
in the mapping degree defined for the  semi-quantum model Hamiltonian, which is viewed as reflecting 
a topological property of the bulk eigenstates of the full quantum system.  
Specifically, there are two mappings assigned, for which changes in mapping degree 
take the values $\pm1$.   The present correspondence is viewed as a bulk-edge correspondence. 
\end{abstract}


\section{Introduction}
A relation between band rearrangement and topological change in quantum states 
or what is called the bulk-edge correspondence have been increasingly attractive  
in condensed matter physics, especially in topological insulators, a review of  which 
is found in \cite{TopInsTopSup, HasKan}, for example.  
It is to be recognized that among different fields of physics similar topological ideas are shared 
under respective nomenclatures such as band rearrangement or energy level redistribution,  
gap closing \cite{TSN}, gap node \cite{GrVo}, gapless excitation, 
contact of the conduction band and the valence 
band at a single point \cite{AbBe}, and Dirac points \cite{TopInsTopSup}, {\it etc}. 
The  authors have been interested in this theme from the viewpoint of molecular physics 
\cite{DIZ, IwaiAnnPhys, IZ2013, IZChemAcc, IZ2017}.  

A setup for pursuing this theme in molecular physics is stated as follows: 
There is a class of molecular systems such that the whole set of dynamical variables can be split into 
subsets associated with low- and high-energy excitations  \cite{FaureLMP}.  
Depending on modelings, the dynamical variables in a molecular system are separated into electronic and 
vibrational variables,  orbital and spin ones,  or rotational and vibrational ones, to cite a few.  
Among such modelings, the rotation-vibrational problem is taken as a reference problem.    
In this modeling, the rotational variables describing high-density energy levels due to low-energy 
excitations can be treated as classical ones which range over a two-dimensional sphere 
on account of the conservation of the angular momentum, but the vibrational variables 
describing high-energy excitations remain to be treated as quantum ones.  
Such construction is called a semi-quantum model and can be naturally generalized 
to quantum systems possessing subsystems of ``slow" and ``fast" variables of different nature. 

Within semi-quantum rotation-vibration models,  
the authors have been studying band rearrangements for isolated molecules in terms of Chern numbers 
in a series of papers \cite{IwaiAnnPhys, Iwai, IZ2013, IZ2017, DIZ}. 
A family of semi-quantum Hamiltonians is defined to be Hermitian matrices on the two-sphere along with 
some control parameters.  Symmetry is often taken into account, and the family of the Hamiltonians 
are supposed to be invariant under a prescribed group which is a continuous or a finite subgroup of 
the orthogonal group $O(3)$.  
For this family of semi-quantum Hamiltonians, the Chern number is defined 
for the complex line bundles associated with each of non-degenerate eigenvalues.      
The parameter space is divided into connected regions to every point of which the same Chern number 
is assigned but the different regions have respective Chern numbers 
(see figures in \cite{IwaiAnnPhys} and \cite{IZ2013} in the presence of $D_3$ and $O$ symmetries, respectively). 
The boundaries between different regions correspond to degeneracy of eigenvalues, which is responsible 
for the modification of band structure. 
The variation in Chern numbers in crossing the boundary between different regions 
is counted by means of related winding numbers and the orbit of the symmetry group, and 
is shown to qualitatively explain the band rearrangement for an isolated molecule \cite{IZ2013, IZ2017}.  
A general description of the stratification of  the set of eigenvalues and 
eigenvectors of a family of Hermitian matrices is found in \cite{Arnoldmodes, ArnoldChern}. 

If the coupling between rotation and vibration is not weak, 
the simple splitting of quantum energy levels is not available \cite{Herz, MolPhys88}. 
Strictly speaking,  rotational and vibrational motions cannot be separated on the level of 
classical mechanics \cite{Guichardet} and hence one needs to set up classical and quantum mechanics 
in terms of fiber bundles by taking into account the non-separability of 
rotation and vibration \cite{IwaiClass, IwaiHolonomy}.  
In this setting, a number of articles have been published \cite{IwaiYama05, IwaiYama,TachIwai}. 
Nevertheless, if a finite set of vibrational levels can be chosen, effective model Hamiltonians 
for rotation-vibration systems are of practical use 
to study a question as to what kinds of rearrangement of band structure are allowed under the variation of 
control parameters in the presence of symmetry \cite{IZChemAcc}.    

At this stage in Introduction, it is significant to compare the idea as to fast- and slow variables in molecular physics 
with bulk-edge correspondence in topological insulator theory.  
In the topological insulators, while the edge state is described quantum mechanically,  
the bulk states are allowed to be described in terms of classical variables, 
specifically, in the momentum variables \cite{FSFF, MS}.  
This procedure can be interpreted from the viewpoint of molecular physics as follows: 
The insulator is viewed as a two-band system, and the bulk-state energy levels are of high density, so that
the bulk states can be treated in terms of classical variables in a semi-quantum model. 
Hence, the Chern number comes into sight as a topological invariant which characterizes the bands.  
Incidentally, a study of topological insulators has been made also 
under the named of bulk boundary principle \cite{Graf-Porta, PSB}.   In particular, in \cite{PSB}, 
bulk and boundary invariants for lattice systems are studied  by means of K-theory. 

Interest in topological study of intersection of potential energy surfaces 
on the basis of Hermitian matrices dates back to \cite{HLH} and 
led to the Berry phase \cite{Berry, ChrJam, WS}.   
Hermitian matrices have been effectively used \cite{SadZhilMonodr, PhysRep2} in semi-quantum 
molecular models as well. 
The relevance of Chern numbers to band rearrangements in isolated molecules was initially suggested 
in \cite{VPVdp} and studied further for formation of more concrete relation between Chern numbers 
and the numbers of states in the band \cite{FaurePRL, FaureLMP, FaureAAM}.   
As is mentioned above, a spectacular appearance of Chern numbers 
is made in the description of quantum Hall effect \cite{AOS, Kohmoto, TKNN}, 
though the base manifold of the relevant line bundle is a two-torus 
in contrast to the two-sphere for rotation-vibration problems for isolated molecules.  
Increasing interest has centered on topological phenomena such as the quantum spin Hall effect 
and topological insulators \cite{KaneMele, QHRZ, HasKan, BHZ, TopInsTopSup}, 
and further attempt has been made for classifying topological phases \cite{Kitaev, RSFL, SchFurLud}. 

In the course of the study on band rearrangement in semi-quantum models \cite{IZ2013, IZ2017}, 
the authors have observed that elementary band rearrangement (or level-crossing) takes place 
between two adjacent bands and can be characterized by a topological invariant  
which can be evaluated through the linearization of the Hamiltonian at a critical point, and they have called 
the  elementary topological change a delta-Chern. 
Further, they pointed out, in previous papers \cite{IZ2015, IZ2016}, that 
the linearization method works well for the study of band rearrangement both in semi-quantum and 
full quantum models.  
In  \cite{IZ2015, IZ2016}, 2D Dirac equations with boundary conditions are intensively solved 
with interest in qualitative modifications of band structures in the full quantum model together with 
interest in analogous modification in the corresponding semi-quantum model. 
It was shown that the delta-Chern \cite{IZ2013,IZ2017} defined for a parametric family of Hamiltonians 
in the semi-quantum model corresponds to the spectral flow or its extension for the 2D Dirac equation 
in question, where the spectral flow of a one-parameter family of self-adjoint operators 
is originally defined to be the net number of eigenvalues passing through zero in the positive direction as the 
parameter runs \cite{APS,Prokh} and then extended in \cite{IZ2016}, 
and where the delta-Chern is a jump in the Chern number which takes place when the 
parameter crosses a critical value.  
The authors have already given non-linear two-dimensional models which exhibits the correspondence 
between spectral flows and changes in Chern numbers \cite{DIZ}. 
The present correspondence is, in principle,  the same as the bulk-edge correspondence for 
topological insulators without time-reversal symmetry \cite{FSFF}.   

The one-parameter family of Dirac operators studied in  \cite{IZ2015, IZ2016} is given by 
\begin{equation} 
\label{2D_Dirac}
  H_t = -i\sum_{k=1}^2 \sigma_k \nabla_k + t\sigma_3, \quad \nabla_k=\partial/\partial q_k, 
\end{equation}
where $t$ is a control parameter and where $\sigma_j,\,j=1,2,3$, are the standard Pauli matrices. 
The eigenvalue problem for $H_{\mu}$ is set on the disk $q_1^2+q_2^2\leq R^2$ 
under the APS and the chiral bag boundary conditions on the circle $q_1^2+q_2^2=R^2$,  
where APS is an abbreviation of Atiyah-Patodi-Singer \cite{APS}. 
The corresponding semi-quantum Hamiltonian is given by replacing the 
operators $-i\nabla_k$ by the real variables $\xi_k$, $k=1,2$.  

A question arises as to whether the correspondence between the spectral flow and the delta-Chern  
is a key, or not, to the understanding of the relation between band rearrangement and topological change,  
independently of a choice of models.  
Put in another way, the present question is as to whether the bulk-edge correspondence holds for 
a wider class of model Hamiltonians or not.  
A further way to approach the present question is to consider higher-dimensional cases. 
If group actions such as a time-reversal transformation with half-integer spin degrees of freedom 
are taken into account, then new Hamiltonians comes to be defined in dimensions larger than two, 
in both quantum and semi-quantum models. 
Another view of the extension of Hamiltonians is taken through gamma matrices. 
If $4\times 4$ gamma matrices are adopted, corresponding Dirac Hamiltonians are to be defined 
on $\mathbb{R}^3$ or $\mathbb{R}^4$.  
The present paper deals with the Dirac Hamiltonian defined on $\mathbb{R}^3$ 
and the corresponding semi-quantum Hamiltonian,  both of which admit the time-reversal and 
the chiral symmetries. 
Since the Chern number is defined for vector bundles over even-dimensional manifolds and since 
the manifold $\mathbb{R}^3$ on which the model Hamiltonians are to be defined is of odd-dimensions, 
another topological invariant should be introduced to describe a topological character. 
In the last section of this paper, remarks will be made on Dirac Hamiltonians defined on $\mathbb{R}^4$ 
together with the corresponding semi-quantum Hamiltonian.     

Among several representations of the gamma matrices, 
the choice made in the present article is  
\begin{equation}
\label{gamma matrices} 
 \gamma_k^{} = \begin{pmatrix} 0 & -i\sigma_k \\ i\sigma_k & 0 \end{pmatrix}, \quad k=1,2,3, \quad 
 \gamma_0^{}= \begin{pmatrix} \1 & 0 \\ 0 & -\1 \end{pmatrix} , 
\end{equation} 
where $\1$ denotes the $2\times 2$ unit matrix. 
Then, the 3D Dirac Hamiltonian takes the form  
\begin{align}
   H_{\mu}  =-i\sum_{k=1}^3 \gamma_k^{}\nabla_k + \mu \gamma_0^{}  
       = \begin{pmatrix} \mu \1 & -i\boldsymbol{\sigma}\cdot \boldsymbol{p} \\
           i\boldsymbol{\sigma}\cdot \boldsymbol{p} & -\mu \1\end{pmatrix}, \quad 
    \boldsymbol{p}=-i\boldsymbol{\nabla}, 
   \label{3D_Dirac_Ham}
\end{align} 
where $\mu$ is a parameter, which originally denotes the mass but is allowed to take negative values 
in the present study. 
In Dirac's textbook \cite{Dirac}, the alpha matrices are used in place of the gamma matrices 
\eqref{gamma matrices}, where $\alpha_k=\sigma_1\otimes \sigma_k,\,k=1,2,3$ 
while $\gamma_k=\sigma_2\otimes \sigma_k$.  

In order to obtain discrete energy eigenvalues for $H_{\mu}$, a boundary condition is required to be posed on 
the boundary of a bounded domain, since no external field is present.   In this paper, 
the eigenvalue problems for $H_{\mu}$ are put on the ball of radius $R$ under 
the APS and the chiral bag boundary conditions.  
It will be shown that the eigenvalues for $H_{\mu}$ with the APS and the chiral bag boundary conditions 
are broken up into bulk- and edge-states eigenvalues, where the edge-state eigenvalues are responsible for 
band rearrangement but the bulk-state eigenvalues form separate two bands, positive and negative. 
The spectral flow and its extension are defined for edge-state eigenvalues 
and has the value $1+(-1)=0$ for both boundary conditions. 
In spite of the zero value, the band rearrangement takes place indeed. 

From the viewpoint of  energy level density, 
it is shown that the difference between adjacent two bulk-state energy levels is inversely proportional 
to the radius of the ball, so that the bulk-state eigenvalues become of high density for a sufficiently large $R$. 
Hence, the bulk-state bands are allowed to be treated in terms of classical variables or in momentum variables. 
Accordingly, the quantum Hamiltonian $H_{\mu}$ turns into a semi-quantum Hamiltonian 
by replacing the classical momentum variables $\boldsymbol{k}$ for the momentum operators $\boldsymbol{p}$. 
The zero spectral flow for the 3D Dirac equation is shown to have a counterpart in the corresponding 
semi-quantum model, which  is described in terms of the mapping degrees  (or winding numbers) of mappings 
to be defined through the projection operator onto each of eigenspaces for the semi-quantum Hamiltonian.  
To each projection operator, there are assigned two mappings which have the mapping degrees of opposite sign.   
Each of the mapping degree changes the sign as the parameter passes the critical value (zero), but the 
net change, the sum of variation in respective mapping degrees, is zero.  
The zero spectral flow and the zero net change in mapping degrees are in keeping with the 
particle-hole symmetry of the quantum and semi-quantum Hamiltonians.  
Thus, the bulk-edge correspondence proves to hold in the present setting.  
  
This paper is organized as follows:  
In Sec.~\ref{setting-up},  the Dirac Hamiltonian \eqref{3D_Dirac_Ham} and the corresponding semi-quantum 
Hamiltonian are characterized from the viewpoint of the time-reversal and the chiral symmetries. 
Section \ref{full quantum models} is concerned with the full quantum model.    
In the beginning of this section, a brief review is made of the total angular momentum operators 
together with the associated spinor spherical harmonics on $S^2$.    
After the 3D Dirac Hamiltonian is described in spherical polar coordinates,  
a boundary operator is introduced on the two-sphere of radius $R$.    
Then, feasible solutions to the Dirac equation are obtained  without referring to the boundary condition, 
which is the first step toward solutions to the boundary value problem.  
Depending on ranges of the parameter, the feasible solutions are classified into three classes. 
After the APS and the chiral bag boundary conditions are specified,  
eigenvalues for the 3D Dirac equation with respective boundary conditions 
are found together with associated eigenstates, 
which are classified into bulk states, edge states, and zero modes or critical states, 
where the radial functions for bulk and edge states are described in terms of Bessel functions and 
modified Bessel functions, respectively, and zero modes or critical states take the form of solid harmonics. 
Discrete symmetry (or pseudo-symmetry) can explain the pattern of eigenvalues as functions of 
the parameter $\mu$. 
In the last subsection of this section, the $R$-dependence of the bulk-state eigenvalues is discussed and 
then it is shown that the semi-quantum Hamiltonian can be viewed as a bulk Hamiltonian in the limit 
as $R\to \infty$. 
In Sec.~\ref{semi-quantum models}, the corresponding semi-quantum Hamiltonian is studied. 
Two mappings $\mathbb{R}^3 \to S^3$ are defined through the projection operators onto the eigenspaces. 
A jump in the mapping degree against the parameter is shown to have a topological meaning.  
Sec.~\ref{Correspondence} presents an answer to the aforementioned question as to the correspondence 
between band rearrangement and topological change or the bulk-edge correspondence.  
In Sec.~\ref{conclusion}, remarks on further study and related fields to the present study are mentioned.  
Appendix \ref{SU(2) representation}  contains calculations for the derivation of spherical spinor harmonics 
and for the description of the Hamiltonian in the polar spherical coordinates.  

\section{Dirac Hamiltonians with discrete symmetry}
\label{setting-up}
In the momentum representation,  the momentum operators $\boldsymbol{p}$ are replaced  
by the classical variables $\boldsymbol{k}$, and then the Dirac Hamiltonian \eqref{3D_Dirac_Ham} 
is brought into 
\begin{equation} 
\label{SQ Dirac Hamiltonian}
     K_{\mu}(\boldsymbol{k})=\begin{pmatrix} 
         \mu \1 & -i\boldsymbol{k}\cdot \boldsymbol{\sigma} \\
        i\boldsymbol{k}\cdot \boldsymbol{\sigma} & -\mu \1 \end{pmatrix} , \quad 
\boldsymbol{k}\in \mathbb{R}^3, 
\end{equation}
which we call a semi-quantum Dirac Hamiltonian.  
In this section, we characterize the Hamiltonian \eqref{SQ Dirac Hamiltonian} 
after the classification scheme for Hermitian matrices by means of discrete symmetries 
such as time-reversal, particle-hole, and chiral symmetries \cite{AZ, HHZ, Zirn}.   
Remarks on the so-called AZ classification will be made in the last section, 
as far as the present Hamiltonian is concerned.  

According to \cite{Mead, KS}, the introduction of time-reversal symmetry 
along with the spin degrees of freedom for spin-$\frac12$ converts a two-level Hamiltonian acting on $\mathbb{C}^2$ 
into a four-level Hamiltonian acting on $\mathbb{C}^2\otimes \mathbb{C}^2$,  
which we denote by $\mathcal{H}$. 
We now require that this Hamiltonian admits the time-reversal and the chiral (or sublattice) symmetries 
by imposing the constraints, 
\begin{subequations} 
 \begin{align}
     (\1\otimes i\sigma_2)\overline{\mathcal{H}}(\1\otimes (-i\sigma_2)) & =\mathcal{H}, 
     \label{TRS} \\
   (\sigma_1\otimes \1) \mathcal{H} (\sigma_1\otimes \1) & =-\mathcal{H},   
     \label{SLS}
 \end{align}
\end{subequations}
respectively, where the overline on $\mathcal{H}$ denotes the complex conjugation. 
The time-reversal symmetry condition \eqref{TRS} renders the Hamiltonian in the form 
\begin{equation}
\label{TRS Ham}
   \mathcal{H}=\begin{pmatrix} c & & a & b \\  & c & -\overline{b} & \overline{a} \\
    \overline{a} & -b & d & \\ \overline{b} & a & & d \end{pmatrix}, \quad a,b\in \mathbb{C}, 
    \: c,d \in \mathbb{R}.    
\end{equation}
Furthermore, the chiral symmetry condition \eqref{SLS} brings the Hamiltonian \eqref{TRS Ham} into 
\begin{equation}
\label{TRS-SLS Ham}  
    \mathcal{H}=\begin{pmatrix} c & & ia & b \\ & c & -\overline{b} & -ia \\
      -ia & -b & -c & \\ \overline{b} & ia &  & -c \end{pmatrix}, \quad 
     a,c\in \mathbb{R},\;b\in \mathbb{C}.    
\end{equation}
We note that the Hermitian matrix of the form \eqref{TRS Ham} is already found in \cite{Mead, KS}. 
If we set $b=b_2+ib_1$ with $b_1,b_2\in \mathbb{R}$ and further rewrite the parameters as 
\begin{equation}
\label{parameter setting}
  b_1=-k_1, \quad b_2=-k_2, \quad a=-k_3, \quad c=\mu, 
\end{equation}
then the Hamiltonian \eqref{TRS-SLS Ham} is put in the form  
\begin{align} 
\label{3D Ham, TR PH}
   \mathcal{H}(\boldsymbol{k},\mu)  =
     \begin{pmatrix} \mu \1 & -i \boldsymbol{k}\cdot \boldsymbol{\sigma} \\
    i\boldsymbol{k}\cdot \boldsymbol{\sigma} & -\mu\1 \end{pmatrix} .   
\end{align} 
 
According to \cite{SchFurLud}, the chiral (or sublattice) symmetry is the product of the time-reversal and 
the particle-hole symmetries.  
In our present case, the product of the time-reversal and the chiral operators is given by  
\begin{equation}
    (\1\otimes i\sigma_2)K\cdot(\sigma_1\otimes \1)=(\sigma_1\otimes i\sigma_2)K,     
\end{equation}
where $K$ denotes the complex conjugation. 
This operator serves as the particle-hole operator. 
As is easily seen, the Hamiltonian admits the particle-hole symmetry, 
\begin{align}
\label{SU(2)action}
   (\sigma_1\otimes i\sigma_2)\overline{\mathcal{H}(\boldsymbol{k},\mu)}(\sigma_1\otimes(- i\sigma_2))
      &  =- \mathcal{H}(\boldsymbol{k},\mu).   
\end{align}

At this stage, $\boldsymbol{k}$ and $\mu$ are merely parameters. 
We now break up the parameters into dynamical variables and a control parameter by 
taking into account the $SU(2)$ action on the Hamiltonian $\mathcal{H}(\boldsymbol{k},\mu)$. 
Under the adjoint action of $\1\otimes g$, $g\in SU(2)$, the $\mathcal{H}(\boldsymbol{k},\mu)$ transforms 
according to 
\begin{equation}
\label{semi-quantum SU(2) action on Ham}  
   (\1\otimes g)\mathcal{H}(\boldsymbol{k},\mu)(\1\otimes g^{-1})=
     \mathcal{H}(G\boldsymbol{k},\mu), \quad g\in SU(2), 
\end{equation}
where $G$ is the rotation group defined through $g\sigma_kg^{-1}=\sum_j \sigma_j G_{jk}$.  
This shows that the $SU(2)$ action is accompanied with the transform 
$\boldsymbol{k}\mapsto G\boldsymbol{k}$ but the $\mu$ is left invariant.  
For this reason, we can interpret $\boldsymbol{k}$ and $\mu$ as dynamical variables and a control parameter, 
respectively. 
With this in mind, we denote $\mathcal{H}(\boldsymbol{k},\mu)$ by $K_{\mu}(\boldsymbol{k})$, 
which is exactly the same as \eqref{SQ Dirac Hamiltonian}.  
Then, Eq.~\eqref{semi-quantum SU(2) action on Ham} means that the Hamiltonian 
$K_{\mu}(\boldsymbol{k})$ defined on $\mathbb{R}^3$ 
is equivariant with respect to the $SU(2)$ action. 
In a summary, we list the conditions that the $K_{\mu}(\boldsymbol{k})$ satisfies 
for the time-reversal, the particle-hole, and the chiral symmetries, in this order,  
\begin{align}
   (\1\otimes i\sigma_2)\overline{K_{\mu}(\boldsymbol{k}})(\1\otimes(- i\sigma_2)) & =
       K_{\mu}(\boldsymbol{k}),      \label{TRS prim}   \\
   (\sigma_1\otimes i\sigma_2)\overline{K_{\mu}(\boldsymbol{k})}(\sigma_1\otimes(- i\sigma_2))
      &  =- K_{\mu}(\boldsymbol{k}),   \label{PHS prim}\\
   (\sigma_1\otimes \1 )K_{\mu}(\boldsymbol{k})(\sigma_1\otimes \1 ) & =
    -K_{\mu}(\boldsymbol{k}).  \label{SLS prim}
\end{align}

We make a further remark on the Hamiltonian $K_{\mu}(\boldsymbol{k})$ from a generic point of view. 
Though we have set the parameters $b_1,b_2,a,c$ according to \eqref{parameter setting}, 
we may set the parameters to take a more generic form. For example, a generic Hamiltonian is expressed 
as $\widetilde{\mathcal{H}}(\boldsymbol{q},\nu)=\sum_{a=1}^3 h_a(\boldsymbol{q})\gamma_a^{}
+h_0(\nu) \gamma_0^{}$, where $\boldsymbol{q}$ and $\nu$ are variables on a three-dimensional 
manifold and a control parameter, respectively.  
If there exist a critical point $\boldsymbol{q}_0$ and a critical parameter value $\nu_0$ 
such that  $h_0(\nu_0)=h_a(\boldsymbol{q}_0)=0, a=1,2,3$, the Hamiltonian $K_{\mu}(\boldsymbol{k})$ 
is viewed as a linearization of $\widetilde{\mathcal{H}}(\boldsymbol{q},\nu)$ at $(\boldsymbol{q}_0,\nu_0)$. 
The linearization of a Hamiltonian at a critical point is effectively used both in the topological insulator theory 
\cite{HasKan} and in the molecular physics \cite{IZ2013,IZ2017}, in the latter of which 
the linearization at a degeneracy point for eigenvalues of a semi-quantum Hamiltonian is rigorously treated 
to obtain ``delta-Cherns," while the base manifold is the unit two-sphere. 
 
In the corresponding full quantum model, the Dirac Hamiltonian $H_{\mu}$ may admit the time-reversal, 
the particle-hole, and the chiral symmetries as well. 
To translate the symmetry equations for $K_{\mu}$ into those for $H_{\mu}$, we need to take into account the 
inversion $\boldsymbol{k}\mapsto -\boldsymbol{k}$.  
This is because in the corresponding full quantum model, 
the momentum operators $\boldsymbol{p}$ undergo the inversion $\boldsymbol{p}\mapsto -\boldsymbol{p}$ when the 
complex conjugation is applied. 
For this purpose, we may use the operator $\sigma_3\otimes \1$, which has 
the transformation property 
\begin{equation}
      (\sigma_3\otimes \1)K_{\mu}(\boldsymbol{k}) (\sigma_3\otimes \1)=K_{\mu}(-\boldsymbol{k}).  
\end{equation} 
Then, by the operation with $\sigma_3\otimes \1$ for Eqs.~\eqref{TRS prim} and \eqref{PHS prim} and 
without the operation for Eq.~\eqref{SLS prim} and further by replacing 
$\boldsymbol{k}$ by $\boldsymbol{p}$,  we find that the Dirac Hamiltonian $H_{\mu}(\boldsymbol{p})$ 
satisfies, respectively, the equations for the time-reversal, the particle-hole, 
and the chiral symmetries, 
\begin{align}
      (\sigma_3 \otimes i\sigma_2)\overline{H_{\mu}(\boldsymbol{p})}(\sigma_3 \otimes(- i\sigma_2)) & =
       H_{\mu}(\boldsymbol{p}),  \label{TRS quant} \\
      (\sigma_2\otimes \sigma_2)\overline{H_{\mu}(\boldsymbol{p})}(\sigma_2\otimes \sigma_2)
      &  =- H_{\mu}(\boldsymbol{p}),   \label{PHS quant}  \\
   (\sigma_1\otimes \1 )H_{\mu}(\boldsymbol{p})(\sigma_1\otimes \1 ) & =
    -H_{\mu}(\boldsymbol{p}).  \label{SLS quant}
\end{align} 

In closing this section, we have to mention the parity.  
For a four-component spinor defined on $\mathbb{R}^3$, the parity transformation is defined 
to be $\Phi(\boldsymbol{x})\mapsto \gamma_0^{}\Phi(-\boldsymbol{x})$, $\boldsymbol{x}\in \mathbb{R}^3$, 
where $\gamma_0^{}$ is given in \eqref{gamma matrices} and expressed also as
 $\gamma_0^{}=\sigma_3\otimes \1$. 
As is well known, the Dirac Hamiltonian $H_{\mu}$ is parity invariant and transforms according to 
\begin{equation} 
   (\sigma_3\otimes \1)H_{\mu}(\boldsymbol{p}) (\sigma_3\otimes \1) = H_{\mu}(-\boldsymbol{p}). 
\end{equation} 

\section{Full quantum Dirac models}
\label{full quantum models}
This section, broken up into several subsections, deals with the eigenvalue problem for the Dirac 
operator \eqref{3D_Dirac_Ham} on the ball of radius $R$ under both the APS and the chiral bag 
boundary conditions. 

\subsection{A review of the total angular momentum operators}
\label{ang-moment}
In order to solve the Dirac equation by the use of rotational symmetry, we need to review 
the total angular momentum operators and the spinor spherical harmonics on $S^2$.  

\subsubsection{The $SU(2)$ symmetry}
As is already adopted in \eqref{semi-quantum SU(2) action on Ham},  
the $SU(2)$ action on $\mathbb{C}^4$ is expressed by the matrix 
\begin{equation}
\label{D(g)}
    D(g):=\1 \otimes g =\begin{pmatrix} g & \\ & g \end{pmatrix}, \quad g\in SU(2).  
\end{equation} 
Let $\Phi$ be a four-component spinor defined on $\mathbb{R}^3$.  
Then, the actions of $SU(2)$ on $\mathbb{R}^3$ and on $\mathbb{C}^4$ are related through the diagram
\begin{equation}
\label{diagram SU(2)} 
 \begin{CD}
      \mathbb{R}^3 @>\Phi >> \mathbb{C}^4 \\
       @V{\rm Ad}_gVV   @VV{D(g)}V \\
     \mathbb{R}^3 @>>U_{g}\Phi> \mathbb{C}^4
 \end{CD} \quad\;\, .     
\end{equation}
Here, the space $\mathbb{R}^3$ is identified with $\mathcal{H}_0(2)=\{\sum x_k\sigma_k\}$, 
the set of trace-less $2\times2$ Hermitian matrices,  
and the ${\rm Ad}_g$ is represented as the $SO(3)$ action such as 
\begin{equation} 
     {\rm Ad}_{e^{-it\sigma_k/2}}=e^{tR(\boldsymbol{e}_k)}, \quad k=1,2,3, 
\end{equation} 
where $R(\boldsymbol{a})$ is a skew-symmetric matrix  defined through 
$R(\boldsymbol{a})\boldsymbol{x}=
\boldsymbol{a}\times \boldsymbol{x},\,\boldsymbol{a},\boldsymbol{x}\in \mathbb{R}^3$ 
and where $\boldsymbol{e}_k$ denote the standard basis vectors of $\mathbb{R}^3$. 

The diagram \eqref{diagram SU(2)} determines the unitary operator $U_g$ acting on spinors by 
\begin{equation} 
      U_g\Phi=D(g)\Phi\circ {\rm Ad}_g^{-1}. 
\end{equation} 
For $e^{-it\sigma_k/2}$, the generator of the unitary operator $U_{e^{-it\sigma_k/2}}$ is 
denoted by $-iJ_k$ and expressed as 
\begin{equation}
\label{J_k} 
   J_k = \begin{pmatrix}  L_k +\frac12 \sigma_k & \\  &  L_k +\frac12 \sigma_k \end{pmatrix} , \quad k=1,2,3.  
\end{equation}
The operators $\boldsymbol{J}=(J_k)$ are called the total angular momentum operators or 
the spin-orbital angular momentum operators, satisfying 
the commutation relations   
\begin{equation} 
       [J_j, J_k]=i\sum\varepsilon_{jk\ell}J_{\ell}. 
\end{equation} 

Like \eqref{semi-quantum SU(2) action on Ham}, the Dirac Hamiltonian $H_{\mu}(\boldsymbol{p})$ 
admits the $SU(2)$ symmetry
\begin{equation} 
    D(g)H_{\mu}(\boldsymbol{p})D(g)^{-1}=H_{\mu}(G\boldsymbol{p}). 
\end{equation} 
For $g=e^{-it\sigma_k/2}$ and $G=e^{tR(\boldsymbol{e}_k)}$, the above equation is differentiated 
with respect to $t$ at $t=0$ to provide 
\begin{equation} 
   [-\frac{i}{2}\1\otimes \sigma_k, H_{\mu}] = [i\1\otimes L_k, H_{\mu}], \quad k=1,2,3, 
\end{equation} 
which implies that the Hamiltonian and the total angular momentum operators commute, 
\begin{equation} 
 [H_{\mu},J_k] =0. 
\end{equation} 
It then follows that the eigenvalue problem $H_{\mu}\Phi=E\Phi$ reduces to subproblems on the 
eigenspaces of $\boldsymbol{J}^2=\sum J_k^2$ or on the representation spaces of $SU(2)$. 

\subsubsection{The representation spaces for $SU(2)$}
We proceed to write out the basis states of the representation spaces for 
the total angular momentum operators. 
In view of the expression \eqref{J_k}, we see that the representation space is decomposed into 
the direct sum of two spaces of two-component spinors. 
We now provide basis spinors of the representation spaces, viewing 
the operator $\boldsymbol{J}=\boldsymbol{L}+\frac12 \boldsymbol{\sigma}$ 
as acting on two-spinors. 
On account of the Clebsch-Gordan theorem applied to the spin and the orbital angular momentum 
coupling,  for the eigenvalues $j(j+1)$ of $\boldsymbol{J}^2$, there are two possibilities of 
constructing the same value of $j$, 
\begin{equation} 
    j=\ell+\frac12, \quad  j=(\ell+1)-\frac12,  \quad \ell=0,1,2,\dots. 
\end{equation}
where $\ell$ is the parameter for the representation of the orbital angular momentums and 
$\pm\frac12$ are spin eigenvalues. 

Basis spinors of the representation spaces for the total angular momentum 
are composed of spherical harmonics and basis vectors for spin matrices 
(see Appendix \ref{SU(2) representation} for the construction of basis spinors). 
The basis spinors are given,  in the case of $j=\ell+\frac12$, by
\begin{equation} 
\label{basis l+1/2,l} 
      \Phi^{j(+)}_{m} = \begin{pmatrix}  \sqrt{\frac{j+m}{2j}} Y_{j-\frac12}^{m-\frac12}  \\
             \sqrt{\frac{j-m}{2j}} Y_{j-\frac12}^{m+\frac12}  \end{pmatrix}, 
      \quad |m|\leq j=\ell+\frac12,
\end{equation}
and in the case of $j=(\ell+1)-\frac12$ by 
\begin{equation}
\label{basis l+1/2,l+1} 
     \Phi^{j(-)}_{m}= \begin{pmatrix}  \sqrt{\frac{j-m+1}{2j+2}} Y_{j+\frac12}^{m-\frac12}  \\
             -\sqrt{\frac{j+m+1}{2j+2}} Y_{j+\frac12}^{m+\frac12}  \end{pmatrix}, 
         \quad |m|\leq j=(\ell+1)-\frac12, 
\end{equation}
where both $\Phi^{j(+)}_{m}$ and  $\Phi^{j(-)}_{m}$ are eigenstates of $J_3$ with the eigenvalue $m$. 
We refer to the two-component spinors $\Phi^{j(+)}_{m}$ and  $\Phi^{j(-)}_{m}$ 
as the spin-up and the spin-down states, respectively.   
From \eqref{basis l+1/2,l} and \eqref{basis l+1/2,l+1}, it turns out that 
the the space of four-component spinors as the representation space of 
the total angular momentum $\boldsymbol{J}^2$ with the eigenvalue $j(j+1)$ is given by  

\begin{equation}
\label{basis l+1/2, l+1/2}
    V_{j}^{(+)}\oplus V_{j}^{(-)}=
   {\rm span}\{ \Phi^{j(+)}_{m};\,|m|\leq j=\ell+\frac12\} \oplus 
   {\rm span}\{\Phi^{j(-)}_{m};\,|m|\leq j=(\ell+1)-\frac12\}.  
\end{equation}

In describing the four-component 
spinors, we may express them as 
\begin{equation}
\label{eigenvectors,Phi,Psi}  
    \Phi^j_m= \begin{pmatrix} a \Phi^{j(+)}_{m} \\ b \Phi^{j(-)}_{m} \end{pmatrix} \quad {\rm or} \quad 
     \Psi^j_m= \begin{pmatrix} a' \Phi^{j(-)}_{m} \\ b' \Phi^{j(+)}_{m} \end{pmatrix}, 
     \quad a,b,a',b' \in \mathbb{C}. 
\end{equation} 
Though these four-component spinors are related by the  chiral operator $\sigma_1\otimes \1$, 
we have to distinguish them. 
In fact, as will be seen later, the APS boundary condition is invariant under the chiral operator, 
but the chiral bag boundary condition is not so. 
A way to distinguish them is to refer to parity. By using a formula about spherical harmonics 
(see Appendix~\ref{SU(2) representation}), 
we see that the  $\Phi^{j}_{m}$ and $\Psi^{j}_{m}$ possess opposite parity, 
\begin{equation} 
    \gamma_0^{}\Phi^j_m(-\boldsymbol{r})=(-1)^{j-\frac12}\Phi^j_m(\boldsymbol{r}), \quad 
    \gamma_0^{}\Psi^j_m(-\boldsymbol{r})=(-1)^{j+\frac12}\Psi^j_m(\boldsymbol{r}), \quad 
    |\boldsymbol{r}|=1. 
\end{equation} 

\subsubsection{Some formulae}
\label{some formulae}
We here give some formulae in advance, which will be used in solving the Dirac equation. 
We first need to know how  $\Phi^j_m$ and $\Psi^j_m$ 
transform under the action of the operators $\boldsymbol{\sigma}\cdot\boldsymbol{L}$. 
A straightforward calculation with the formula \eqref{up-down-still} given in Appendix 
\ref{SU(2) representation} provides 
\begin{subequations}
\label{sigma L l+1/2}
\begin{align}
   \boldsymbol{\sigma}\cdot \boldsymbol{L}\Phi^{j(+)}_{m} & = (j-\frac12)\Phi^{j(+)}_{m}, \\
    \boldsymbol{\sigma}\cdot \boldsymbol{L}\Phi^{j(-)}_{m} & = -(j+\frac32)\Phi^{j(-)}_{m}. 
\end{align}
\end{subequations} 
On introducing the operator 
\begin{equation} 
\label{parity-like}
     P =\begin{pmatrix}  \boldsymbol{\sigma}\cdot \boldsymbol{L} + \1 &  \\
           &  -(\boldsymbol{\sigma}\cdot \boldsymbol{L} + \1) \end{pmatrix},  
\end{equation} 
two types of the four-spinors given in \eqref{eigenvectors,Phi,Psi} 
can be  distinguished by the property 
\begin{equation}
\label{S+,S-} 
    P \Phi^j_m=(j+\frac12) \Phi^j_m, \quad P\Psi^j_m=-(j+\frac12)\Psi^j_m. 
\end{equation}
The operator $P$ was introduced in \cite{Dirac} and shown to commute with the Hamiltonian 
of the form $\boldsymbol{\alpha}\cdot \boldsymbol{p} +\mu \beta$. 
Likewise, the Hamiltonian $H_{\mu}$ is shown to commute with $P$; $[H_{\mu}, P]=0$, 
so that each of the eigenvalues $\kappa=\pm(j+\frac12)$ serves as a quantum number. 

We proceed to discuss the transformation of the basis states under the action of 
$\sigma_r=\boldsymbol{\sigma}\cdot \boldsymbol{r}/|\boldsymbol{r}|$.   
A straightforward calculation with the recursion formulae for the associated Legendre functions,  
which are given in Appendix \ref{SU(2) representation}, provides 
\begin{equation}
\label{sigma r l+1/2} 
   \sigma_r \Phi^{j(+)}_{m} = \Phi^{j(-)}_{m}, \quad \sigma_r \Phi^{j(-)}_{m} = \Phi^{j(+)}_{m}. 
\end{equation}
These formulae show that the action of $\sigma_r$ exchanges 
the spin-up and the spin-down states.    

\subsection{Dirac Hamiltonian in spherical polar coordinates}
\label{setting-up in polar coordinates}
Because of the $SU(2)$ symmetry, the Hamiltonian $H_{\mu}$ can be expressed in terms of 
the polar spherical coordinates $(r,\theta,\phi)$ as 
\begin{align} 
   H_{\mu} = -i\gamma_r \partial_r -\frac{i}{r}(\gamma_{\theta}^{}X_{\theta}+\gamma_{\phi}X_{\phi}) 
   +\mu\gamma_0^{}, 
    \label{Dirac_Ham_spherical}
\end{align}
where 
\begin{equation} 
   \partial_r=\frac{\partial}{\partial r}, \quad X_{\theta}=\frac{\partial}{\partial \theta}, \quad 
      X_{\phi}=\frac{1}{\sin \theta}\frac{\partial}{\partial \phi}, 
\end{equation}
and where 
\begin{subequations}
 \begin{align} 
   \gamma_r^{} & = 
       \begin{pmatrix} 0 & -i\sigma_r \\ i\sigma_r & 0 \end{pmatrix}, \quad 
          \sigma_r =\begin{pmatrix} \cos \theta & e^{-i\phi} \sin \theta \\
                  e^{i\phi} \sin \theta & -\cos \theta \end{pmatrix}, \\ 
   \gamma_{\theta}^{} & = 
       \begin{pmatrix} 0 & -i\sigma_{\theta} \\ i\sigma_{\theta} & 0 \end{pmatrix}, \quad 
     \sigma_{\theta}  = \begin{pmatrix} -\sin \theta & e^{-i\phi} \cos \theta \\
                  e^{i\phi} \cos \theta & \sin \theta \end{pmatrix},  \\  
   \gamma_{\phi} & = 
       \begin{pmatrix} 0 & -i\sigma_{\phi} \\ i\sigma_{\phi} & 0 \end{pmatrix}, \quad 
      \sigma_{\phi}  = \begin{pmatrix}  0  & -i e^{-i\phi} \\ i e^{i\phi}  &  0 \end{pmatrix}.  
 \end{align} 
\end{subequations}
Furthermore,  we can show that 
\begin{equation}
\label{polar decomposition}
   H_{\mu}=\gamma_r^{}\Bigl(-i\partial_r I 
    +\frac{i}{r}\begin{pmatrix} \boldsymbol{\sigma}\cdot \boldsymbol{L} &  
                      \mu   \boldsymbol{\sigma}\cdot \boldsymbol{r}     \\ 
     \mu  \boldsymbol{\sigma}\cdot \boldsymbol{r}  &  \boldsymbol{\sigma}\cdot \boldsymbol{L} 
      \end{pmatrix} \Bigr),
\end{equation}
where $I$ denotes the $4\times 4$ identity matrix (see Appendix \ref{SU(2) representation} for the proof).     

\subsection{A boundary operator on the sphere}
\label{boundary operator}
In order to pose the APS boundary condition on the sphere of radius $R$,  we need a boundary operator. 
By restricting the $H_{\mu}$ to the sphere of radius $R$,  we define the operator $A_{\mu}$. 
The boundary operator $B_{\mu}$ is then defined 
to be $B_{\mu}=i\gamma_r A_{\mu}$ after \cite{ABPP} and written out, 
on account of \eqref{polar decomposition}, as 
\begin{equation}
\label{Boundary K(mu)}
  B_{\mu}  = i\gamma_r A_{\mu} = 
  \frac{-1}{R}\begin{pmatrix} 
    \boldsymbol{\sigma}\cdot \boldsymbol{L}  & \mu \boldsymbol{\sigma}\cdot \boldsymbol{r} \\
    \mu \boldsymbol{\sigma}\cdot \boldsymbol{r} & \boldsymbol{\sigma}\cdot \boldsymbol{L} 
    \end{pmatrix} , \quad |\boldsymbol{r}|=R.  
\end{equation} 

In order to see a property of $B_{\mu}$, we use the formula 
\begin{equation}
\label{gamma K} 
     B_{\mu}\gamma_r^{}+\gamma_r^{} B_{\mu}=\frac2R \gamma_r^{}, 
\end{equation} 
which can be verified in a straightforward manner. 
On account of the formula \eqref{gamma K}, we show that the gamma matrix 
$\gamma_r^{}$ has the exchanging property for the $(\pm)$-eigenstates of $B_{\mu}$. 
Let $\Phi^{(+)}$ be an eigenstate of $B_{\mu}$ associated with a positive eigenvalue $\kappa^{(+)}$, 
that is, $B_{\mu} \Phi^{(+)}=\kappa^{(+)}\Phi^{(+)}$. 
Operating this equation with $\gamma_r^{}$ and using the relation \eqref{gamma K}, we obtain 
\begin{equation} 
\label{exchanging sign}
    B_{\mu} \gamma_r^{}\Phi^{(+)}=-\bigl(\kappa^{(+)}-\frac2R\bigr)\gamma_r^{}\Phi^{(+)}. 
\end{equation}
If $\kappa^{(+)}-\frac2R>0$, this equation implies that $\gamma_r \Phi^{(+)}$ is an eigenstate associated with 
a negative eigenvalue of $B_{\mu}$. 
This fact will play a key role to posing the APS boundary condition. 
The $B_{\mu}$ is in principle the same as that used in \cite{APS}.  

\subsection{Feasible solutions to the Dirac equation}
\label{Feasible solutions}
We set out to solve the Dirac equation $H_{\mu}\Phi=E\Phi$ 
in the spherical polar coordinates.  
The APS and the chiral bag boundary conditions will be applied  in 
Sec.~\ref{Eigenvalues} and in Sec.~\ref{Eigenvalues, chiral bag}, respectively,  
to the feasible solutions obtained in this section.    
On account of \eqref{polar decomposition} together with $\gamma_r^2=I$, 
we put the eigenvalue equation in the form 
\begin{equation} 
  \label{gamma r Dirac} 
       i\gamma_r^{}H_{\mu}\Phi = iE \gamma_r^{}\Phi. 
\end{equation}
In applying the separation of variable method, Eq.~\eqref{eigenvectors,Phi,Psi} shows that 
there are two ways to express unknown states, 
\begin{equation}
\label{Phi, Psi} 
    \Phi^j_m=\begin{pmatrix} f(r)\Phi^{j(+)}_m \\ g(r)\Phi^{j(-)}_m \end{pmatrix}, \quad 
     \Psi^j_m=\begin{pmatrix} f(r)\Phi^{j(-)}_m \\ g(r)\Phi^{j(+)}_m \end{pmatrix}, 
\end{equation}
where $f,g$ are unknown functions.  
Writing out Eq.~\eqref{gamma r Dirac} for  $\Phi=\Phi^j_m$, the left one of the two unknown states given 
in \eqref{Phi, Psi},  
and using the formulae about the actions of $\boldsymbol{\sigma}\cdot\boldsymbol{L}$ and $\sigma_r$ 
on $\Phi^{j(+)}_{m}, \Phi^{j(-)}_{m}$ given in Sec.~\ref{some formulae}, 
we obtain the radial equations 
\begin{subequations}
\label{rad eq l+1/2}
  \begin{align} 
       \frac{df}{dr}-\frac{\ell}{r}f & =(E+\mu)g, \\
       \frac{dg}{dr}+\frac{\ell+2}{r}g & = (-E+\mu)f. 
 \end{align} 
\end{subequations} 
In a similar manner, Eq.~\eqref{gamma r Dirac} with $\Phi=\Psi^j_m$ reduces to  
the following radial equations  
\begin{subequations}
\label{rad eq l-1/2}
  \begin{align} 
       \frac{df}{dr}+\frac{\ell+2}{r}f & =(E+\mu)g, \\
       \frac{dg}{dr}-\frac{\ell}{r}g & = (-E+\mu)f. 
 \end{align} 
\end{subequations}   

We are to find solutions to respective radial equations \eqref{rad eq l+1/2} and \eqref{rad eq l-1/2}.  
In what follows, the procedure for solving those equations is divided into three, according to 
$|E|<|\mu|$, $|E|>|\mu|$, and $|E|=|\mu|$, and each of cases is further broken up into two, 
according to whether the unknown state is of the form $\Phi^j_m$ or $\Psi^j_m$ (see \eqref{Phi, Psi}). 

\subsubsection{The case of $|E|<|\mu|$} 
The coupled first-order differential equations \eqref{rad eq l+1/2} for $j=\ell+\frac12$ 
are put together to give rise to the uncoupled second-order differential equations 
\begin{subequations}
\label{decoupled rad eq, l+1/2}
  \begin{align} 
      \frac{d^2f}{dr^2}+\frac{2}{r}\frac{df}{dr}-\bigl(\mu^2-E^2+\frac{\ell(\ell+1)}{r^2}\bigr)f & =0, \\
      \frac{d^2g}{dr^2}+\frac{2}{r}\frac{dg}{dr}-\bigl(\mu^2-E^2+\frac{(\ell+1)(\ell+2)}{r^2}\bigr)g & =0. 
  \end{align}
\end{subequations}
For $\mu^2-E^2>0$, these equations are modified spherical Bessel equations. 
We set 
\begin{equation} 
     \varepsilon = \sqrt{\mu^2-E^2}, \quad |E|<|\mu|.  
\end{equation} 
Then, solutions to \eqref{decoupled rad eq, l+1/2}  take the form 
\begin{equation} 
   f(r)=\frac{C_1}{\sqrt{\varepsilon r}}I_{\ell+\frac12}(\varepsilon r), \quad 
   g(r)=\frac{C_2}{\sqrt{\varepsilon r}}I_{\ell+\frac32}(\varepsilon r), 
\end{equation} 
where $C_1,C_2$ are constants and the $I_{\ell+\frac12},I_{\ell+\frac32}$ are modified Bessel functions 
of the first kind, and where modified Bessel functions of the second kind have been deleted on account of 
the regularity of solutions at the origin $r=0$. 
The constants $C_1,C_2$ are to be related through \eqref{rad eq l+1/2}. 
By using the recursion formula for the modified spherical Bessel functions, 
\begin{subequations}
  \begin{align} 
    \Bigl(\frac{d}{dr}-\frac{\ell}{r}\Bigr)\frac{1}{\sqrt{\varepsilon r}}I_{\ell+\frac12}(\varepsilon r) & = 
          \frac{\varepsilon}{\sqrt{\varepsilon r}}I_{\ell+\frac32}(\varepsilon r), \\
   \Bigl(\frac{d}{dr}+\frac{\ell+2}{r}\Bigr)\frac{1}{\sqrt{\varepsilon r}}I_{\ell+\frac32}(\varepsilon r) & = 
          \frac{\varepsilon}{\sqrt{\varepsilon r}}I_{\ell+\frac12}(\varepsilon r),
  \end{align}
\end{subequations}
we find that 
\begin{equation} 
\label{C1 C2 I}
   \frac{C_1}{E+\mu} = \frac{C_2}{\sqrt{\mu^2-E^2}}. 
\end{equation} 
Since the region defined by $\mu^2-E^2>0$ in the $(E,\mu)$-space consists of two connected components 
distinguished by the sign of $\mu$ (see Fig.~\ref{Et_plane}), this relation leads to 
\begin{subequations}
\label{C1C2 mu+-} 
 \begin{align}
      \frac{C_1}{\sqrt{\mu+E}} & =\frac{C_2}{\sqrt{\mu-E}} \quad {\rm for} \quad \mu>0, \\
       \frac{C_1}{-\sqrt{|\mu+E|}} & =\frac{C_2}{\sqrt{|\mu-E|}} \quad {\rm for} \quad \mu<0. 
 \end{align} 
\end{subequations} 
It turns out that solutions to the Dirac equation for $\Phi^j_m$ take the form 
\begin{subequations}
\label{I, l+1/2} 
 \begin{align} 
   \Phi^j_m = 
    c \begin{pmatrix} \frac{\sqrt{\mu+E}}{\sqrt{\varepsilon r}}I_{\ell+\frac12}(\varepsilon r) \Phi^{j(+)}_{m} \\ 
       \frac{\sqrt{\mu-E}}{\sqrt{\varepsilon r}}I_{\ell+\frac32}(\varepsilon r) \Phi^{j(-)}_{m} \end{pmatrix} 
     \quad   {\rm for} \quad \mu>0,  \label{I, l+1/2, +}\\
   \Phi^j_m = 
    c' \begin{pmatrix} -\frac{\sqrt{|\mu+E|}}{ \sqrt{\varepsilon r}} I_{\ell+\frac12}(\varepsilon r) \Phi^{j(+)}_{m} \\ 
          \frac{\sqrt{|\mu-E|}}{ \sqrt{\varepsilon r}}I_{\ell+\frac32}(\varepsilon r) \Phi^{j(-)}_{m} \end{pmatrix} 
      \quad  {\rm for} \quad \mu<0.    \label{I, l+1/2, -}
 \end{align} 
\end{subequations} 

We proceed to \eqref{rad eq l-1/2}. 
Carrying out the same procedure, 
we eventually find that solutions to the Dirac equation for $\Psi^j_m$ take the form 
\begin{subequations} 
\label{I, l-1/2}
 \begin{align} 
  \Psi^j_m = c \begin{pmatrix} \frac{\sqrt{\mu+E}}{\sqrt{\varepsilon r}}I_{\ell+\frac32}(\varepsilon r) \Phi^{j(-)}_{m} \\ 
      \frac{\sqrt{\mu-E}}{\sqrt{\varepsilon r}}I_{\ell+\frac12}(\varepsilon r) \Phi^{j(+)}_{m} \end{pmatrix} \quad 
      {\rm for} \quad \mu>0,     \label{I, l-1/2, +}  \\
 \Psi^j_m= c' \begin{pmatrix} -\frac{\sqrt{|\mu+E|}}{ \sqrt{\varepsilon r}} I_{\ell+\frac32}(\varepsilon r) \Phi^{j(-)}_{m} \\ 
       \frac{\sqrt{|\mu-E|}}{ \sqrt{\varepsilon r}}I_{\ell+\frac12}(\varepsilon r) \Phi^{j(+)}_{m} \end{pmatrix}
    \quad    {\rm for} \quad \mu<0.  \label{I, l-1/2, -}
 \end{align} 
\end{subequations}

\subsubsection{The case of $|E|>|\mu|$}
The coupled first-order differential equations \eqref{rad eq l+1/2} for $j=\ell+\frac12$ 
are put together to provide the uncoupled second-order differential equations 
\begin{subequations}
\label{decoupled rad eq, l+1/2 I}
  \begin{align} 
      \frac{d^2f}{dr^2}+\frac{2}{r}\frac{df}{dr}+\bigl(E^2-\mu^2 -\frac{\ell(\ell+1)}{r^2}\bigr)f & =0, \\
      \frac{d^2g}{dr^2}+\frac{2}{r}\frac{dg}{dr}+\bigl(E^2-\mu^2 -\frac{(\ell+1)(\ell+2)}{r^2}\bigr)g & =0. 
  \end{align}
\end{subequations}
For $E^2-\mu^2>0$, these equations are spherical Bessel equations. 
On introducing the parameter
\begin{equation} 
     \beta = \sqrt{E^2-\mu^2}, \quad |E|>|\mu|,   
\end{equation} 
solutions to \eqref{decoupled rad eq, l+1/2 I} are put in the form 
\begin{equation} 
   f(r)=\frac{C_1}{\sqrt{\varepsilon r}}J_{\ell+\frac12}(\beta r), \quad 
   g(r)=\frac{C_2}{\sqrt{\varepsilon r}}J_{\ell+\frac32}(\beta r), 
\end{equation} 
where $C_1,C_2$ are constants and the $J_{\ell+\frac12},J_{\ell+\frac32}$ are Bessel functions,  
and where  Neumann functions have been deleted on account of the regularity of solutions 
at the origin $r=0$. 
The constants $C_1,C_2$ are related through \eqref{rad eq l+1/2}. 
By using the recursion formula for the spherical Bessel functions, 
\begin{subequations} 
 \begin{align} 
    \Bigl(\frac{d}{dr}+\frac{\ell+2}{r}\Bigr)\frac{1}{\sqrt{\beta r}}J_{\ell+\frac32}(\beta r) & = 
     \frac{\beta}{\sqrt{\beta r}} J_{\ell+\frac12}(\beta r), \\
    \Bigl(\frac{d}{dr}-\frac{\ell}{r}\Bigr)\frac{1}{\sqrt{\beta r}}J_{\ell+\frac12}(\beta r) & = 
     -\frac{\beta}{\sqrt{\beta r}} J_{\ell+\frac32}(\beta r). 
 \end{align} 
\end{subequations}
we find from \eqref{rad eq l+1/2} that 
\begin{equation} 
\label{C1 C2}
   \frac{C_1}{E+\mu} = -\frac{C_2}{\sqrt{E^2-\mu^2}}. 
\end{equation} 
Since the region defined by $E^2-\mu^2>0$ in the $(E,\mu)$-space is  broken up into two, 
according to the sign of $E$ (see Fig.~\ref{Et_plane}), this relation leads to 
\begin{subequations} 
 \begin{align}
      \frac{C_1}{\sqrt{E+\mu}} & =-\frac{C_2}{\sqrt{E-\mu}} \quad {\rm for} \quad E>0, \\
       \frac{C_1}{\sqrt{|E+\mu|}} & =\frac{C_2}{\sqrt{|E-\mu|}} \quad {\rm for} \quad E<0. 
 \end{align} 
\end{subequations} 
It then follows that solutions to the Dirac equation for $\Phi^j_m$ take the form 
\begin{subequations} 
\label{sol, j=l+1/2} 
 \begin{align} 
   \Phi^j_m = c \begin{pmatrix} -\frac{\sqrt{E+\mu}}{\sqrt{\beta r}}J_{\ell+\frac12}(\beta r) \Phi^{j(+)}_{m} \\ 
                    \frac{\sqrt{E-\mu}}{\sqrt{\beta r}} J_{\ell+\frac32}(\beta r) \Phi^{j(-)}_{m} \end{pmatrix} 
         \quad  {\rm for} \quad E>0,  \label{sol, j=l+1/2, E>0}  \\
   \Phi^j_m= c' \begin{pmatrix} \frac{\sqrt{|E+\mu|}}{ \sqrt{\beta r}} J_{\ell+\frac12}(\beta r) \Phi^{j(+)}_{m} \\ 
                    \frac{\sqrt{|E-\mu|}}{ \sqrt{\beta r}} J_{\ell+\frac32}(\beta r) \Phi^{j(-)}_{m} \end{pmatrix} 
         \quad  {\rm for} \quad E<0.  \label{sol, j=l+1/2, E<0}
 \end{align} 
\end{subequations}

We turn to the radial equations \eqref{rad eq l-1/2}. 
Following a similar procedure,  
we verify that solutions to the Dirac equation for $\Psi^j_m$ take the form 
\begin{subequations} 
\label{sol, j=l-1/2}
 \begin{align} 
   \Psi^j_m = c \begin{pmatrix} \frac{\sqrt{E+\mu}}{\sqrt{\beta r}}J_{\ell+\frac32}(\beta r) \Phi^{j(-)}_{m} \\ 
                    \frac{\sqrt{E-\mu}}{\sqrt{\beta r}}J_{\ell+\frac12}(\beta r) \Phi^{j(+)}_{m} \end{pmatrix} \quad 
      {\rm for} \quad E>0,    \label{sol, j=l-1/2, E>0}    \\
   \Psi^j_m = c' \begin{pmatrix} -\frac{\sqrt{|E+\mu|}}{ \sqrt{\beta r}} J_{\ell+\frac32}(\beta r) \Phi^{j(-)}_{m} \\ 
                    \frac{\sqrt{|E-\mu|}}{ \sqrt{\beta r}} J_{\ell+\frac12}(\beta r) \Phi^{j(+)}_{m} \end{pmatrix}
    \quad    {\rm for} \quad E<0.     \label{sol, j=l-1/2, E<0} 
 \end{align} 
\end{subequations}

\subsubsection{The case of $|E|=|\mu|$} 
Equations \eqref{decoupled rad eq, l+1/2} and \eqref{decoupled rad eq, l+1/2 I} are valid for $|E|=|\mu|$ 
and thereby reduce to 
\begin{subequations}
\label{decoupled rad eq, l+1/2,critical}
  \begin{align} 
      \frac{d^2f}{dr^2}+\frac{2}{r}\frac{df}{dr}-\frac{\ell(\ell+1)}{r^2}f & =0, \\
      \frac{d^2g}{dr^2}+\frac{2}{r}\frac{dg}{dr}-\frac{(\ell+1)(\ell+2)}{r^2}g & =0. 
  \end{align}
\end{subequations}
Solving these equations with the boundary condition that $f$ and $g$ be bounded as $r\to 0$, we obtain 
\begin{equation}
\label{f,g,j=l+1/2}
     f(r)=ar^{\ell}, \quad g(r)=br^{\ell+1}, 
\end{equation} 
where $a$ and $b$ are constants. 
These constants should be related through \eqref{rad eq l+1/2}.   
If we impose the condition that $E=-\mu$, Eq.~\eqref{rad eq l+1/2}  reduces to 
\begin{subequations}
\label{rad eq l+1/2,critical}
  \begin{align} 
       \frac{df}{dr}-\frac{\ell}{r}f & =0, \\
       \frac{dg}{dr}+\frac{\ell+2}{r}g & =2\mu f. 
 \end{align} 
\end{subequations}
Eqs.~\eqref{f,g,j=l+1/2} and \eqref{rad eq l+1/2,critical} are put together to give 
the ratio of $a$ and $b$, and thereby 
the critical solution for $\Phi^j_m$ are shown to be  
\begin{equation} 
\label{cr,Phi_jm}
    \Phi_{\rm cr}=c\begin{pmatrix} (2\ell+3)r^{\ell}\Phi^{j(+)}_{m} \\ 
         2\mu r^{\ell+1}\Phi^{j(-)}_{m} \end{pmatrix} \quad {\rm for} \quad E=-\mu, 
\end{equation} 
where $c$ is a constant. 
In contrast to this, if we impose the condition that $E=\mu$, Eq.~\eqref{rad eq l+1/2} 
reduces to 
\begin{subequations} 
\label{rad eq l+1/2,critical,E=mu}
     \begin{align} 
       \frac{df}{dr}-\frac{\ell}{r}f & =2\mu g, \\
       \frac{dg}{dr}+\frac{\ell+2}{r}g & =0.  
    \end{align} 
\end{subequations}
Eqs.~\eqref{f,g,j=l+1/2} and \eqref{rad eq l+1/2,critical,E=mu} are put together to provides the solution 
\begin{equation} 
\label{Phi'}
     \Phi'_{\rm cr}=c'\begin{pmatrix}  r^{\ell}\Phi^{j(+)}_m \\ 0 \end{pmatrix} \quad {\rm for} \quad E=\mu. 
\end{equation}

If we start with \eqref{rad eq l-1/2}, then we eventually find the following critical solutions, 
\begin{equation} 
\label{cr,Psi_jm}
     \Psi_{\rm cr}= c \begin{pmatrix} 2\mu r^{\ell+1} \Phi^{j(-)}_{m} \\ 
          (2\ell+3)r^{\ell} \Phi^{j(+)}_{m} \end{pmatrix} 
       \quad {\rm for} \quad E=\mu, 
\end{equation} 
and 
\begin{equation}
\label{Psi'} 
    \Psi'_{\rm cr}= c' \begin{pmatrix} 0 \\ r^{\ell} \Phi^{j(+)}_m \end{pmatrix} \quad {\rm for} \quad E=-\mu, 
\end{equation} 
where $c$ and $c'$ are constants.  

In particular, in the case of $E=\mu=0$, the critical solutions obtained above 
reduce to \eqref{Phi'} and \eqref{Psi'}, which are called zero modes.

\begin{figure}[htbp]  
\begin{center}  
\includegraphics[width=0.4\columnwidth]{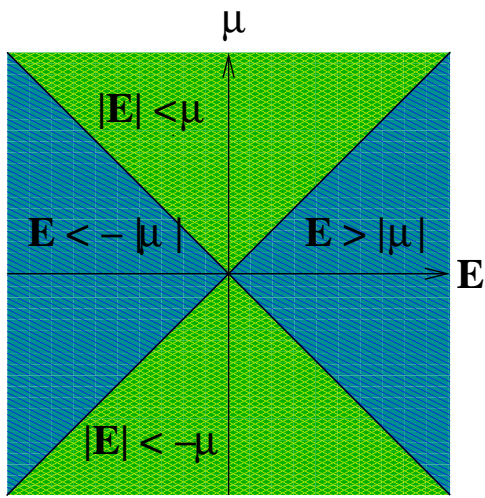}  
\end{center}  
\caption{Splitting of the $E$-$\mu$ plane into domains with $|E|>|\mu|$  
and with $|E|<|\mu|$ and the boundary lines. 
Each  domain consists of two connected components.    
\label{Et_plane} }  
\end{figure}  

So far we have found feasible solutions to the Dirac equation without referring to boundary conditions, 
which are listed in the following table: 
\begin{equation} 
 \begin{array}{c|c|c|c|}
       & |E|<|\mu| & |E|>|\mu| & |E|=|\mu|   \\ \hline 
  \Phi^j_m &  \eqref{I, l+1/2} & \eqref{sol, j=l+1/2} &  \eqref{cr,Phi_jm} / \eqref{Phi'}  \\  
  \Psi^j_m  &  \eqref{I, l-1/2}  &   \eqref{sol, j=l-1/2}   &  \eqref{cr,Psi_jm} / \eqref{Psi'}  
 \end{array}
\end{equation} 
These  solutions are assigned to each of connected domains and the boundary lines 
shown in Fig.~\ref{Et_plane}.

\subsection{The APS boundary condition on the sphere}
\label{APS bdy cond}
To start with, we give a basic equation on which boundary conditions are set up.   
Let $V$ be a domain bounded by a surface $S$. 
The inner products for multi-component functions on $V$ and on $S$ are defined as usual to be   
\begin{equation}  
    \langle \Phi, \Psi \rangle_V=\int_V \sum \overline{\Phi}_{\alpha}\Psi_{\alpha}dV, \quad   
    \langle \phi, \psi \rangle_S=\int_S \sum \overline{\phi}_{\alpha}\psi_{\alpha}dS,   
\end{equation}  
respectively, where $\overline{\Phi}_{\alpha}, \overline{\phi}_{\alpha}$ are the complex conjugates 
of $\Phi_{\alpha}$ and $\phi_{\alpha}$, respectively.    
The Green formula for the Dirac operator $H_{\mu}$ is put in the form    
\begin{equation}  
\label{Dirac-Green}   
    \langle \Phi, H_{\mu} \Psi\rangle_V - \langle H_{\mu} \Phi, \Psi \rangle_V =   
    -i \langle \phi, \boldsymbol{\gamma}\cdot \boldsymbol{n} \,\psi \rangle_S,   
\end{equation}   
where $\phi=\Phi|_S, \psi=\Psi|_S$ and where $\boldsymbol{n}$ denotes the outward unit normal vector 
to $S$ \cite{IZ2015}.   
In general, boundary conditions should be determined so that the boundary integral, the right-hand side of 
the above equation, may vanish \cite{ABPP}.   

If we can find a boundary operator $B$ on $S$ such that (i) $B$ has no zero eigenvalue and 
(ii) the $\boldsymbol{\gamma}\cdot \boldsymbol{n}$ transforms eigenstates associated with positive 
eigenvalues of $B$ to those associated with negative ones, and vice versa, and if 
the boundary functions $\phi=\Phi|_S, \psi=\Psi|_S$ are required to belong to the eigenspace 
associated with either positive or negative eigenvalues, then 
the right-hand side of \eqref{Dirac-Green} vanishes,  
so that the $H_{\mu}$ is a symmetric operator. 
Furthermore, with some Sobolev conditions, the $H_{\mu}$ becomes self-adjoint \cite{IPP}. 
The operator $B_{\mu}$ we have already defined in \eqref{Boundary K(mu)} is such an operator.  
In fact, we have shown in Eq.~\eqref{exchanging sign} that 
the $\gamma_{r}^{}=\boldsymbol{\gamma}\cdot \boldsymbol{n}$ exchanges eigenstates associated with 
positive and negative eigenvalues, if $\kappa^{(+)}-\frac{2}{R}>0$.  
We will soon show that $\gamma_r^{}$ indeed exchanges eigenstates associated with positive and negative 
eigenvalues (see \eqref{gamma +-}).     

We shall solve eigenvalue problem for $B_{\mu}$ in order to describe the APS boundary condition in an explicit form. 
Since the initial Dirac operator $H_{\mu}$ admits an $SU(2)$ symmetry, 
the $B_{\mu}$, which is defined by restricting the $H_{\mu}$ to the sphere of radius $R$ 
and by multiplying the matrix $i\gamma_r^{}$,  is expected to admit the same symmetry. 
In fact, a straightforward calculation shows that $B_{\mu}$ commutes with $J_k$; 
\begin{equation} 
 [B_{\mu}, J_k] =0. 
\end{equation}
On account of the $SU(2)$ symmetry of $B_{\mu}$, 
the eigenvalue problem for $B_{\mu}$ reduces to subproblems on the eigenspaces of $\boldsymbol{J}^2$.

\subsubsection{Eigenspaces of $B_{\mu}$ with $\mu\neq 0$}
Any state associated with the eigenvalues $J_3=m$ and $\boldsymbol{J}^2=j(j+1)$ is put in the form 
given in \eqref{eigenvectors,Phi,Psi}. 
We first take the left of \eqref{eigenvectors,Phi,Psi}. 
Then, by the use of \eqref{Boundary K(mu)} together with \eqref{sigma L l+1/2} and \eqref{sigma r l+1/2}, 
the eigenvalue equation $B_{\mu}\Phi^j_m=\lambda \Phi^j_m$ reduces to the matrix eigenvalue equation 
\begin{equation} 
    \frac{-1}{R}\begin{pmatrix} \ell & \mu R \\ \mu R & -(\ell+2) \end{pmatrix}  \begin{pmatrix} a \\ b \end{pmatrix} 
    = \lambda \begin{pmatrix} a \\ b \end{pmatrix}, \quad j=\ell+\frac12. 
\end{equation} 
From this equation, the eigenvalues are easily found to be  
\begin{equation}
    \lambda^{(\pm)}_{j}=\frac{1}{R}(1\pm \sqrt{(\ell+1)^2+(\mu R)^2}),
\end{equation} 
and the associated eigenvectors are easily obtained, and thereby   
the associated eigenstates of $B_{\mu}$ are expressed, within constant multiples, as 
\begin{subequations}
\label{Phi(+,-)}
 \begin{align} 
    \Phi^{(+)}_{m,j} &  =\begin{pmatrix}  
       -\mu R \Phi^{j(+)}_{m} \\ (\ell+1 +\sqrt{(\ell+1)^2+(\mu R)^2})\Phi^{j(-)}_{m} \end{pmatrix}
     \; {\rm for} \;\,
         \lambda^{(+)}_{j}=\frac{1}{R}(1+\sqrt{(\ell+1)^2+(\mu R)^2}), \label{Phi,k(+),l+1/2} \\
  \Phi^{(-)}_{m,j}  & = \begin{pmatrix}  
       -\mu R \Phi^{j(+)}_{m} \\ (\ell+1 -\sqrt{(\ell+1)^2+(\mu R)^2})\Phi^{j(-)}_{m} \end{pmatrix} 
     \; {\rm for} \;\,
         \lambda^{(-)}_{j}=\frac{1}{R}(1-\sqrt{(\ell+1)^2+(\mu R)^2}).  \label{Phi,k(-),l+1/2}
 \end{align}
\end{subequations}
It is to be noted that 
\begin{equation} 
     \lambda^{(+)}_{j}>0, \quad \lambda^{(-)}_{j}<0, \quad {\rm if}\quad \mu\neq 0. 
\end{equation} 
Though $\Phi^{(\pm)}_{m,j}$ are not normalized, they are orthogonal to each other, 
as is easily verified; 
\begin{equation} 
    \langle \Phi^{(+)}_{m,j}, \Phi^{(-)}_{m,j}\rangle_{S^2}=0. 
\end{equation}

We proceed to treat $\Psi^j_m$, the right of Eq.~\eqref{eigenvectors,Phi,Psi}. 
Then, in a similar manner to the above,  
the eigenvalue equation $B_{\mu}\Psi^j_m=\kappa \Psi^j_m$ reduces to 
\begin{equation} 
    \frac{-1}{R}\begin{pmatrix} -(\ell+2) & \mu R \\ \mu R & \ell \end{pmatrix}  
    \begin{pmatrix} a' \\ b' \end{pmatrix} 
    = \kappa \begin{pmatrix} a' \\ b' \end{pmatrix}, \quad j=\ell+\frac12, 
\end{equation} 
from which one  easily obtains the eigenvalues 
\begin{equation}
    \kappa^{(\pm)}_{j}=\frac{1}{R}\bigl(1\pm \sqrt{(\ell+1)^2+(\mu R)^2}\bigr), 
\end{equation} 
and the associated eigenvectors, and thereby the associated eigenstates  
of $B_{\mu}$ are expressed as 
\begin{subequations}
\label{Psi(+,-)}
 \begin{align} 
    \Psi^{(+)}_{m,j} &  =\begin{pmatrix}  
       -\mu R \Psi^{j(-)}_{m} \\ (-(\ell+1) +\sqrt{(\ell+1)^2+(\mu R)^2})\Phi^{j(+)}_{m} \end{pmatrix}
     \; {\rm for} \;\,
         \kappa^{(+)}_{j}=\frac{1}{R}(1+\sqrt{(\ell+1)^2+(\mu R)^2}),    \label{Psi,k(+),l-1/2}     \\
  \Phi^{(-)}_{m,j}  & = \begin{pmatrix}  
       -\mu R \Phi^{j(-)}_{m} \\ (-(\ell+1) -\sqrt{(\ell+1)^2+(\mu R)^2})\Phi^{j(+)}_{m} \end{pmatrix} 
     \; {\rm for} \;\,
         \kappa^{(-)}_{j}=\frac{1}{R}(1-\sqrt{(\ell+1)^2+(\mu R)^2}),    \label{Psi,k(-),l-1/2}       
 \end{align}
\end{subequations}
where 
\begin{equation} 
    \kappa^{(+)}_{j}>0, \quad \kappa^{(-)}_{j}<0.  
 \end{equation}
Though these eigenstates are not normalized, they are orthogonal to each other 
\begin{equation} 
    \langle \Psi^{(+)}_{m,j},   \Psi^{(-)}_{m,j}\rangle_{S^2}=0. 
\end{equation} 
Furthermore, the four-component spinors $\Phi^{(\pm)}_{m,j}$ and $\Psi^{(\pm)}_{m,j}$ 
are mutually orthogonal; 
\begin{equation} 
   \langle  \Phi^{(+)}_{m,j}, \Psi^{(\pm)}_{m,j}\rangle_{S^2}= 
   \langle  \Phi^{(-)}_{m,j}, \Psi^{(\pm)}_{m,j}\rangle_{S^2}=0. 
\end{equation}

So far, for $\mu\neq 0$, we have obtained eigenspaces of $B_{\mu}$ 
associated with the eigenvalues $\lambda^{(\pm)}_{j}$ and $\kappa^{(\pm)}_j$,  
\begin{subequations}
 \begin{align} 
      \mathcal{H}^{(+)}_{j}(S^2) & ={\rm span}\{ \Phi^{(+)}_{m,j};\, |m|\leq j=\ell+\frac12 \} 
       \quad {\rm with} \quad \lambda^{(+)}_{j},  \\
      \mathcal{K}^{(+)}_{j}(S^2) & ={\rm span}\{ \Psi^{(+)}_{m,j};\, |m|\leq j=\ell+\frac12 \} 
       \quad {\rm with} \quad \kappa^{(+)}_{j},  \\  
      \mathcal{H}^{(-)}_{j}(S^2) & ={\rm span}\{ \Phi^{(-)}_{m,j};\, |m|\leq j=\ell+\frac12 \} 
       \quad {\rm with} \quad \lambda^{(-)}_{j},  \\
      \mathcal{K}^{(-)}_{j}(S^2) & ={\rm span}\{ \Psi^{(-)}_{m,j};\, |m|\leq j=\ell+\frac12 \} 
       \quad {\rm with} \quad \kappa^{(-)}_{j}. 
 \end{align}
\end{subequations}

\subsubsection{The APS boundary condition with $\mu\neq 0$}
The Hilbert space of four-component spinors on the sphere $S^2$  
is decomposed into the direct sum of the subspaces  
associated with positive and negative eigenvalues of $B_{\mu}$.   We now show that 
the $(\pm)$-eigenspaces mutually exchange under the action of $\gamma_r^{}$; 
\begin{equation}
\label{gamma +-} 
     \gamma_r^{}\mathcal{H}^{(+)}_{j}(S^2)=\mathcal{H}^{(-)}_{j}(S^2), \quad 
     \gamma_r^{}\mathcal{K}^{(+)}_{j}(S^2)=\mathcal{K}^{(-)}_{j}(S^2). 
\end{equation} 
To this end, we have only to show that 
\begin{equation}
\label{exchanging property +-}  
   \gamma_r^{}\Phi^{(+)}_{m,j}={\rm const}\,\Phi^{(-)}_{m,j}, \quad 
   \gamma_r^{}\Psi^{(+)}_{m,j}={\rm const}\,\Psi^{(-)}_{m,j}, 
\end{equation}
and we can easily verify these equations by using \eqref{sigma r l+1/2}.  

Now we are in a position to explicitly describe the APS boundary condition with $\mu\neq 0$: 
According to whether feasible solutions are of the form $\Phi^j_m$ or $\Psi^j_m$, 
the APS boundary condition is given by 
\begin{equation}
\label{APS(+)} 
    \Phi|_{r=R}\in \mathcal{H}^{(+)}_{j}(S^2) \quad {\rm or} \quad 
     \Psi|_{r=R}\in \mathcal{K}^{(+)}_{j}(S^2), 
\end{equation} 
or  by 
\begin{equation} 
\label{APS(-)}
    \Phi|_{r=R}\in \mathcal{H}^{(-)}_{j}(S^2) \quad {\rm or} \quad 
  \Psi|_{r=R}\in \mathcal{K}^{(-)}_{j}(S^2). 
\end{equation}

\subsubsection{Eigenspaces of $B_0$}
The remaining task is to deal with the boundary condition for $\mu=0$. 
The boundary operator $B_{\mu}$ with $\mu=0$ takes the form 
\begin{equation} 
    B_0=\frac{-1}{R}\begin{pmatrix} \boldsymbol{\sigma}\cdot \boldsymbol{L} & 0 \\
           0 & \boldsymbol{\sigma}\cdot \boldsymbol{L}  \end{pmatrix}. 
\end{equation} 
For a spinor given  in the left of \eqref{eigenvectors,Phi,Psi},  
we write out the eigenvalue equation $B_0\Phi^j_m=\lambda\Phi^j_m$ with $j=\ell+\frac12$ to obtain 
\begin{equation} 
   \frac{-1}{R}\begin{pmatrix} \ell & 0 \\ 0 & -(\ell+2) \end{pmatrix} \begin{pmatrix} a \\ b \end{pmatrix} = 
    \lambda \begin{pmatrix} a \\ b \end{pmatrix}. 
\end{equation} 
The eigenvalues and eigenvectors are easily obtained, and thereby 
the eigenvalues and the associated eigenstates of $B_0$ with $j=\ell+\frac12$ are expressed as 
\begin{subequations} 
\label{eigenstates K0, l+1/2}
 \begin{align} 
    \Phi^{(-)}_0 & = c\begin{pmatrix}  \Phi^{j(+)}_{m} \\ 0 \end{pmatrix} \quad {\rm for} \quad 
              \lambda^{(-)}_+ =-\frac{\ell}{R}, \\
    \Phi^{(+)}_0 & =c'\begin{pmatrix}  0 \\ \Phi^{j(-)}_{m} \end{pmatrix} \quad {\rm for} \quad 
              \lambda^{(+)}_+  = \frac{\ell+2}{R}, 
 \end{align} 
\end{subequations} 
where $\lambda^{(\pm)}_+$ are the limits of $\lambda^{(\pm)}_{\ell+\frac12}$ as $\mu\to 0$, respectively.  
Comparison between \eqref{eigenstates K0, l+1/2} with \eqref{Phi'} shows that 
if evaluated on the boundary, the zero mode $\Phi'_{\rm cr}$ is an eigenstate associated 
with the eigenvalue $\lambda^{(-)}_+$. 

For a spinor given in the right of \eqref{eigenvectors,Phi,Psi}, 
the eigenvalue equation $B_0\Psi^j_m =\kappa \Psi^j_m$ is solved in a similar manner, and eventually 
the eigenstates of $B_0$ are shown to be expressed as 
\begin{subequations}
\label{eigenstates K0, l-1/2}
 \begin{align} 
    \Psi^{(+)}_0 & = c\begin{pmatrix}  \Phi^{j(-)}_{m} \\ 0 \end{pmatrix} \quad {\rm for} \quad 
              \kappa^{(+)}_- =\frac{\ell+2}{R}, \\
    \Psi^{(-)}_0 & =c'\begin{pmatrix}  0 \\ \Phi^{j(+)}_{m} \end{pmatrix} \quad {\rm for} \quad 
              \kappa^{(-)}_-  = \frac{-\ell}{R}, 
 \end{align} 
\end{subequations}
where $\kappa^{(\pm)}_-$ are the limits of $\kappa^{(\pm)}_{j}$ as $\mu\to 0$, respectively.    
Comparison of  \eqref{eigenstates K0, l-1/2} with \eqref{Psi'} shows that 
if evaluated on the boundary, the zero mode $\Psi'_{\rm cr}$ is an eigenstate associated 
with the eigenvalue $\kappa^{(-)}_-$.

\subsection{The chiral bag boundary condition on the sphere}
We now make a review of the chiral bag boundary condition, according to \cite{ABPP}. 
Any four-component spinor is decomposed into the sum of chiral components, 
\begin{equation} 
     \Phi=\Phi_+ + \Phi_-, \quad     \Phi_{\pm}:=\frac12( I\pm  \boldsymbol{\gamma}\cdot \boldsymbol{n})\Phi. 
\end{equation} 
As is easily seen, these quantities satisfy    
\begin{equation} 
   \boldsymbol{\gamma}\cdot \boldsymbol{n}\,\Phi_+=\Phi_+, \quad 
   \boldsymbol{\gamma}\cdot \boldsymbol{n}\,\Phi_-=-\Phi_- , 
    \quad        \langle \Psi_+, \Phi_-\rangle  = 0, 
\end{equation} 
where $\Psi_+$ is defined in a similar manner. 
Using these properties, we can put the right-hand side of \eqref{Dirac-Green} in the form 
\begin{align} 
  -i\langle \phi, \boldsymbol{\gamma}\cdot \boldsymbol{n}\,\psi \rangle_S 
  &  =     -i\langle \phi_+, \psi_+ \rangle_S +  i\langle \phi_-, \psi_- \rangle_S. 
  \label{chiral decomp}
\end{align} 
If the chiral components $\psi_{\pm}$ of  $\psi=\Psi|_S$ and $\phi_{\pm}$ of $\phi=\Phi|_S$ 
are related by 
\begin{equation}
\label{chiral bag bdy cond} 
     \psi_-=U\gamma^{}_{0}\psi_+, \quad  \phi_-=U\gamma^{}_{0}\phi_+, 
\end{equation}
respectively, where $U$ is any unitary operator acting on spinors defined on the boundary and 
further commutes with $\boldsymbol{\gamma}\cdot \boldsymbol{n}$, then those components satisfy  
\begin{align} 
 \langle \phi_-, \psi_- \rangle_S  = \langle \phi_+, \psi_+ \rangle_S, 
\end{align} 
so that the right-hand side of \eqref{chiral decomp} vanishes. 
Eq.~\eqref{chiral bag bdy cond} is called the chiral bag boundary condition. 

For the sake of simplicity, we may assume that the unitary operator $U$ is a local one, which is 
expressed as a finite order matrix acting fiberwise on spinors.  
A very simple unitary matrix is given by 
\begin{equation} 
    U=e^{2i\arctan e^{\lambda}} I,
\end{equation}
where the $\lambda$ is a real parameter. 
Then, the chiral bag boundary condition \eqref{chiral bag bdy cond}  for $\psi=\Psi|_S$ reads 
\begin{equation}
\label{chiral bag bdy cond lambda} 
   \boldsymbol{\gamma}\cdot \boldsymbol{n}\,\psi = -ie^{\lambda \gamma_{0}^{}}\gamma_{0}^{}\psi. 
\end{equation}

\subsection {Currents on the boundary}   
\label{current}
In this section, we show that the APS and the chiral bag boundary conditions adopted in the present paper 
involves physically reasonable consequences for currents on the boundary. 
As is well known, staring with the time-dependent Dirac equation in a natural unit system ($c=\hbar=1$), 
one finds the continuity equation, after a straightforward calculation,
\begin{equation} 
\label{conservation law}
    \frac{\partial}{\partial t}(\psi^{\dag}\psi) + 
     \sum_{k=1}^3 \frac{\partial}{\partial x_k}(\psi^{\dag}\gamma^{}_k \psi) =0. 
\end{equation} 
The quantities $\psi^{\dag}\gamma^{}_k \psi$ are interpreted as components of the current vector.  
In particular, the radial component of the current vector is given by 
\begin{equation} 
    \psi^{\dag} \gamma_{r}^{} \psi. 
\end{equation} 
Since we work with the time-independent states, the continuity equation reduces to 
${\rm div}(\psi^{\dag} \boldsymbol{\gamma} \psi)=0$. 
Our objective is to show that 
$\psi^{\dag} \gamma_{r}^{} \psi=0$ on the boundary under both the APS and the chiral 
bag boundary conditions. 

We first deal with the APS boundary condition. 
According to \cite{AkBe, ABPP, McC-F}, a boundary operator $B$ used for describing  
a boundary condition is required to anticommute with 
$\gamma_r^{}=\boldsymbol{\gamma}\cdot \boldsymbol{n}$, {\it i.e.}, $B\gamma_r^{}+\gamma_r^{} B=0$, 
where $\boldsymbol{n}$ denotes the outward unit normal to the boundary.  
This requirement seems to be adopted in order that the current normal to the 
boundary vanishes.   
However, as was shown in \eqref{gamma K}, our boundary operator $B_{\mu}$ does not 
anticommute with $\gamma_r^{}$.  
We can show that the current normal to the boundary 
vanishes under the APS boundary condition in spite of the non-anticommutation of $B_{\mu}$ 
with $\gamma_r^{}$.  
We recall that the eigenstates of $H_{\mu}$ evaluated on the boundary are proportional to 
the eigenstates of $B_{\mu}$, which are given in \eqref{Phi(+,-)} and \eqref{Psi(+,-)}. 
With this in mind, we work with the current of the eigenstates $\Phi^{(\pm)}_{m,j}$ 
in the direction of the outward normal to the boundary.  
A straightforward calculation provides 
\begin{equation}
\label{r-current}
 \Phi^{(\pm)\dag}_{m,j} \gamma_r^{} \Phi^{(\pm)}_{m,j} = 
 i\mu R\bigl(\ell+1\pm \sqrt{(\ell+1)^2+(\mu R)^2}\bigr)
   \bigl(\Phi^{j(+)\dag}_{m} \Phi^{j(+)}_{m} - \Phi^{j(-)\dag}_{m} \Phi^{j(-)}_{m}\bigr). 
\end{equation}
By using the formulae \eqref{sigma r l+1/2}, we can show that 
\begin{equation}
\label{r-recurrence}
   \Phi^{j(-)\dag}_{m} \Phi^{j(-)}_{m} = \Phi^{j(+)\dag}_{m} \Phi^{j(+)}_{m}, 
\end{equation}
so that Eq.~\eqref{r-current} becomes  
\begin{equation} 
\label{r-current phi=0}
    \Phi^{(\pm)}_{m,j} \gamma_r^{} \Phi^{(\pm)}_{m,j} =0.  
\end{equation}
In a similar manner,  we can verify that 
\begin{equation}
\label{r-current psi=0} 
    \Psi^{(\pm)\dag}_{m,j} \gamma_r^{} \Psi^{(\pm)}_{m,j} =0.  
\end{equation}
Equations \eqref{r-current phi=0} and \eqref{r-current psi=0}  
imply that the current normal to the boundary vanishes for any eigenstate associated 
with the eigenvalues $\lambda^{(\pm)}_{j}$ and $\kappa^{(\pm)}_j$, and hence vanishes for every eigenstate  
satisfying the APS boundary condition. 

As for the chiral bag boundary condition, it is rather easy to see  
that  the normal component of the current vector $\psi^{\dag}\boldsymbol{\gamma}\psi$ 
on the boundary sphere vanishes.  
In fact, from \eqref{chiral bag bdy cond lambda}, one can easily verify that 
\begin{equation}
     \psi^{\dag}(\boldsymbol{\gamma}\cdot \boldsymbol{n})\psi =0. 
\end{equation}

\subsection{The eigenvalues of the Dirac operator with the APS boundary condition}
\label{Eigenvalues}
We are now in a position to find the eigenvalues of the Dirac operator with the APS boundary condition. 
As we have already obtained feasible solutions, our task is to apply the APS boundary condition to 
those solutions. 
According to the classification of  feasible solutions into three classes depending on (i) $|E|<|\mu|$, 
(ii)  $|E|>|\mu|$,  and (iii) $|E|=|\mu|$,    
we refer to the eigenstates satisfying the APS boundary condition 
as (i) edge states, (ii) bulk states, and (iii) zero modes, respectively. 
A reason for the nomenclature will be given in respective subsections. 

\subsubsection{Edge states}
As the APS boundary conditions \eqref{APS(-)} and \eqref{APS(+)} are to be imposed on  
feasible solutions of the form $\Phi^j_m$ and $\Psi^j_m$, respectively, we work with them separately. 
We first apply the APS boundary condition \eqref{APS(-)}  to 
the feasible solutions \eqref{I, l+1/2}.  
For $\mu>0$, the boundary condition applied to \eqref{I, l+1/2, +} is written out and arranged as 
\begin{equation}
 \label{bdy cond l+1/2,mu>0} 
 \mu\sqrt{\frac{\mu+E}{\mu-E}}I_{\ell+\frac12}(\varepsilon R)= 
  \Bigr(\frac{\ell+1}{R}+\sqrt{\bigl(\frac{\ell+1}{R}\bigr)^2+\mu^2}\Bigr)I_{\ell+\frac32}(\varepsilon R). 
\end{equation}
If $E<0$,  one has, for $\ell\geq 0$, 
\begin{equation} 
   0 <\mu  \sqrt{\frac{\mu+E}{\mu-E}}<\mu, \quad 0<\mu < \frac{\ell+1}{R}+\sqrt{\bigl(\frac{\ell+1}{R}\bigr)^2+\mu^2}. 
\end{equation} 
Since $I_{\ell+\frac12}(\varepsilon R)> I_{\ell+\frac32}(\varepsilon R)$, Eq.~\eqref{bdy cond l+1/2,mu>0} 
may have a solution. 
If $E>0$, one has 
\begin{equation} 
   0 <\mu <\mu  \sqrt{\frac{\mu+E}{\mu-E}}, \quad 0<\mu < \frac{\ell+1}{R}+\sqrt{\bigl(\frac{\ell+1}{R}\bigr)^2+\mu^2}.
\end{equation} 
Since $I_{\ell+\frac12}(\varepsilon R)> I_{\ell+\frac32}(\varepsilon R)$, Eq.~\eqref{bdy cond l+1/2,mu>0} 
may have no solution for $E$.  
We turn to the case of $\mu<0$.  
From \eqref{I, l+1/2, -}, the boundary condition is written out and arranged as  
\begin{equation}
 \label{bdy cond l+1/2,mu<0} 
 |\mu|\sqrt{\frac{|\mu+E|}{|\mu-E|}}I_{\ell+\frac12}(\varepsilon R)= 
  \Bigr(\frac{\ell+1}{R}+\sqrt{\bigl(\frac{\ell+1}{R}\bigr)^2+\mu^2}\Bigr)I_{\ell+\frac32}(\varepsilon R). 
\end{equation}
For $E>0$, one has 
\begin{equation} 
   |\mu|\sqrt{\frac{|\mu+E|}{|\mu-E|}}<|\mu|, \quad 
   |\mu|< \frac{\ell+1}{R}+\sqrt{\bigl(\frac{\ell+1}{R}\bigr)^2+\mu^2}. 
\end{equation}
Since $I_{\ell+\frac12}(\varepsilon R)> I_{\ell+\frac32}(\varepsilon R)$, Eq.~\eqref{bdy cond l+1/2,mu<0} 
may have a solution.  Contrary to this, for $E<0$, one has 
\begin{equation} 
   0 <|\mu| <|\mu|  \sqrt{\frac{|\mu+E|}{|\mu-E|}}, \quad 0<|\mu| < \frac{\ell+1}{R}+
\sqrt{\bigl(\frac{\ell+1}{R}\bigr)^2+\mu^2}.
\end{equation} 
Since $I_{\ell+\frac12}(\varepsilon R)> I_{\ell+\frac32}(\varepsilon R)$, Eq.~\eqref{bdy cond l+1/2,mu<0} 
may have no solution.  

We proceed to apply the APS boundary condition \eqref{APS(-)} to feasible solutions 
given in \eqref{I, l-1/2}. 
For $\mu>0$, one obtains from \eqref{I, l-1/2, +} the condition  
\begin{equation}
\label{boundary cond, l-1/2, mu>0} 
   \mu \sqrt{\frac{\mu-E}{\mu+E}} I_{\ell+\frac12}(\varepsilon R) = 
  \Bigl( \frac{\ell+1}{R} +\sqrt{\bigl(\frac{\ell+1}{R}\bigr)^2 +\mu^2} \Bigr) I_{\ell +\frac32}(\varepsilon R). 
\end{equation} 
If $E<0$, one has 
\begin{equation} 
   0<\mu<\mu \sqrt{\frac{\mu-E}{\mu+E}}, \quad 
   0 < \mu <\frac{\ell+1}{R} +\sqrt{\bigl(\frac{\ell+1}{R}\bigr)^2 +\mu^2}. 
\end{equation} 
Since $I_{\ell+\frac12}(\varepsilon R)> I_{\ell +\frac32}(\varepsilon R)$, 
Eq.~\eqref{boundary cond, l-1/2, mu>0} may have no solution for $E$. 
If $E>0$, one has 
\begin{equation} 
   0<\mu \sqrt{\frac{\mu-E}{\mu+E}}<\mu, \quad 
   0 < \mu <\frac{\ell+1}{R} +\sqrt{\bigl(\frac{\ell+1}{R}\bigr)^2 +\mu^2}. 
\end{equation}
Since $I_{\ell+\frac12}(\varepsilon R)> I_{\ell +\frac32}(\varepsilon R)$, 
Eq.~\eqref{boundary cond, l-1/2, mu>0} may have a solution. 
We turn to the case of $\mu<0$. 
For \eqref{I, l-1/2, -}, the APS boundary condition \eqref{APS(-)} yields 
\begin{equation}
\label{boundary cond, l-1/2, mu<0} 
   |\mu| \sqrt{\frac{|\mu-E|}{|\mu+E|}} I_{\ell+\frac12}(\varepsilon R) = 
  \Bigl( \frac{\ell+1}{R} +\sqrt{\bigl(\frac{\ell+1}{R}\bigr)^2 +\mu^2} \Bigr) I_{\ell +\frac32}(\varepsilon R). 
\end{equation} 
If $E<0$, one has 
\begin{equation} 
    0<  |\mu| \sqrt{\frac{|\mu-E|}{|\mu+E|}}< |\mu|, \quad 
   |\mu| < \frac{\ell+1}{R} +\sqrt{\bigl(\frac{\ell+1}{R}\bigr)^2 +\mu^2}. 
\end{equation} 
Since  $I_{\ell+\frac12}(\varepsilon R)> I_{\ell +\frac32}(\varepsilon R)$, 
Eq.~\eqref{boundary cond, l-1/2, mu<0} may have a solution for $E$. 
On the contrary, if $E>0$, one has 
\begin{equation} 
    0< |\mu| < |\mu| \sqrt{\frac{|\mu-E|}{|\mu+E|}}, \quad 
   |\mu| < \frac{\ell+1}{R} +\sqrt{\bigl(\frac{\ell+1}{R}\bigr)^2 +\mu^2}. 
\end{equation} 
Since  $I_{\ell+\frac12}(\varepsilon R)> I_{\ell +\frac32}(\varepsilon R)$, 
Eq.~\eqref{boundary cond, l-1/2, mu<0} may have no solution.  

\begin{figure}[htbp]  
\begin{center}
\includegraphics[width=0.5\columnwidth]{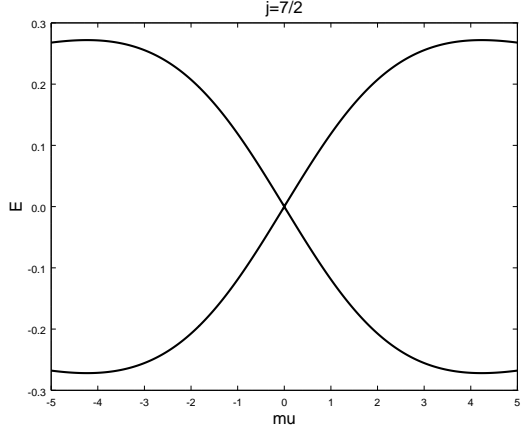} 
\end{center}  
\caption{The eigenvalues for edge states with $j=\frac72$ are plotted against the parameter $\mu$. 
As is seen from \eqref{eqs for edge eigenvalues}, 
the graphs going from the upper left downward to the lower right and from the lower left upward 
to the upper right are associated with the eigenstates of the form $\Phi^j_m$ and 
$\Psi^j_m$, respectively, where $\Phi^j_m$ and $\Psi^j_m$ are associated with a positive and a 
negative eigenvalues of $P$, respectively. }   
\label{eigenvalue graphs of edge states} 
\end{figure}

The functional equations obtained above to determine edge state eigenvalues are listed as follows: 
\renewcommand{\multirowsetup}{\centering}
\hfill
\begin{equation}
\label{eqs for edge eigenvalues} 
  \begin{array}{c|c|c}
   {\rm edg}(-)  & \Phi^j_m  & \Psi^j_m  \\ \hline 
  \mu>0 \;\vline\; E>0 & \mbox{no sol.} & \eqref{boundary cond, l-1/2, mu>0} \\
  \mu>0 \;\vline\; E<0 & \eqref{bdy cond l+1/2,mu>0} & \mbox{no sol.} \\ \hline 
  \mu<0 \;\vline\; E>0 & \eqref{bdy cond l+1/2,mu<0} &  \mbox{no sol.} \\
  \mu<0 \;\vline\; E<0 &  \mbox{no sol.}  & \eqref{boundary cond, l-1/2, mu<0} \\ \hline
 \end{array}
\end{equation}
\noindent 
Here, the symbol ${\rm edg}(-)$ indicates that the boundary conditions corresponding to the 
eigenvalue $\lambda^{(-)}_{j}$ and $\kappa^{(-)}_j$ (see \eqref{APS(-)}) are adopted. 
The functional equations \eqref{bdy cond l+1/2,mu>0}, \eqref{bdy cond l+1/2,mu<0}, 
\eqref{boundary cond, l-1/2, mu>0}, and \eqref{boundary cond, l-1/2, mu<0}  can be numerically 
solved to give graphs of energy eigenvalues as functions of the parameter $\mu$.  
For $j=\frac72$ and $R=1$, the graphs of energy eigenvalues are given 
in Fig.~\ref{eigenvalue graphs of edge states}, which shows that the edge state eigenvalues are 
responsible for band rearrangement, where the eigenstate for $\mu=0$ will be discussed later. 
Further, the density $\psi^{\dag}\psi$ for each of the edge eigenstates leans to the boundary. 
For these reasons, the nomenclature ``edge state" has been adopted.

In the rest of this subsection, we treat the APS boundary condition \eqref{APS(+)}.  
We show that the boundary condition \eqref{APS(+)} does not provide eigenvalues for either $\Phi^j_m$ 
or $\Psi^j_m$. 
If $\Phi|_{r=R}$ for \eqref{I, l+1/2}  is required to belong to $\mathcal{H}^{(+)}_{j}(S^2)$,  
the boundary condition is expressed as 
\begin{subequations}
 \begin{align} 
    \frac{\sqrt{\mu+E}I_{\ell+\frac12}(\varepsilon R)/\sqrt{\varepsilon R}}{-\mu R} & = 
     \frac{\sqrt{\mu-E}I_{\ell+\frac32}(\varepsilon R)/\sqrt{\varepsilon R}}{\ell+1 +\sqrt{(\ell+1)^2+(\mu R)^2}} 
        \quad {\rm for} \quad \mu>0,   \\
    \frac{\sqrt{|\mu+E|}I_{\ell+\frac12}(\varepsilon R)/\sqrt{\varepsilon R}}{\mu R} & = 
     \frac{\sqrt{|\mu-E|}I_{\ell+\frac32}(\varepsilon R)/\sqrt{\varepsilon R}}{\ell+1 +\sqrt{(\ell+1)^2+(\mu R)^2}} 
    \quad {\rm for} \quad \mu<0. 
 \end{align}
\end{subequations} 
Since the left-hand sides of the above equations are negative, but the right-hand sides are positive, 
so that no solution exists for $E$.   
It then turns out that the boundary condition $\Phi|_{r=R}\in \mathcal{H}^{(+)}_{j}(S^2)$ is ill. 
  
If $\Psi|_{r=R}$ for \eqref{I, l-1/2} is required to belong to $\mathcal{K}^{(+)}_{j}(S^2)$, 
the boundary condition are brought into  
\begin{subequations} 
 \begin{align}
   \frac{\sqrt{\mu+E}I_{\ell+\frac32}(\varepsilon R)/\sqrt{\varepsilon R}}{-\mu R} & = 
  \frac{\sqrt{\mu-E}I_{\ell+\frac12}(\varepsilon R)/\sqrt{\varepsilon R}}
          {-(\ell+1) +\sqrt{(\ell+1)^2+(\mu R)^2}} \quad {\rm for} \quad \mu>0, \\
   \frac{-\sqrt{|\mu+E|}I_{\ell+\frac32}(\varepsilon R)/\sqrt{\varepsilon R}}{-\mu R} & = 
  \frac{\sqrt{|\mu-E|}I_{\ell+\frac12}(\varepsilon R)/\sqrt{\varepsilon R}}
          {-(\ell+1) +\sqrt{(\ell+1)^2+(\mu R)^2}} \quad {\rm for} \quad \mu<0. 
 \end{align}
\end{subequations} 
Since the left-hand sides of the above two equations are negative but the right-hand sides are positive, 
the boundary condition $\Psi|_{r=R}\in \mathcal{K}^{(+)}_{j}(S^2)$ is ill too. 
  
\subsubsection{Bulk states}
\begin{figure}[htbp]  
\begin{center}
\includegraphics[width=0.35\columnwidth]{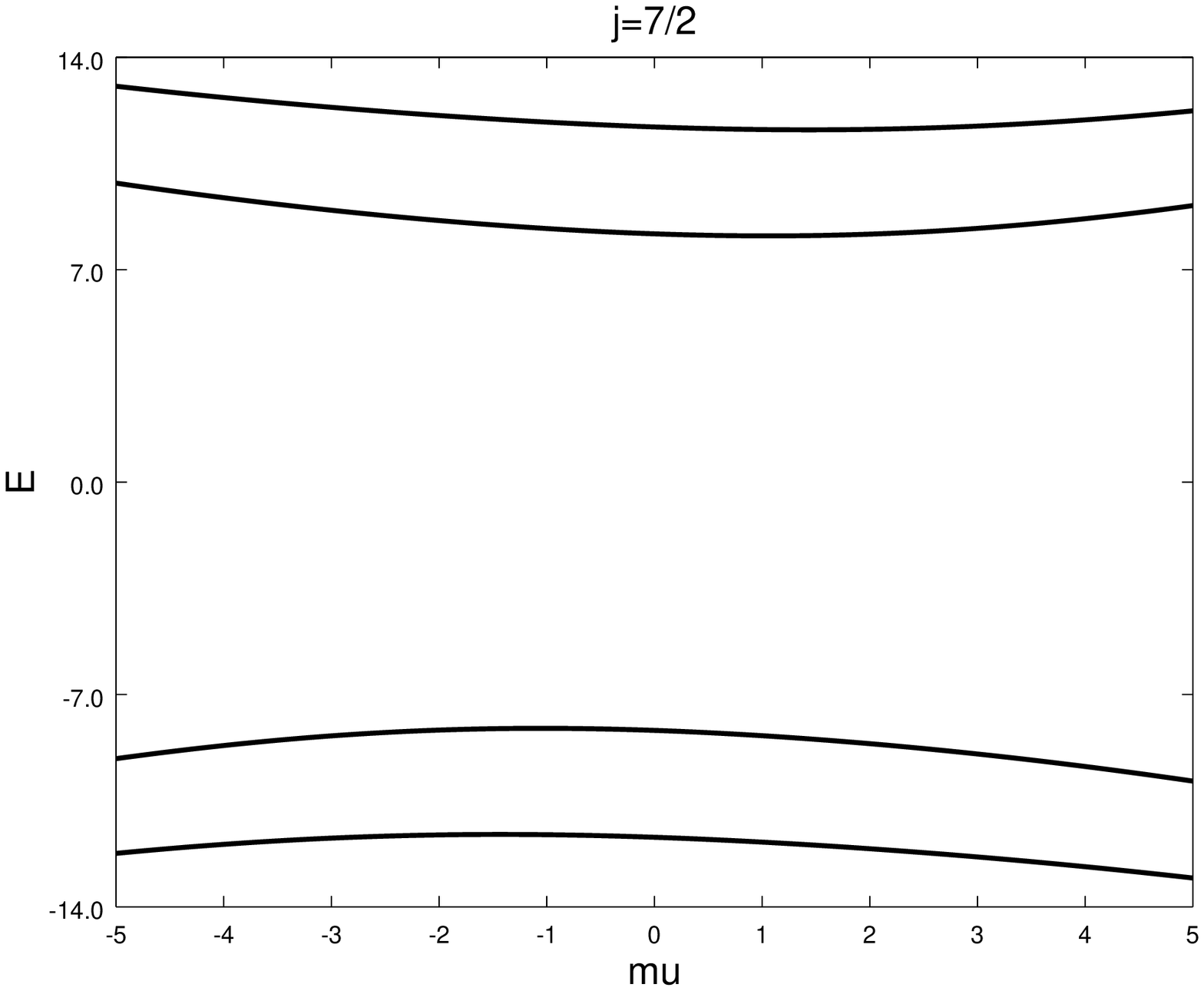} \quad 
\includegraphics[width=0.35\columnwidth]{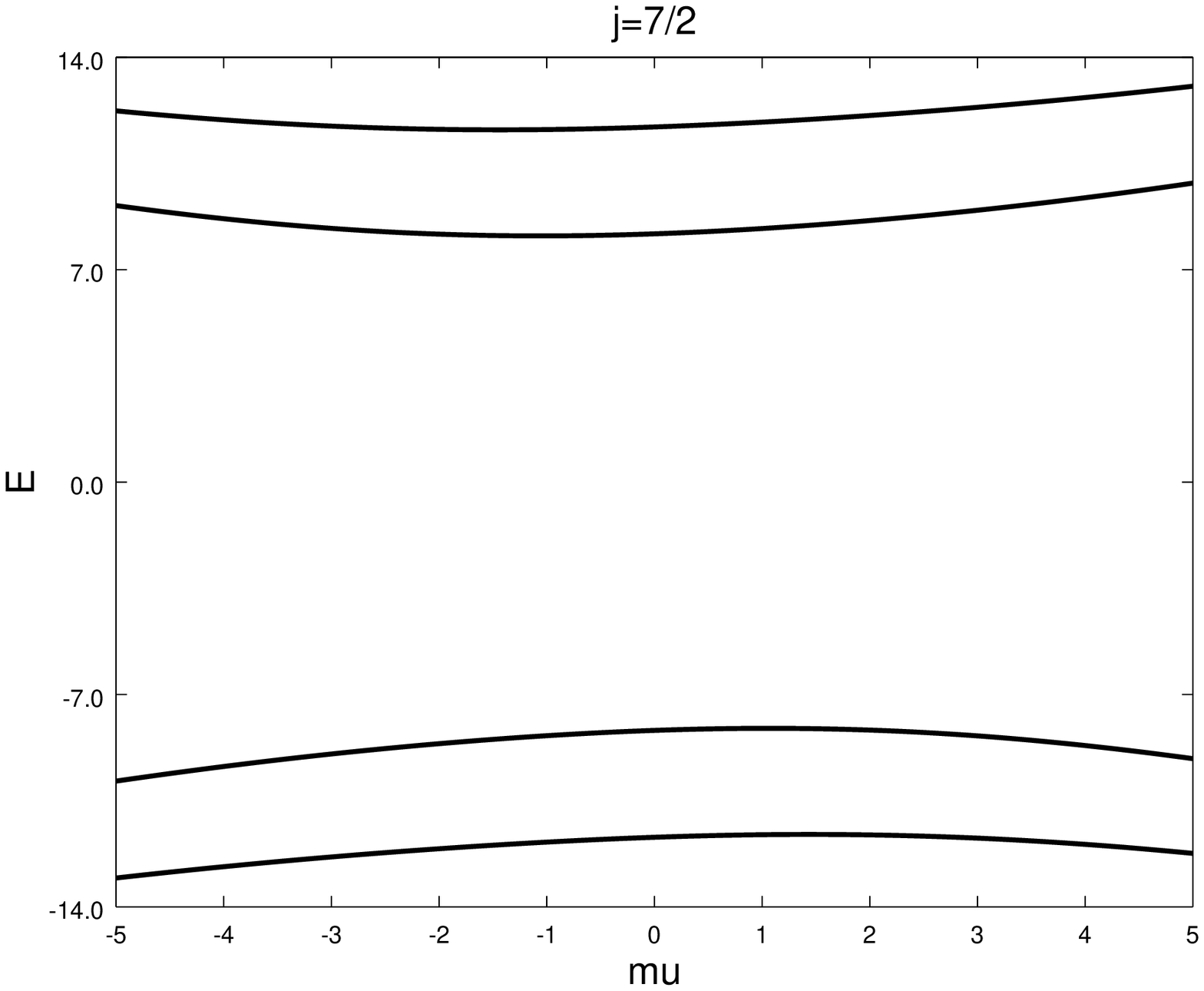}
\end{center}  
\caption{For $j=\frac72$ and $R=1$, the eigenvalues for bulk states under the boundary conditions 
$\Phi|_R \in \mathcal{H}^{(-)}_{j}(S^2)$ (left panel) and  $\Psi|_R \in \mathcal{K}^{(-)}_{j}(S^2)$ 
(right panel) are plotted against the parameter $\mu$. If reflected with respect to the horizontal axis 
$E=0$, the graph in the left panel is transformed into the graph in the right panel, which is a consequence 
of the chiral symmetry of the present eigenvalue problem. }
\label{eigenvalue graphs of bulk states (-)} 
\end{figure}

In a similar manner to the manipulation of edge states, we apply 
the APS boundary conditions \eqref{APS(-)} and \eqref{APS(+)} 
to feasible solutions \eqref{sol, j=l+1/2} and \eqref{sol, j=l-1/2} separately. 
We first treat the APS boundary condition \eqref{APS(-)} corresponding to $\lambda^{(-)}_{j}$. 
From  \eqref{sol, j=l+1/2} and \eqref{Phi,k(-),l+1/2}, the APS boundary condition reads 
\begin{subequations}
 \begin{align}
 \label{bdy cond l+1/2,E>0} 
 -\mu\sqrt{\frac{E+\mu}{E-\mu}} J_{\ell+\frac12}(\beta R) & = 
  \Bigr(\frac{\ell+1}{R}+\sqrt{\bigl(\frac{\ell+1}{R}\bigr)^2+\mu^2}\Bigr)J_{\ell+\frac32}(\beta R)
   \quad {\rm for} \quad  E>0, \\
 \label{bdy cond l+1/2, E<0} 
 \mu \sqrt{\frac{|E+\mu|}{|E-\mu|}} J_{\ell+\frac12}(\beta R) & = 
  \Bigr(\frac{\ell+1}{R}+\sqrt{\bigl(\frac{\ell+1}{R}\bigr)^2+\mu^2}\Bigr)J_{\ell+\frac32}(\beta R) 
  \quad {\rm for} \quad E<0. 
 \end{align}
\end{subequations}
We turn to the APS boundary condition \eqref{APS(-)} corresponding to $\kappa^{(-)}_j$. 
From \eqref{sol, j=l-1/2} and \eqref{Psi,k(-),l-1/2}, the APS boundary condition is put in the form   
\begin{subequations}
 \begin{align}
\label{boundary cond, l-1/2, E>0} 
   \mu \sqrt{\frac{E-\mu}{E+\mu}} J_{\ell+\frac12}(\beta R) &= 
  \Bigl( \frac{\ell+1}{R} +\sqrt{\bigl(\frac{\ell+1}{R}\bigr)^2 +\mu^2} \Bigr) J_{\ell +\frac32}(\beta R) 
  \quad {\rm for}\quad E>0, \\
\label{boundary cond, l-1/2, E<0} 
  -\mu  \sqrt{\frac{|E-\mu|}{|E+\mu|}} J_{\ell+\frac12}(\beta R) & = 
  \Bigl( \frac{\ell+1}{R} +\sqrt{\bigl(\frac{\ell+1}{R}\bigr)^2 +\mu^2} \Bigr) J_{\ell +\frac32}(\beta R)
   \quad  {\rm for} \quad E<0. 
 \end{align}
\end{subequations} 
The functional equations obtained so far to determine eigenvalues for bulk states are listed as follows: 
\begin{equation} 
    \begin{array}{c|c|c}
      {\rm reg}(-) & \Phi^j_m & \Psi^j_m \\ \hline 
      E>0 & \eqref{bdy cond l+1/2,E>0} &  \eqref{boundary cond, l-1/2, E>0} \\ \hline 
      E<0 & \eqref{bdy cond l+1/2, E<0} & \eqref{boundary cond, l-1/2, E<0} 
    \end{array} 
\end{equation}
Here the symbol ${\rm reg}(-)$ indicates that the boundary conditions corresponding to 
$\lambda^{(-)}_{j}$ and $\kappa^{(-)}_j$ are adopted.  
Equations \eqref{bdy cond l+1/2,E>0}, \eqref{bdy cond l+1/2, E<0}, \eqref{boundary cond, l-1/2, E>0}, 
and \eqref{boundary cond, l-1/2, E<0} are numerically solved to give eigenvalues as functions of 
the parameter $\mu$, which form bands, as is shown in Fig.~\ref{eigenvalue graphs of bulk states (-)}. 
It is here to be noted that there are two states with the same quantum number $j$, which are 
distinguished by using the operator $P$ (see \eqref{S+,S-}). 

\begin{figure}[htbp]  
\begin{center}
\includegraphics[width=0.35\columnwidth]{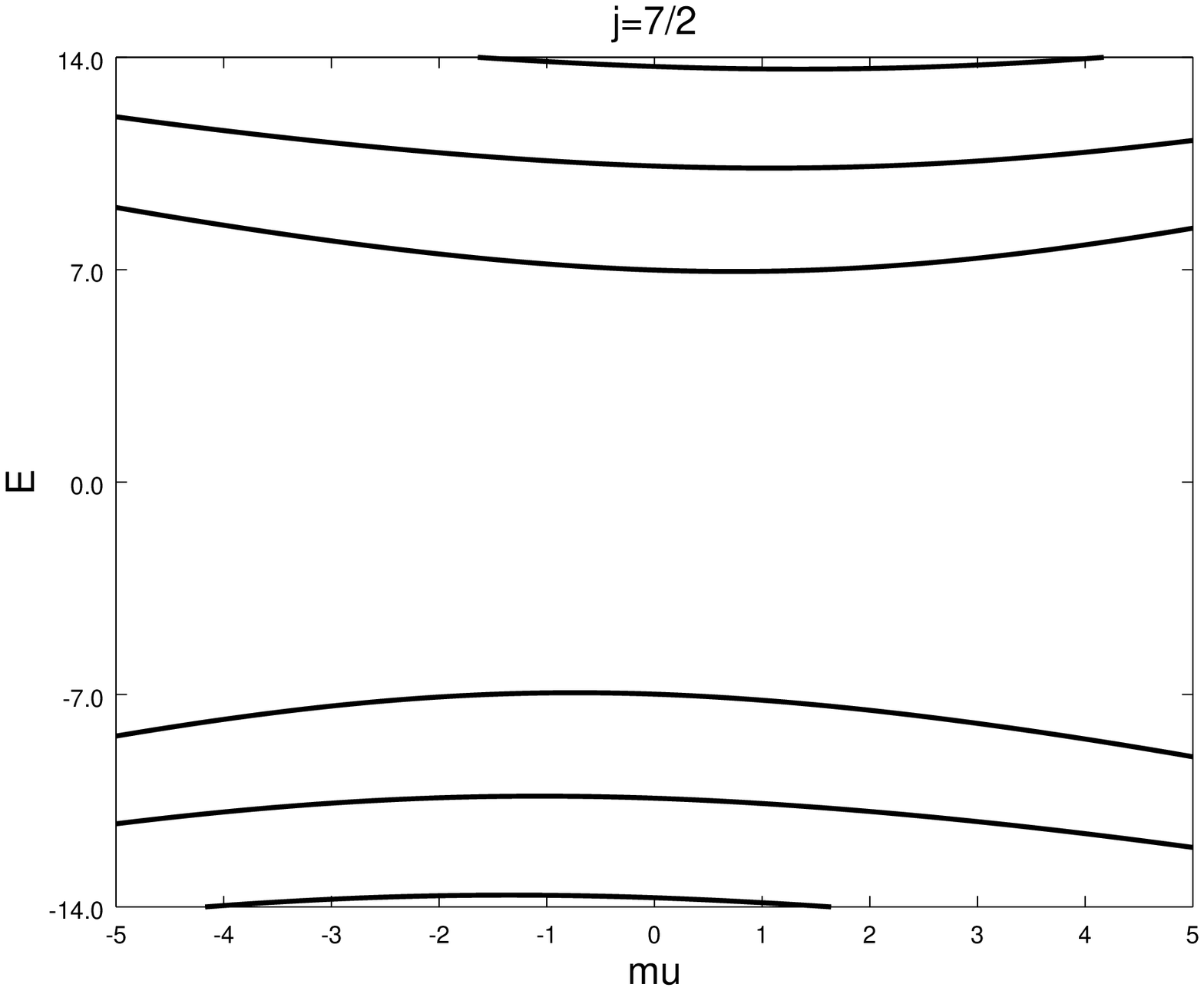} \quad 
\includegraphics[width=0.35\columnwidth]{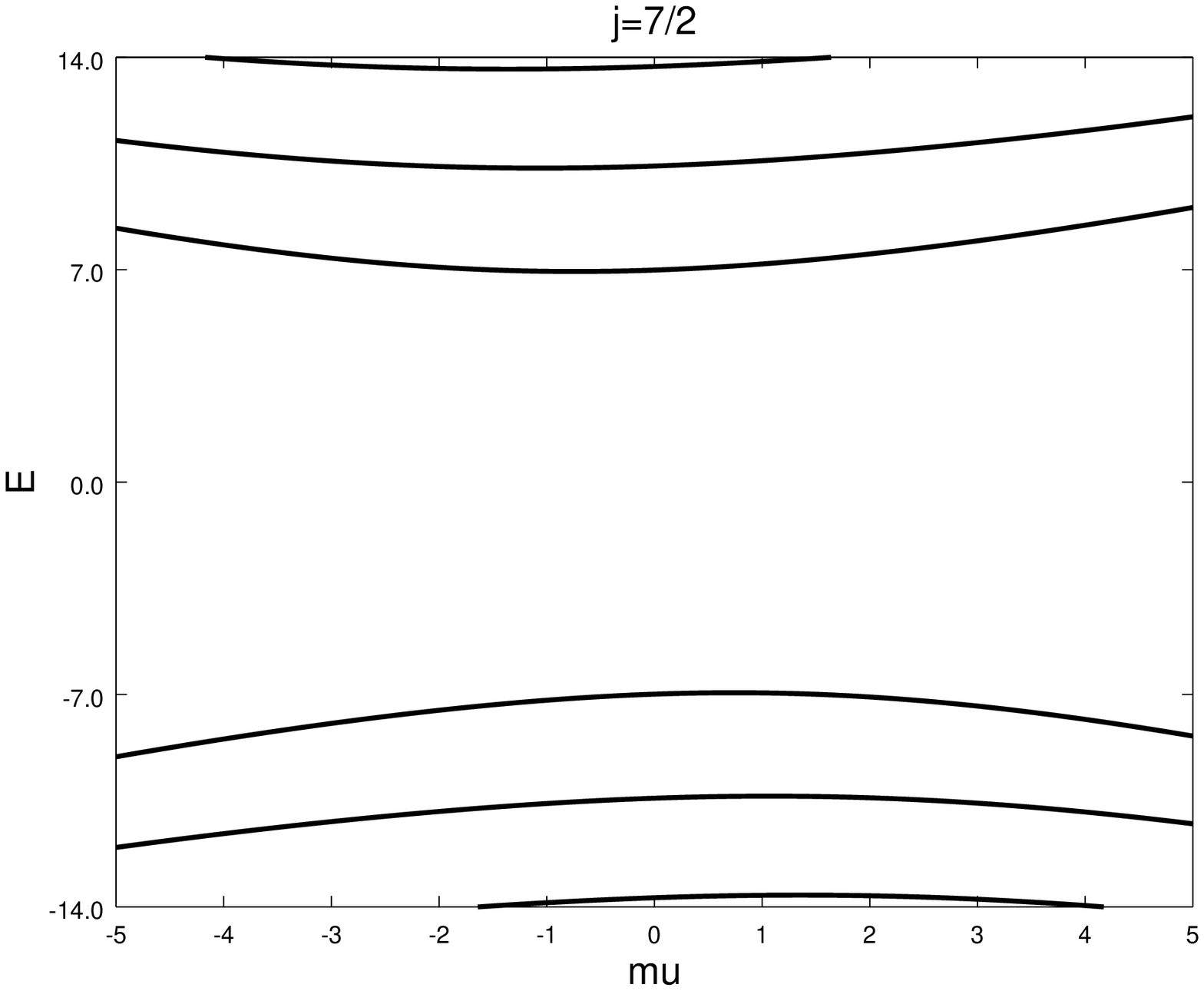}
\end{center}  
\caption{The eigenvalues for bulk states under the boundary condition 
$\Phi|_R \in \mathcal{H}^{(+)}_{j}(S^2)$  (left panel) and $\Psi|_R \in  \mathcal{K}^{(+)}_{j}(S^2)$ 
(right panel) are plotted against the parameter $\mu$ with $j=\frac72$ and $R=1$.  
Like Fig.~\ref{eigenvalue graphs of bulk states (-)}, this figure also exhibits the chiral symmetry 
of the present eigenvalue problem. }
\label{eigenvalue graphs of bulk states (+)} 
\end{figure}

We proceed to the APS boundary condition \eqref{APS(+)}, which 
requires that the boundary values belong to 
the eigenstates associated with $\lambda^{(+)}_{j}$ or $\kappa^{(+)}_{j}$. 
If $\Phi|_{r=R}$ is required to belong to the eigenspace associated with 
the eigenvalue $\lambda^{(+)}_{j}$, then 
from \eqref{sol, j=l+1/2} and \eqref{Phi,k(+),l+1/2}, 
the boundary condition is expressed and arranged as 
\begin{subequations} 
 \begin{align}
  \label{k(+),l+1/2,E>0} 
    \mu  \sqrt{\frac{E-\mu}{E+\mu}} J_{\ell+\frac32}(\beta R) & = 
  \Bigl( \frac{\ell+1}{R} +\sqrt{\bigl(\frac{\ell+1}{R}\bigr)^2 +\mu^2} \Bigr) J_{\ell +\frac12}(\beta R) 
   \quad {\rm for} \quad E>0, \\
 \label{k(+),l+1/2,E<0}  
  -\mu  \sqrt{\frac{|E-\mu|}{|E+\mu|}} J_{\ell+\frac32}(\beta R) & = 
  \Bigl( \frac{\ell+1}{R} +\sqrt{\bigl(\frac{\ell+1}{R}\bigr)^2 +\mu^2} \Bigr) J_{\ell +\frac12}(\beta R) 
   \quad {\rm for} \quad  E<0.
 \end{align}
\end{subequations}  
If $\Psi|_{r=R}$ is required to belong to the  eigenspace associated with $\kappa^{(+)}_{j}$, 
from \eqref{sol, j=l-1/2} and \eqref{Psi,k(+),l-1/2}, 
the boundary condition for $E>0$ reads 
\begin{subequations}
 \begin{align}
 \label{k(+),l-1/2,E>0} 
 -\mu \sqrt{\frac{E+\mu}{E-\mu}} J_{\ell+\frac32}(\beta R) & = 
    \Bigl(\frac{\ell+1}{R}+\sqrt{\bigl(\frac{\ell+1}{R}\bigr)^2+\mu^2}\Bigr) J_{\ell+\frac12}(\beta R) 
   \quad {\rm for}\quad E>0, \\
 \label{k(+),l-1/2,E<0}  
  \mu \sqrt{\frac{|E+ \mu|}{|E-\mu|}} J_{\ell+\frac32}(\beta R) & = 
   \Bigl(\frac{\ell+1}{R} +\sqrt{\bigl(\frac{\ell+1}{R}\bigr)^2+\mu^2} \Bigr)  J_{\ell-\frac12}(\beta R)
  \quad {\rm for} \quad E<0. 
 \end{align}
\end{subequations}
The functional equations obtained above to determine eigenvalues for bulk states are 
listed as follows: 
\begin{equation} 
    \begin{array}{c|c|c} 
      {\rm reg}(+)  & \Phi^j_m & \Psi^j_m \\ \hline 
      E>0 &  \eqref{k(+),l+1/2,E>0} &  \eqref{k(+),l-1/2,E>0}  \\ \hline 
      E<0 &  \eqref{k(+),l+1/2,E<0} & \eqref{k(+),l-1/2,E<0} 
  \end{array} 
\end{equation} 
Here the symbol  ${\rm reg}(+)$ indicates that the boundary conditions corresponding to 
$\lambda^{(+)}_{j}$ and $\kappa^{(+)}_j$  are adopted. 
Equations \eqref{k(+),l+1/2,E>0}, \eqref{k(+),l+1/2,E<0}, \eqref{k(+),l-1/2,E>0}, and 
\eqref{k(+),l-1/2,E<0} are numerically solved to give eigenvalues as functions of 
the  parameter $\mu$.  Solutions to respective equations form bands, which are shown 
in Fig.~\ref{eigenvalue graphs of bulk states (+)}. 

Figures \ref{eigenvalue graphs of bulk states (-)} and \ref{eigenvalue graphs of bulk states (+)} 
show that the bulk state eigenvalues form two bands, positive and negative, which have nothing 
to do with band rearrangement. 

\subsubsection{Zero modes}
We now deal with critical solutions in the case of $|E|=|\mu|$. 
Among feasible solutions  \eqref{cr,Phi_jm}, \eqref{Phi'}, \eqref{cr,Psi_jm}, and \eqref{Psi'}, 
the only solutions which satisfy the APS boundary conditions are \eqref{Phi'} and \eqref{Psi'}, 
which are associated with the eigenvalue $E=0$ and are called zero modes. 
These eigenstates correspond to eigenvalues $\lambda^{(-)}_{+}$ and $\kappa^{(-)}_-$ of the 
boundary operator with $\mu=0$, 
which are limits of $\lambda^{(-)}_{j}$ and $\kappa^{(-)}_j$, respectively, as $\mu\to 0$.  
Let us be reminded of the fact that the APS boundary condition \eqref{APS(+)} never 
admits edge states but the APS boundary condition \eqref{APS(-)} yields edge states, which 
are associated with the eigenvalues $\lambda^{(-)}_{j}$ and $\kappa^{(-)}_j$ 
of the boundary operator $B_{\mu}$. 
Furthermore, both of the conditions \eqref{APS(+)} and \eqref{APS(-)} give rise to bulk states, 
which are away from zero modes. 
Figure \ref{eigenvalue graphs of edge states} shows that zero eigenvalue is realized as limits of 
edge state eigenvalues as $\mu\to 0$, and Figs.~\ref{eigenvalue graphs of bulk states (-)} and 
\ref{eigenvalue graphs of bulk states (+)} show that zero eigenvalue is never realized as limits of 
bulk state eigenvalue as $\mu \to 0$.  

\subsubsection{Zero modes as transient states}
We show that the edge eigenstate indeed approaches the zero mode as 
$E(\mu)\to 0$ along with $\mu\to 0$.  
To this end, we introduce the power series $I_{\nu}^P(z)$ through 
\begin{equation}
\label{I power series}
     I_{\nu}(z)=\bigl(\frac{z}{2}\bigr)^{\nu} I^P_{\nu}(z), \quad    
       I^P_{\nu}(z)=\sum_{n=0}^{\infty}\frac{1}{n!\Gamma(\nu +n+1)} \bigl(\frac{z}{2}\bigr)^{2n}.  
\end{equation}  
For $\mu>0$, we take the edge eigenstate \eqref{I, l+1/2, +} with $E=E(\mu)$ determined by 
\eqref{bdy cond l+1/2,mu>0}.  
By using the symbols $I^P_{\ell+\frac12}$ and 
$I^P_{\ell+\frac32}$,  we rewrite  \eqref{I, l+1/2, +} within a constant factor as 
\begin{equation}
   \begin{pmatrix} \frac{\sqrt{\mu+E}}{\sqrt{\varepsilon r}}\bigl(\frac{\varepsilon r}{2}\bigr)^{\ell+\frac12} 
     I^P_{\ell+\frac12}(\varepsilon r) \Phi^{j(+)}_{m} \\ 
       \frac{\sqrt{\mu-E}}{\sqrt{\varepsilon r}}\bigl(\frac{\varepsilon r}{2}\bigr)^{\ell+\frac32} 
     I^P_{\ell+\frac32}(\varepsilon r) \Phi^{j(-)}_{m} \end{pmatrix}  = 
   \frac{\varepsilon^{\ell} \sqrt{\mu+E}}{2^{\ell+\frac12}} 
     \begin{pmatrix} r^{\ell}   I^P_{\ell+\frac12}(\varepsilon r) \Phi^{j(+)}_{m} \\
    \frac{\mu-E}{2} r^{\ell+1}  I^P_{\ell+\frac32}(\varepsilon r) \Phi^{j(-)}_{m} \end{pmatrix} . 
\end{equation}  
Deleting the scalar factor $\varepsilon^{\ell} \sqrt{\mu+E}/2^{\ell+\frac12}$, 
which vanishes as $E(\mu)\to 0$ along with $\mu\to 0$, we introduce an edge eigenstate 
\begin{equation}
\label{red edg eigenstate, I +}  
   \tilde{\Phi}_{\rm edg}^{(+)} =  
     \begin{pmatrix} r^{\ell}   I^P_{\ell+\frac12}(\varepsilon r) \Phi^{j(+)}_{m} \\
    \frac{\mu-E}{2} r^{\ell+1}  I^P_{\ell+\frac32}(\varepsilon r) \Phi^{(-)}_{m} \end{pmatrix}
    \quad {\rm for} \quad \mu>0,  
\end{equation} 
which remains to be associated with the same eigenvalue $E(\mu)$. 
Letting $\mu\to 0$ and $E(\mu)\to 0$, we obtain the limit 
\begin{equation} 
\label{edg limit(mu>0)} 
  \tilde{\Phi}_{\rm edg}^{(+)} \to  \frac{1}{\Gamma(\ell +\frac32)} \begin{pmatrix} 
     r^{\ell}\Phi^{j(+)}_{m} \\   0  \end{pmatrix},   
\end{equation} 
which is a zero mode (see  \eqref{Phi'}).  
We take \eqref{I, l+1/2, -} in turn, where $E=E(\mu)$ is determined by \eqref{bdy cond l+1/2,mu<0}.   
Rewriting \eqref{I, l+1/2, -} as 
\begin{equation}
  \begin{pmatrix} -\frac{\sqrt{|\mu+E|}}{ \sqrt{\varepsilon r}} I_{\ell+\frac12}(\varepsilon r) \Phi^{j(+)}_{m} \\
       \frac{\sqrt{|\mu-E|}}{ \sqrt{\varepsilon r}}I_{\ell+\frac32}(\varepsilon r) \Phi^{j(-)}_{m} \end{pmatrix} = 
 \frac{\varepsilon^{\ell} \sqrt{|\mu+E|}} {2^{\ell+\frac12}}  \begin{pmatrix} 
    -r^{\ell} I^P_{\ell+\frac12}(\varepsilon r) \Phi^{j(+)}_{m} \\
   \frac{|\mu-E|}{2} r^{\ell+1} I^P_{\ell+\frac32} (\varepsilon r) \Phi^{j(-)}_{m} \end{pmatrix} ,  
\end{equation} 
we introduce 
\begin{equation} 
  \tilde{\Phi}^{(-)}_{\rm edg} = \begin{pmatrix} 
    r^{\ell} I^P_{\ell+\frac12}(\varepsilon r) \Phi^{j(+)}_{m} \\
   -\frac{|\mu-E|}{2} r^{\ell+1} I^P_{\ell+\frac32} (\varepsilon r) \Phi^{j(-)}_{m} \end{pmatrix} 
  \quad {\rm for} \quad \mu<0. 
\end{equation} 
Letting $E(\mu)\to 0$ together with $\mu\to 0$,  we obtain the limit 
\begin{equation}
 \label{edg limit(mu<0)}  
    \tilde{\Phi}^{(-)}_{\rm edg} \to 
   \frac{1}{\Gamma(\ell +\frac32)} \begin{pmatrix} 
     r^{\ell}\Phi^{j(+)}_{m} \\   0  \end{pmatrix},  
\end{equation} 
the same limit as \eqref{edg limit(mu>0)}.   
Equations \eqref{edg limit(mu>0)} and \eqref{edg limit(mu<0)} are put together 
to show that the $\mu$-dependent edge states  $\tilde{\Phi}^{(-)}_{\rm edg}$ and  
$\tilde{\Phi}^{(+)}_{\rm edg}$ are continuously connected together through a zero mode at $\mu=0$.    

Starting with \eqref{I, l-1/2, +}, we introduce     
\begin{equation} 
   \tilde{\Psi}_{\rm edg}^{(+)} = \begin{pmatrix} 
    \frac{\mu+E}{2} r^{\ell+1} I^P_{\ell+\frac32}(\varepsilon r) \Phi^{j(-)}_{m} \\
    r^{\ell} I^P_{\ell+\frac12}(\varepsilon r) \Phi^{j(+)}_{m} \end{pmatrix} \quad 
   {\rm for} \quad \mu>0, 
\end{equation} 
and let $E(\mu)\to 0$ along with $\mu\to 0$ to get the limit 
\begin{equation}
\label{edge limit, l-1/2(mu>0)}  
    \tilde{\Psi}_{\rm edg}^{(+)} \to \frac{1}{\Gamma(\ell+\frac32)} \begin{pmatrix} 
    0 \\  r^{\ell}  \Phi^{j(+)}_{m} \end{pmatrix}, 
\end{equation} 
which is a zero mode (see \eqref{Psi'}). 
Starting with \eqref{I, l-1/2, -} in turn, 
we introduce 
\begin{equation} 
  \tilde{\Psi}^{(-)}_{\rm edg} = \begin{pmatrix} 
    -\frac{|\mu+E|}{2} r^{\ell+1} I^P_{\ell+\frac12}(\varepsilon r) \Phi^{j(-)}_{m} \\
  r^{\ell} I^P_{\ell+\frac12}(\varepsilon r) \Phi^{j(+)}_{m} \end{pmatrix} \quad 
   {\rm for} \quad \mu<0. 
\end{equation} 
Letting $E(\mu)\to 0$ together with $\mu\to 0$, we obtain the limit 
\begin{equation}
\label{edge limit, l-1/2(mu<0)}  
  \tilde{\Psi}^{(-)}_{\rm edg} \to \frac{1}{\Gamma(\ell+\frac32)} \begin{pmatrix} 
    0 \\  r^{\ell}\Phi^{j(+)}_{m} \end{pmatrix}, 
\end{equation} 
the same limit as \eqref{edge limit, l-1/2(mu>0)}.  
Put together, Eqs.~\eqref{edge limit, l-1/2(mu>0)} and \eqref{edge limit, l-1/2(mu<0)} 
show that the $\mu$-dependent edge states  $\tilde{\Psi}^{(-)}_{\rm edg}$ and  
$\tilde{\Psi}^{(+)}_{\rm edg}$ are continuously connected together through a zero mode 
at $\mu=0$.

\subsection{The eigenvalues of the Dirac operator with the chiral bag boundary condition}
\label{Eigenvalues, chiral bag}
We treat the chiral bag boundary condition \eqref{chiral bag bdy cond lambda} on the sphere of 
radius $R$.  On setting $\gamma_r^{}=\boldsymbol{\gamma}\cdot \boldsymbol{n}$, the condition 
reads 
\begin{equation} 
\label{chiral bag bdy cond(4-spinor)}
   \psi = -i \gamma_r e^{\lambda \gamma_0^{}} \gamma_0^{} \psi. 
\end{equation}  
We apply this condition to the feasible solutions in each of the cases 
(i) $|E|<|\mu|$, (ii) $|E|>|\mu|$, (iii) $|E|=|\mu|$, in a similar manner to the APS boundary condition. 

\subsubsection{Edge states}
For feasible solutions \eqref{I, l+1/2}, 
the boundary condition  \eqref{chiral bag bdy cond(4-spinor)} is written out and arranged 
for $\mu>0$ as 
\begin{equation}
\label{chiral bdy cond, j=l+1/2}  
 \sqrt{\frac{\mu+E}{\mu-E}} I_{\ell+\frac12}(\varepsilon R) = e^{-\lambda} I_{\ell+\frac32}(\varepsilon R). 
\end{equation}
where $\varepsilon=\sqrt{\mu^2-E^2}$. 
If $\lambda>0$, one obtains 
$I_{\ell+\frac12}(\varepsilon R)>e^{-\lambda}I_{\ell+\frac32}(\varepsilon R)$ 
from the fact that $I_{\ell+\frac12}(\varepsilon R)>I_{\ell+\frac32}(\varepsilon R)$.  
Since $\frac{\mu+E}{\mu-E}>1$ for $E>0$ with $\mu>E$, 
Eq.~\eqref{chiral bdy cond, j=l+1/2} implies that there are no solution for $E>0$.  
If $E<0$, Eq.~\eqref{chiral bdy cond, j=l+1/2} may have a solution.  
For $\mu<0$, the boundary condition \eqref{chiral bag bdy cond(4-spinor)} 
applied to \eqref{I, l+1/2}  yields 
\begin{equation} 
   -\frac{\sqrt{|\mu+E|}}{\sqrt{|\mu-E|}} I_{\ell+\frac12}(\varepsilon R)=
   e^{-\lambda} I_{\ell+\frac32}(\varepsilon R). 
\end{equation} 
Though the left-hand side is negative, the right-hand side is positive, so that this equation 
has no solution. 

We proceed to the feasible solutions \eqref{I, l-1/2}. 
For $\mu>0$,  the boundary condition \eqref{chiral bag bdy cond(4-spinor)} brings about 
\begin{equation} 
\label{chiral bdy cond, Psi}
 \sqrt{\frac{\mu+E}{\mu-E}}  I_{\ell+\frac32}(\varepsilon R) =e^{-\lambda} I_{\ell+\frac12}(\varepsilon R).  
\end{equation} 
Since $I_{\ell+\frac12}(\varepsilon R)<I_{\ell-\frac12}(\varepsilon R)$, this equation may have a solution 
for $E>0$ with $\lambda>0$ fixed, and has no solution for $E<0$, if 
$|E|$ is relatively large within the restriction $|E|<|\mu|$.  
For $\mu<0$, the boundary condition \eqref{chiral bag bdy cond(4-spinor)} 
gives rise to 
\begin{equation} 
   -\sqrt{\frac{|\mu+E|}{|\mu-E|}} I_{\ell+\frac32}(\varepsilon R) = e^{-\lambda} I_{\ell+\frac12}(\varepsilon R). 
\end{equation} 
The left-hand side of this equation is negative but the right-hand side positive, and hence the above equation 
has no solution. 

Among four equations obtained above, only two equations, Eqs.~\eqref{chiral bdy cond, j=l+1/2} 
and \eqref{chiral bdy cond, Psi}, may have solutions, which can be numerically solved to give 
eigenvalues as functions of the parameter $\mu$. 
Examples will be given later together with examples of eigenvalues for bulk states.  

\subsubsection{Bulk states}
\label{chjral bag, bulk}
We apply the chiral bag boundary condition to the feasible solutions \eqref{sol, j=l+1/2}.   
The boundary condition \eqref{chiral bag bdy cond(4-spinor)} provides 
\begin{subequations}
 \begin{align} 
     -\sqrt{\frac{E+\mu}{E-\mu}} J_{\ell+\frac12}(\beta R) & = e^{-\lambda} J_{\ell+\frac32}(\beta R) 
    \quad {\rm for} \quad E>0, \\
    \label{chiral bag, reg E<0}
      \sqrt{\frac{|E+\mu|}{|E-\mu|}} J_{\ell+\frac12}(\beta R) & = e^{-\lambda} J_{\ell+\frac32}(\beta R) 
    \quad {\rm for} \quad E<0. 
 \end{align}
\end{subequations} 

Applying the chiral bag boundary condition to the feasible solutions \eqref{sol, j=l-1/2}, 
we find the functional equations for $E>0$ and for $E<0$, 
\begin{subequations}
 \begin{align}
   \label{chiral bag, E >0, Psi}
     \sqrt{\frac{E+\mu}{E-\mu}} J_{\ell+\frac32}(\beta R) & = e^{-\lambda} J_{\ell+\frac12}(\beta R)
    \quad {\rm for} \quad E>0, \\
      -\sqrt{\frac{|E+\mu|}{|E-\mu|}} J_{\ell+\frac32}(\beta R) & = e^{-\lambda} J_{\ell+\frac12}(\beta R) 
    \quad {\rm for} \quad E<0. 
 \end{align}
\end{subequations}

All functional equations obtained in Sec.~\ref{chjral bag, bulk} can be numerically solved.  
Examples are given in Fig.~\ref{reg & edge states, chiral bag}. 
A characteristic of this figure is that two of eigenvalues for bulk states are connected with 
eigenvalues for edge sates, crossing the lines determined by $|E|=|\mu|$. 
The upper and lower eigenvalues crossing the lines $E=\mu$ and $E=-\mu$ are associated 
with a negative and a positive eigenvalues of $P$, respectively.  
The crossing points give critical eigenvalues, which will be evaluated in the following subsection. 

\begin{figure}[htbp]  
\begin{center}
\includegraphics[width=0.55\columnwidth]{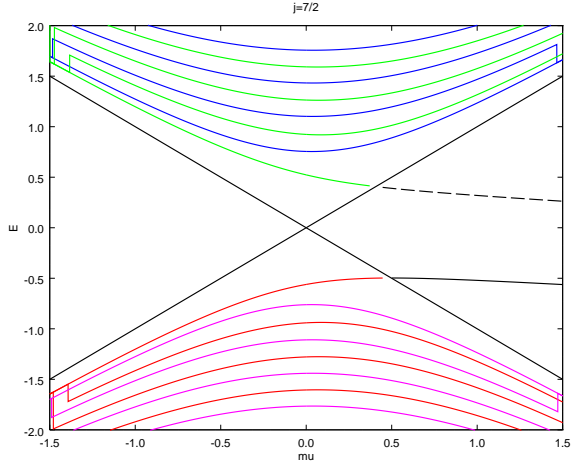} 
\end{center}  
\caption{Eigenvalues against the parameter $\mu$ under the chiral bag boundary condition 
with $j=\frac72, R=10, \lambda=0.1$ The black solid lines of X figure form the boundary of bulk and 
edge regions defined by $|E|>|\mu|$ and $|E|<|\mu|$, respectively.  Some numerical errors are 
contained in graphs for the parameter values about $|\mu|=1.5$, which are negligible. }
\label{reg & edge states, chiral bag} 
\end{figure}

\subsubsection{Singular states}
We discuss the singular eigenstates in the case of $|E|=|\mu|$. 
Applying the chiral bag boundary condition \eqref{chiral bag bdy cond(4-spinor)} to the 
feasible solutions \eqref{cr,Phi_jm} and \eqref{cr,Psi_jm},  we obtain the eigenvalues 
\begin{equation}
\label{E=-mu} 
    E=-\mu=-\frac{(2\ell+3)e^{\lambda}}{2R}
\end{equation} 
and 
\begin{equation}
\label{E=mu}
    E=\mu=\frac{(2\ell+3)e^{-\lambda}}{2R},   
\end{equation} 
respectively. 
At the same time, Eqs.~\eqref{cr,Phi_jm} and \eqref{cr,Psi_jm} prove to be respective eigenstates. 

It is to be noted that since the eigenvalues given above are not reflection symmetric 
with respect to the $E=0$ axis, which means that the chiral bag boundary condition 
does not admit the chiral symmetry, while the Hamiltonian admits the chiral symmetry. 
We will discuss discrete symmetries for the eigenvalue problems in Sec.~\ref{discrete symmetry}.  

\subsubsection{Singular states as transient states}
As is indicated in Fig.~\ref{reg & edge states, chiral bag}, 
two of bulk eigenstates are connected with edge eigenstates through the critical eigenstates. 
In the following, we show that this is the case. 
We rewrite \eqref{I, l+1/2} for $\mu>0$, 
by employing the functions $I^P_{\ell+\frac12}$ and  $I^P_{\ell+\frac32}$ (see \eqref{I power series}), as 
\begin{equation} 
  \Phi^j_m= 
   \frac{\varepsilon^{\ell} \sqrt{\mu+E}}{2^{\ell+\frac12}} 
     \begin{pmatrix} r^{\ell}   I^P_{\ell+\frac12}(\varepsilon r) \Phi^{j(+)}_{m} \\
    \frac{\mu-E}{2} r^{\ell+1}  I^P_{\ell+\frac32}(\varepsilon r) \Phi^{j(-)}_{m} \end{pmatrix} , 
\end{equation} 
where $E=E(\mu)$ is determined by \eqref{chiral bdy cond, j=l+1/2}. 
Deleting the scalar factor $\varepsilon^{\ell} \sqrt{\mu+E}$, which vanishes as $E(\mu)\to -\mu$, 
we introduce 
\begin{equation}
\label{red edg eigenstate, I +}  
   \tilde{\Phi}_{\rm edg} =  \frac{1}{2^{\ell+\frac12}} 
     \begin{pmatrix} r^{\ell}   I^P_{\ell+\frac12}(\varepsilon r) \Phi^{j(+)}_{m} \\
    \frac{\mu-E}{2} r^{\ell+1}  I^P_{\ell+\frac32}(\varepsilon r) \Phi^{j(-)}_{m} \end{pmatrix} ,  
\end{equation} 
which remains to be an edge eigenstate associated with the same eigenvalue $E(\mu)$. 
Letting $E(\mu)\to -\mu$ in the above equation, we obtain, as a limit,  
\begin{equation}
\label{edg limit} 
  \tilde{\Phi}_{\rm edg} \to \frac{1}{2^{\ell+\frac12}} \begin{pmatrix} 
     \frac{r^{\ell}}{\Gamma(\ell +\frac32)}\Phi_m^{j(+)} \\
     \frac{\mu r^{\ell+1}}{\Gamma(\ell+\frac52)} \Phi^{j(-)}_{m} \end{pmatrix} 
 =\frac{1}{2^{\ell+\frac32}\Gamma(\ell+\frac52)} \begin{pmatrix} 
     (2\ell+3) r^{\ell} \Phi^{j(+)}_{m} \\ 2\mu r^{\ell+1} \Phi^{j(-)}_{m} \end{pmatrix}=\tilde{\Phi}_{\rm cri}. 
\end{equation} 
Within a constant factor, the right-hand side of the above equation is equal to the critical 
eigenstate \eqref{cr,Phi_jm} with $E(\mu)$ given in \eqref{E=-mu}. 
 
We proceed to show that a bulk eigenstate can approach a critical eigenstate as $E(\mu) \to -\mu$. 
To this end, we introduce the power series $J^P_{\nu}$ through 
\begin{equation}
\label{J power series}
    J_{\nu}(z)=\bigl(\frac{z}{2}\bigr)^{\nu} J^P_{\nu}(z), \quad    
       J^P_{\nu}(z)=\sum_{n=0}^{\infty}\frac{(-1)^n}{n!\Gamma(\nu +n+1)} \bigl(\frac{z}{2}\bigr)^{2n}. 
\end{equation}
We take the bulk eigenstate \eqref{sol, j=l+1/2} for $E<0$ and rewrite it, 
by using $J^P_{\nu}$, as 
\begin{equation}  
   \Phi^j_m = 
  \frac{\beta^{\ell}\sqrt{|E+\mu|}}{2^{\ell+\frac32}} \begin{pmatrix}  
     2 r^{\ell} J^P_{\ell+\frac12}(\beta r) \Phi^{j(+)}_{m} \\
     |E-\mu| r^{\ell+1} J^P_{\ell+\frac32}(\beta r) \Phi^{j(-)}_{m} \end{pmatrix}, 
\end{equation}
where $E=E(\mu)$ is determined by \eqref{chiral bag, reg E<0}. 
Deleting the scalar factor $\beta\sqrt{|E+\mu|}$, which vanishes as $E(\mu)\to -\mu$, 
in the right-hand side of the above equation, we introduce 
\begin{equation}
\label{red reg eigenstate, J -}
  \tilde{\Phi}_{\rm reg} = \frac{1}{2^{\ell+\frac32}}  \begin{pmatrix}  
     2 r^{\ell} J^P_{\ell+\frac12}(\beta r) \Phi^{j(+)}_{m} \\
      |E-\mu| r^{\ell+1} J^P_{\ell+\frac32}(\beta r) \Phi^{j(-)}_{m} \end{pmatrix}, 
\end{equation} 
which remains to be a bulk eigenstates. 
Letting $E(\mu)\to -\mu$ with $\mu>0$ in the right-hand side of the above equation, we obtain 
\begin{equation}
\label{reg limit} 
   \tilde{\Phi}_{\rm reg} \to \frac{1}{2^{\ell+\frac32}}  \begin{pmatrix}  
     \frac{2 r^{\ell}}{\Gamma(\ell+\frac32)} \Phi^{j(+)}_{m} \\
     \frac{2\mu r^{\ell+1}}{\Gamma(\ell+\frac52)} \Phi^{j(-)}_{m} \end{pmatrix} = 
  \frac{1}{2^{\ell+\frac32} \Gamma(\ell+\frac52)} \begin{pmatrix} 
   (2\ell+3) r^{\ell} \Phi^{j(+)}_{m} \\ 2\mu r^{\ell+1} \Phi^{j(-)}_{m} \end{pmatrix}=\tilde{\Phi}_{\rm cri}. 
\end{equation} 
From \eqref{edg limit} and \eqref{reg limit}, we see that the edge and the bulk eigenstates, 
$\tilde{\Phi}_{\rm edg}$ and $\tilde{\Phi}_{\rm reg}$,  approach the same critical eigenstate 
$\tilde{\Phi}_{\rm cri}$  as $E(\mu)\to -\mu$. 
Put another way, bulk and edge eigenstates, $\tilde{\Phi}_{\rm edg}$ and $\tilde{\Phi}_{\rm reg}$, 
can change into each other through the critical eigenstates $\tilde{\Phi}_{\rm cri}$ 
accompanying the variation in the parameter $\mu$. 

We turn to the eigenstates \eqref{I, l-1/2} for $\mu>0$, where $E(\mu)$ is determined by 
\eqref{chiral bdy cond, Psi} and assumed to be compatible with 
the limiting procedure $E(\mu)\to \mu$. 
Like $\tilde{\Phi}_{\rm edg}$, we introduce     
\begin{equation} 
   \tilde{\Psi}_{\rm edg}=\frac{1}{2^{\ell+\frac32}} \begin{pmatrix} 
    (\mu+E) r^{\ell+1} I^P_{\ell+\frac32}(\varepsilon r) \Phi^{j(-)}_{m} \\
    2r^{\ell} I^P_{\ell+\frac12}(\varepsilon r) \Phi^{j(+)}_{m} \end{pmatrix},
\end{equation} 
and let $E(\mu)\to \mu$ to get 
\begin{equation}
\label{edg limit,-} 
  \tilde{\Psi}_{\rm edg} \to \frac{1}{2^{\ell+\frac32}} \begin{pmatrix} 
    \frac{2\mu r^{\ell}}{\Gamma(\ell+\frac52)} \Phi^{j(-)}_{m} \\ 
     \frac{2r^{\ell-1}}{\Gamma(\ell+\frac32)} \Phi^{j(+)}_{m} \end{pmatrix} = 
   \frac{1}{2^{\ell+\frac32}\Gamma(\ell+\frac52)} \begin{pmatrix} 
     2\mu r^{\ell+1} \Phi^{j(-)}_{m} \\ (2\ell +3) r^{\ell} \Phi^{j(+)}_{m} \end{pmatrix}=\tilde{\Psi}_{\rm cri} . 
\end{equation} 
The right-hand side of the above equation is the same, within a constant factor, 
as the critical eigenstate \eqref{cr,Psi_jm} with $E(\mu)$ determined by \eqref{E=mu}. 
 
Like \eqref{reg limit}, a bulk eigenstates can be shown to approach the critical eigenstate $\tilde{\Psi}_{\rm cri}$.  
We take the bulk eigenstate  \eqref{sol, j=l-1/2} for $E>0$, 
where $E=E(\mu)$ is determined by \eqref{chiral bag, E >0, Psi} and 
assumed to be compatible with the limiting procedure $E(\mu)\to \mu$. 
Like \eqref{red reg eigenstate, J -}, we  introduce, 
\begin{equation} 
   \tilde{\Psi}_{\rm reg} = \frac{1}{2^{\ell+\frac32}} \begin{pmatrix} 
    (E+\mu)r^{\ell+1} J^P_{\ell+\frac32}(\beta r) \Phi^{j(-)}_{m} \\ 
     2r^{\ell} J^P_{\ell+\frac12}(\beta r) \Phi^{j(+)}_{m}  \end{pmatrix}. 
\end{equation} 
Letting $E(\mu)\to \mu$, we verify that 
\begin{equation} 
\label{reg limit,-}
  \tilde{\Psi}_{\rm reg} \to \frac{1}{2^{\ell+\frac32}} \begin{pmatrix} 
   \frac{2\mu r^{\ell+1}}{\Gamma(\ell+\frac52)} \Phi^{j(-)}_{m} \\
   \frac{2r^{\ell}}{\Gamma(\ell+\frac32)} \Phi^{j(+)}_{m} \end{pmatrix} = 
    \frac{1}{2^{\ell+\frac32} \Gamma(\ell+\frac52)} \begin{pmatrix}
     2\mu r^{\ell+1} \Phi^{j(-)}_{m} \\ (2\ell+3) r^{\ell} \Phi^{j(+)}_{m} \end{pmatrix}=\tilde{\Psi}_{\rm cri}, 
\end{equation} 
which is the same limit as \eqref{edg limit,-}. 
Eqs.~\eqref{edg limit,-} and \eqref{reg limit,-} are put together to imply that the edge and the bulk eigenstates, 
$\tilde{\Psi}_{\rm edg}$ and $\tilde{\Psi}_{\rm reg}$,  
change into each other through the critical eigenstate $\tilde{\Psi}_{\rm cri}$, 
like $\tilde{\Phi}_{\rm edg}$ and $\tilde{\Phi}_{\rm reg}$.

\subsection{Discrete symmetries}
\label{discrete symmetry}
We have already characterized in  Sec.~\ref{setting-up} the Dirac Hamiltonian \eqref{3D_Dirac_Ham} 
in terms of discrete symmetries. 
We now observe that discrete symmetries are realized in 
the pattern of eigenvalues as functions of the parameter $\mu$ 
(see Figs.~\ref{eigenvalue graphs of edge states}, \ref{eigenvalue graphs of bulk states (-)}, 
\ref{eigenvalue graphs of bulk states (+)}, and \ref{reg & edge states, chiral bag}).  
To this end, we need to discuss the invariance of the APS and the chiral bag boundary conditions.  

We first take up the chiral operator $\sigma_1\otimes \1$.    
A straightforward calculation with $\sigma_1\otimes \1$ provides 
\begin{subequations}
\begin{align}
    \label{anti-invariance H}  
    (\sigma_1\otimes \1) H_{\mu} (\sigma_1\otimes \1) & = -H_{\mu}, \\ 
      \label{invariance of B(mu)} 
     (\sigma_1\otimes \1) B_{\mu} (\sigma_1\otimes \1) & = B_{\mu}, \\
    (\sigma_1\otimes \1) J_k (\sigma_1\otimes \1)  & = J_k,  \label{J_k gamma_5} \\
    (\sigma_1\otimes \1) P (\sigma_1\otimes \1)  & = - P,  
    \label{spin-inversion gamma5}
\end{align} 
\end{subequations}
where $H_{\mu}$ and $B_{\mu}$ are the Hamiltonian and the boundary operators, respectively, 
and where $J_k$ and $P$ are the total angular momentum operators of $4\times 4$ matrix form and 
the operator distinguishing the spinor type $\Phi^j_m$ or $\Psi^j_m$ (see \eqref{S+,S-}).   
The chiral symmetry \eqref{anti-invariance H} may be called an energy-reflecting symmetry, 
{\it i.e.}, if $E$ is an energy eigenvalue, then so is $-E$.  
Eq.~\eqref{invariance of B(mu)}  means that the APS boundary condition is chiral-invariant. 
Eq.~\eqref{J_k gamma_5} implies that the total angular momentum quantum number $j$ and 
the eigenvalue $m$ of $J_3$ are invariant, 
but  Eq.~\eqref{spin-inversion gamma5} shows that the spinor type is inverted under the chiral transformation.  
It then turns out that if an eigenstate $\Phi$ is assigned by the quantum numbers $(E,j,m,\pm)$ then 
$(\sigma_1\otimes \1)\Phi$ becomes an eigenstate assigned by the quantum numbers $(-E,j,m,\mp)$,  
where the signs $+$ and $-$ indicate that the eigenvalue of $P$ is positive and negative, respectively. 
This fact can explain the pattern symmetry of Figs.~\ref{eigenvalue graphs of edge states}, 
\ref{eigenvalue graphs of bulk states (-)}, and \ref{eigenvalue graphs of bulk states (+)} 
for the eigenvalues under the APS boundary condition: 
The two curves of eigenvalue functions shown in Fig.~\ref{eigenvalue graphs of edge states} are 
reflection symmetric with respect to the $\mu$-axis in the $(\mu,E)$-plane, 
if the inversion of $P$-eigenvalue is taken into account. 
The two panels in Fig.~\ref{eigenvalue graphs of bulk states (-)} are related by $\sigma_1\otimes \1$:  
If we reflect the graph of the left panel with respect to the $\mu$-axis the resultant graph 
coincides with that of the right panel. 
The same procedure can be performed with Fig.~\ref{eigenvalue graphs of bulk states (+)} 
to observe the same pattern symmetry. 

In contrast to this, the pattern of eigenvalues observed in Fig.~\ref{reg & edge states, chiral bag}  
for the Dirac equation under the chiral bag boundary condition is quite different from the pattern 
under the APS boundary condition.  Though the Hamiltonian admits the chiral symmetry, the 
pattern of energy eigenvalues does not exhibit the energy-reflection symmetry.  
This is because the chiral bag boundary condition is not invariant under the action of the chiral operator. 
In fact, if $\psi$ is subject to $\psi=-i\gamma_r^{}e^{\lambda \gamma_0^{}} \gamma_0^{}\psi$ then 
$(\sigma_1\otimes \1)\psi$ is subject to 
\begin{equation}
   (\sigma_1\otimes \1)\psi=-i\gamma_r^{}e^{-\lambda \gamma_0^{}}\gamma_0^{} (\sigma_1\otimes \1)\psi, 
\end{equation}
where $\lambda$ has changed into $-\lambda$. 

We turn to the particle-hole symmetry.  We easily verify that 
\begin{subequations}
\label{particle-hole}
 \begin{align}
    \label{particle-hole Ham} 
     (\sigma_2 \otimes \sigma_2)\overline{H}_{\mu}(\sigma_2 \otimes \sigma_2) & =-H_{\mu}, \\
    \label{particle-hole Bdy}
    (\sigma_2 \otimes \sigma_2)\overline{B}_{\mu}(\sigma_2 \otimes \sigma_2) & =B_{\mu}, \\
     \label{particle-hole Jk}
    (\sigma_2 \otimes \sigma_2)\overline{J}_k (\sigma_2 \otimes \sigma_2)  & = -J_k, \\
    (\sigma_2 \otimes \sigma_2)\overline{\boldsymbol{J}}^2 (\sigma_2 \otimes \sigma_2) & = 
      \boldsymbol{J}^2,  \label{particle-hole J^2}  \\
     (\sigma_2 \otimes \sigma_2)\overline{P}(\sigma_2 \otimes \sigma_2) & = -P.  
    \label{particle-hole spin}
 \end{align}
\end{subequations} 
The above equations imply that if $\Phi$ is an eigenstate associated with the quantum numbers 
$(E,j,m,\pm)$ under the APS boundary condition then $(\sigma_2\otimes \sigma_2)K\Phi$ is 
an eigenstate associated with the quantum numbers $(-E,j,-m,\mp)$, where $K$ denotes the complex 
conjugation.  
Since the inversion $m\mapsto -m$ makes no appearance in the eigenvalue pattern,     
the chiral and the particle-hole symmetries make no difference in the eigenvalue pattern 
under the APS boundary condition, 
but the chiral operator is unitary while the particle-hole operator is anti-unitary.  

Contrary to this, the chiral bag boundary condition is not invariant under the operator 
$(\sigma_2\otimes \sigma_2)K$.   In fact, we easily verify that if $\psi$ is subject to 
$\psi=-i\gamma_r^{}e^{\lambda \gamma_0^{}} \gamma_0^{}\psi$ then  $(\sigma_2\otimes \sigma_2)K\psi$ is 
subject to 
\begin{equation} 
    (\sigma_2\otimes \sigma_2)K\psi =
     i\gamma_r^{}e^{-\lambda \gamma_0^{}} \gamma_0^{}(\sigma_2\otimes \sigma_2)K\psi. 
\end{equation} 

We now deal with the time-reversal symmetry. 
A straightforward calculation with $(\sigma_3\otimes i\sigma_2)K$ provides 
\begin{subequations}
\label{time-reversal-like}
 \begin{align}
    \label{time-reversal Ham} 
    (\sigma_3\otimes i\sigma_2)\overline{H}_{\mu}(\sigma_3\otimes (-i\sigma_2)) & 
     =H_{\mu}, \\
    \label{time-reversal Bdy}
    (\sigma_3 \otimes i\sigma_2)\overline{B_{\mu}}(\sigma_3\otimes (-i\sigma_2)) & 
    =B_{\mu}, \\
  \label{inversion Jk}
   (\sigma_3 \otimes i\sigma_2)\overline{J}_k (\sigma_3\otimes (-i\sigma_2)) & = -J_k, \\
  (\sigma_3 \otimes i\sigma_2)\overline{\boldsymbol{J}}^2 (\sigma_3 \otimes (-i\sigma_2))  & 
    =  \boldsymbol{J}^2,  \label{invariant J^2}  \\
    (\sigma_3 \otimes i\sigma_2)\overline{P}(\sigma_3 \otimes (-i\sigma_2)) & 
    =   P . 
 \end{align}
\end{subequations} 
Equations \eqref{time-reversal-like} means that if $\Phi$ is an eigenstate assigned by quantum 
numbers $(E,j,m,\pm)$ under the APS boundary condition 
then $(\sigma_3 \otimes i\sigma_2)K\Phi$ becomes an eigenstates assigned 
by the quantum numbers $(E,j,-m, \pm)$,   
where the Kramers degeneracy is realized as the inversion $m \mapsto -m$ with $E$ kept invariant.  

Unlike the chiral and the particle-hole operators, the time-reversal operator leaves 
the chiral bag boundary condition invariant. In fact, one verifies that 
if $\psi$ is subject to $\psi=-i\gamma_r^{}e^{\lambda \gamma_0^{}} \gamma_0^{}\psi$ then  
$(\sigma_3\otimes i\sigma_2)K\psi$ is also subject to 
\begin{equation} 
    (\sigma_3\otimes i\sigma_2)K\psi = 
    -i\gamma_r^{}e^{\lambda \gamma_0^{}} \gamma_0^{} (\sigma_3\otimes i\sigma_2)K\psi .
\end{equation} 
Hence, the Kramers degeneracy takes place for energy eigenvalues under the 
chiral bag boundary condition, like energy eigenvalues under the APS boundary condition. 

\subsection{Spectral flows under the APS and the chiral bag boundary conditions}
\label{Spectral flow}
While the bulk state eigenvalues have nothing to do with band rearrangement, 
the edge state eigenvalues are responsible for the band rearrangement under both of the APS and 
the chiral bag boundary conditions, as was seen in the behavior 
of eigenvalues against the parameter $\mu$ (see 
Figs.~\ref{eigenvalue graphs of edge states} and \ref{reg & edge states, chiral bag}). 
We now single out the eigenvalues responsible for the band rearrangement 
to draw illustrative figures (see Fig.~\ref{transient eigenvalue curve}). 
 
\unitlength=0.60mm
\begin{figure}[htbp]
\begin{center}
\begin{minipage}[c][50mm][t]{50mm}
 \begin{center}
 \begin{picture}(65,65)
\put(0,35){\vector(1,0){70}}
\put(35,0){\vector(0,1){70}}
\put(35,35){\line(1,1){30}}
\put(35,35){\line(1,-1){30}}
\put(35,35){\line(-1,1){30}}
\put(35,35){\line(-1,-1){30}}
\put(37,69){$E$}
\put(66,68){$E=\mu$}
\put(66,2){$E=-\mu$}
\put(56,42){$(P<0)$}
\put(56,26){$(P>0)$}
\bezier{150}(20,57)(28,46)(60,38)
\bezier{150}(20,14)(28,29)(60,32)
\put(71,35){$\mu$}
 \end{picture}
 \end{center}
\end{minipage}
\hspace{5mm}
\begin{minipage}[c][50mm][t]{50mm}
 \begin{center}
 \begin{picture}(65,65)
\put(0,35){\vector(1,0){70}}
\put(35,0){\vector(0,1){70}}
\put(35,35){\line(1,1){30}}
\put(35,35){\line(1,-1){30}}
\put(35,35){\line(-1,1){30}}
\put(35,35){\line(-1,-1){30}}
\put(37,69){$E$}
\bezier{150}(15,42)(26,37)(60,28)
\bezier{150}(15,28)(26,33)(60,42)
\put(71,35){$\mu$}
\put(66,68){$E=\mu$}
\put(66,2){$E=-\mu$}
\put(56,44){$(P<0)$}
\put(56,23){$(P>0)$}
 \end{picture}
 \end{center}
\end{minipage}
\end{center}
\caption{A schematic view of transient eigenvalue curves.  
The left and the right panels are for the chiral bag and the APS boundary conditions, respectively, 
where $P>0$ and $P<0$ mean that the eigenvalue of $P$ is positive and negative, respectively, 
and correspondingly, the eigenstates are of the type $\Phi^j_m$ and $\Psi^j_m$, respectively. 
In the left panel, the parameter is chosen as $\lambda=0$ for simplicity. 
\label{transient eigenvalue curve} }  
\end{figure}
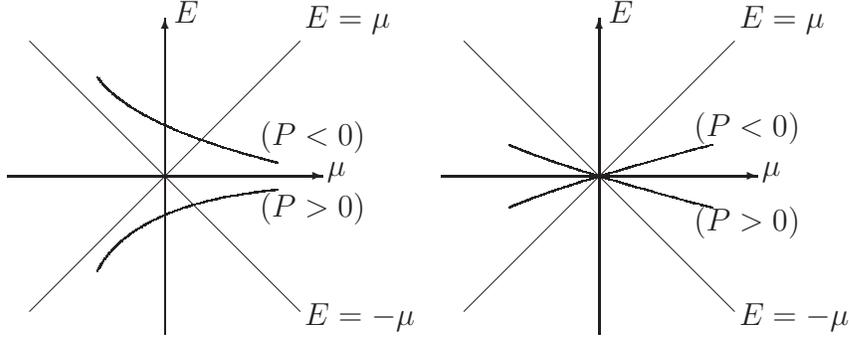

The spectral flow for a one-parameter family of operators is defined to be 
the net number of eigenvalues passing through zero in the positive direction as the parameter runs 
\cite{Prokh}. 
For the eigenvalues under the APS boundary condition, the spectral flow is found to be 
$(-1)+1=0$, where $+1$ is assigned to the the edge state eigenvalue with $P<0$ and $-1$ to 
that with $P>0$, where $P<0$ and $P>0$ mean that the eigenvalue of $P$ is negative and positive, 
respectively. 

In the case of the chiral bag boundary condition, there exist  transient eigenvalue curves  
which cross  one of the boundary lines $E=\pm \mu$, depending on whether the eigenvalues of $P$ is 
positive or negative. 
In this case, the original definition of spectral flow cannot apply, since the crossing of the $E=0$ value 
has no meaning in band rearrangement. 
However, we can extend the notion of spectral flow by assigning 
$-1$ and $+1$ to the crossing of the boundary lines $E=-\mu$ and $E=\mu$, respectively.  
Put another way, the numbers  $+1$ and $-1$ are alloted to 
the transient eigenvalue curves associated with $P<0$ and $P>0$, respectively. 
Then, the extended spectral flow under the chiral bag boundary condition is $(-1)+1=0$, 
the same as that under the APS boundary condition.  

\subsection{$R$-dependence of the bulk-state eigenvalues}
\label{R-dependence}
Though the bulk-state eigenvalues have nothing to do with band rearrangement, 
they are under the influence of band rearrangement. 
To see this, we start by studying the $R$-dependence of the bulk-state eigenvalues. 
For the sake of simplicity, we first consider the bulk-state eigenvalues determined 
by \eqref{bdy cond l+1/2,E>0} with $\mu=0$. 
The eigenvalues are then determined by the zeros of the Bessel function $J_{\ell+\frac32}$.  
Let $\eta_n$ be the zeros of $J_{\ell+\frac32}$, where $0<\eta_1<\eta_2<\cdots$ with $\eta_n 
\to \infty$.    
Then, we obtain $\beta R=\eta_n$ with $\beta=E$, so that the eigenvalues are given by 
$E_n=\eta_n/R$. As is seen form the asymptotic behavior of the Bessel function, 
\begin{equation}
 \label{asymptotic J_nu} 
    J_{\nu}(x)\sim \sqrt{\frac{2}{\pi x}} \cos\bigl(x-\frac{2\nu+1}{4}\pi \bigr), 
\end{equation}
the difference, $\eta_{n+1}-\eta_{n}$, between consecutive pair of zeros is approximately constant in $n$. 
Accordingly, the difference $E_{n+1}-E_{n}$ is approximately inversely proportional to $R$.  
If $\mu\neq 0$, a similar consequence will be brought about for small $\mu$. 
Equation \eqref{bdy cond l+1/2,E>0} for $E>0$ with $\mu \neq 0$ shows that the solutions 
to this equation are obtained from the intersection points of the the graphs of the left- and 
the right-hand sides as functions of $E$.  
As is well known, there lies one and only one zero of $J_{\nu+1}$ between any 
two consecutive zeros of $J_{\nu}$ \cite{WW}. 
Since the factor $\sqrt{(E+\mu)/(E-\mu)}$ plays the role of an amplitude factor, 
the intersection points of the graphs of the left- and the right-hand sides of \eqref{bdy cond l+1/2,E>0}  
take place once and only once over the interval between any consecutive zeros of $J_{\ell+\frac12}$. 
Let $\xi_n$ denote the projection of each intersection point to the axis of the independent variable, 
$0<\xi_1<\xi_2<\cdots$ with $\xi_n\to \infty$ as $n\to \infty$. 
Then, one has $\beta R=\xi_n$.  
The same reasoning applies to Eq.~\eqref{boundary cond, l-1/2, E<0} to give negative eigenvalues. 
It then turns out that the bulk-state eigenvalues determined by \eqref{bdy cond l+1/2,E>0} and 
\eqref{boundary cond, l-1/2, E<0} are expressed as  
\begin{equation}
 \label{bulk-eigenvalue, positive-negative}
   E_n=\pm \sqrt{(\xi_n/R)^2+\mu^2}.   
\end{equation}
The other bulk-state eigenvalues are expressed in the same manner.  
For $\mu$ small, $E_n>0$ is approximated as $E_n\approx \frac{\xi_n}{R}(1+\frac{\mu^2 R^2}{2\xi_n^2})$. 
If $\xi_n/R$ are sufficiently larger than $\mu$,  
the difference $E_{n+1}-E_n$ is approximately 
\begin{equation} 
     E_{n+1}-E_n \approx 
   \frac{\xi_{n+1}-\xi_n}{R} + \frac{\mu^2}{2}\bigl(\frac{R}{\xi_{n+1}}-\frac{R}{\xi_n}\bigr).
\end{equation}
As is seen from the asymptotic behavior \eqref{asymptotic J_nu} and 
the functional equation \eqref{bdy cond l+1/2,E>0}, 
the difference $\xi_{n+1}-\xi_n$ can be  approximately estimated, from the zeros of 
$J_{\nu+1}(x)-J_{\nu}(x)$, to be nearly constant (within a vicinity of $\pi$) in $n$.  
Hence, $E_{n+1}-E_n$ is approximately proportional to $1/R$.  
A similar reasoning can apply to other defining equations  for bulk-state eigenvalues 
under the chiral bag boundary condition.    

It then turns out that if $R$, which is, physically speaking, looked on as representing the system size, 
is sufficiently large, then the energy levels become of high density, so that 
the bands of bulk-states eigenvalues are allowed to be treated in terms of classical variables.  
Like transition from Fourier series to Fourier integral along with the period tending to infinity, 
the momentum operators $-i\partial/\partial q_k$ in the Hamiltonian $H_{\mu}$ can be replaced 
by the classical momentum variables, like a consequence of the Fourier transform.  
Accordingly, the Hamiltonian $H_{\mu}$ turns into a semi-quantum 
Hamiltonian, which we denote by $K_{\mu}$ and has been given in Eq.~\eqref{SQ Dirac Hamiltonian}.  
In the following section, we will study the ``bulk" Hamiltonian $K_{\mu}$ to observe 
a topological change corresponding to the spectral flow attributed to the edge-state eigenvalues. 
As will be soon mentioned in the next section, in the present procedure the discrete variable $\xi_n/R$ 
may be viewed as changing into a continuous radial variable $k$ in the momentum space. 
However, we have to look on $k=0$ as a singular value with respect to the procedure $R\to \infty$,  
and hence the singularity may correspond to the edge-state eigenvalues and be concerned with 
corresponding topological change.   
 
\section{Semi-quantum Dirac models}
\label{semi-quantum models}
We are interested in a correspondence between the energy level transfer for the 3D Dirac equation 
studied in Sec.~\ref{Eigenvalues} and a topological change to be observed in 
the corresponding semi-quantum Hamiltonian.   
In the two-dimensional case, a similar correspondence has been already found in \cite{IZ2016}. 

\subsection{Projection operators onto eigenspaces}
\label{Projection}
Now we work with the semi-quantum Hamiltonian \eqref{SQ Dirac Hamiltonian} in detail. 
Since 
\begin{equation}
\label{characteristic polynomial}
    \det(\lambda I - K_{\mu})=(\lambda^2-(\mu^2+k^2))^2, \quad k^2=k_1^2+k_2^2+k_3^2, 
\end{equation}  
the eigenvalues of $K_{\mu}$ are given by 
\begin{equation} 
        \lambda^{\pm}(\boldsymbol{k}) =\pm \sqrt{\mu^2+k^2}. 
\end{equation} 
Here, we note that each of $\lambda^{\pm}(\boldsymbol{k})$ is doubly degenerate, 
a consequence of the time-reversal symmetry.  Further, the chiral symmetry results in 
the fact that $\lambda^-(\boldsymbol{k})=-\lambda^+(\boldsymbol{k})$.  
In addition, each of the eigenvalues $\lambda^{\pm}(\boldsymbol{k})$ may be viewed as a limit 
of the bulk-state eigenvalues \eqref{bulk-eigenvalue, positive-negative} under the procedure 
in which the discrete variable $\xi_n/R,\,n=1,2,\dots,$ changes into the radial variable $k$ in the momentum space, 
as $R$ tends to infinity.   

For the eigenvalue $\lambda^+_{}(\boldsymbol{k})$, we have two expressions, ``up" and ``down",  
of associated orthonormalized eigenvectors, which are 
\begin{equation}
\label{eigen-vec u+ up} 
   |u^+_{\rm up}(\boldsymbol{k})_1\rangle =\frac{1}{N^+_{\rm up}}
    \begin{pmatrix} k_3 \\ k_1+ik_2 \\ i(\lambda^+ -\mu) \\ 0 \end{pmatrix}, \quad 
   |u^+_{\rm up}(\boldsymbol{k})_2\rangle =\frac{1}{N^+_{\rm up}}
    \begin{pmatrix} k_1-ik_2 \\ -k_3 \\ 0  \\ i(\lambda^+ -\mu),  \end{pmatrix}, 
\end{equation}
\begin{equation}
\label{eigen-vec u+ down}
   |u^+_{\rm down}(\boldsymbol{k})_1\rangle  =\frac{1}{N^+_{\rm down}}
     \begin{pmatrix} -i(\lambda^+ +\mu) \\ 0 \\ k_3 \\ k_1+ik_2 \\ \end{pmatrix} , \quad 
   |u^+_{\rm down}(\boldsymbol{k})_2\rangle =\frac{1}{N^+_{\rm down}}
     \begin{pmatrix}  0  \\ -i(\lambda^+ +\mu) \\ k_1-ik_2 \\ -k_3  \end{pmatrix} ,  
\end{equation} 
where 
\begin{equation} 
   N^+_{\rm up}=\sqrt{2\lambda^+(\lambda^+-\mu)},   \quad   
   N^+_{\rm down}=\sqrt{2\lambda^+(\lambda^+ +\mu)} . 
\end{equation} 
We here notice that the words ``up" and ``down" used in the semi-quantum model do not allude to spin.  
For $\mu>0$, the eigenvectors $|u^+_{\rm up}(\boldsymbol{k})_a\rangle,\,a=1,2,$ are defined on 
$\mathbb{R}^3-\{0\}$, but for $\mu<0$ they are defined on the whole $\mathbb{R}^3$. 
In contrast with this, for $\mu>0$, the eigenvectors $|u^+_{\rm down}(\boldsymbol{k})_a\rangle,\,a=1,2,$ 
are defined on $\mathbb{R}^3$, but for $\mu<0$ they are defined on $\mathbb{R}^3-\{0\}$. 
We call a point $\boldsymbol{k}$ an exceptional point, if the eigenvector in question is not defined at the point 
$\boldsymbol{k}$, For example, the origin $\boldsymbol{k}=0$ is an exceptional point of  
$|u^+_{\rm up}(\boldsymbol{k})_a\rangle,\,a=1,2,$ for $\mu>0$. 

The normalized eigenvectors associated with the eigenvalue $\lambda^-(\boldsymbol{k})$ are expressed, 
in the same manner, as 
\begin{equation}
\label{eigen-vec u- up} 
   |u^-_{\rm up}(\boldsymbol{k})_1\rangle  =\frac{1}{N^-_{\rm up}}
    \begin{pmatrix} k_3 \\ k_1+ik_2 \\ i(\lambda^- -\mu) \\ 0 \end{pmatrix}, \quad 
   |u^-_{\rm up}(\boldsymbol{k})_2\rangle =\frac{1}{N^-_{\rm up}}
    \begin{pmatrix} k_1-ik_2 \\ -k_3 \\ 0  \\ i(\lambda^- -\mu) \end{pmatrix}, 
\end{equation}
\begin{equation}
\label{eigen-vec u- down} 
   |u^-_{\rm down}(\boldsymbol{k})_1\rangle  =\frac{1}{N^-_{\rm down}}
     \begin{pmatrix} -i(\lambda^- +\mu) \\ 0 \\ k_3 \\ k_1+ik_2 \\ \end{pmatrix} , \quad 
   |u^-_{\rm down}(\boldsymbol{k})_2\rangle =\frac{1}{N^-_{\rm down}}
     \begin{pmatrix}  0  \\ -i(\lambda^- +\mu) \\ k_1-ik_2 \\ -k_3  \end{pmatrix} ,  
\end{equation} 
where 
\begin{equation} 
\begin{array}{l}
   N^-_{\rm up}=\sqrt{2\lambda^-(\lambda^- -\mu)}= \sqrt{2|\lambda^-|(|\lambda^-|+\mu)}=N^+_{\rm down},   \\
   N^-_{\rm down}=\sqrt{2\lambda^-(\lambda^- +\mu)}= \sqrt{2|\lambda^-|(|\lambda^-| -\mu)}=N^+_{\rm up}. 
\end{array}
\end{equation} 
The exceptional point for  $|u^-_{\rm up/down}(\boldsymbol{k})_a\rangle$ 
can be identified in a similar manner to that for $|u^+_{\rm up/down}(\boldsymbol{k})_a\rangle$. 
Accompanying the variation of the parameter $\mu$, the origin $\boldsymbol{k}=0$ is  attached 
as an exceptional points to eigenvectors associated with either positive or negative eigenvalue, 
which is summed up as follows; 
\begin{equation}
\label{change in exc pt}  
   \begin{array}{c|c|c|c}
   \boldsymbol{k}=0  & \mu<0 & \mu=0 & \mu>0 \\ \hline
   \mbox{ex. pt. for}\; |u^{\pm}_{\rm up}(\boldsymbol{k})_a\rangle & -  & \mbox{deg. pt.} & + \\ \hline 
   \mbox{ex. pt. for}\; |u^{\pm}_{\rm down}(\boldsymbol{k})_a\rangle & +  & \mbox{deg. pt.} & -\\ \hline 
   \end{array}
\end{equation} 

The eigenvectors $|u^+_{\rm up}(\boldsymbol{k})_a\rangle$ and $|u^+_{\rm down}(\boldsymbol{k})_b\rangle$ 
are related  on the intersection of respective domains through  
\begin{equation}
\label{U+ def}
      |u^+_{\rm down}(\boldsymbol{k})_{b}\rangle=\sum |u^+_{\rm up}(\boldsymbol{k})_{a}\rangle U_{ab}^+ .
\end{equation}
Calculating the matrix elements $\langle u^+_{\rm up}(\boldsymbol{k})_a|u^+_{\rm down}(\boldsymbol{k})_{b}\rangle$,  
we find the explicit expression of $U^+=(U^+_{ab})$ as 
\begin{equation} 
\label{transf U}
   U^+=\begin{pmatrix} -i\frac{k_3}{k} & -i\frac{k_1-ik_2}{k} \\ 
        -i\frac{k_1+ik_2}{k}  & i\frac{k_3}{k} \end{pmatrix} = -i\frac{1}{k}\boldsymbol{k}\cdot \boldsymbol{\sigma}. 
\end{equation} 
The transformation matrix $U^-=(U^-_{ab})$ to be determined in a similar manner to \eqref{U+ def} 
is found to be given by 
\begin{equation} 
\label{transf U^-}
   U^-=\begin{pmatrix} i\frac{k_3}{k} & i\frac{k_1-ik_2}{k} \\ 
        i\frac{k_1+ik_2}{k}  & -i\frac{k_3}{k} \end{pmatrix} = i\frac{1}{k}\boldsymbol{k}\cdot \boldsymbol{\sigma}. 
\end{equation}
 
We now calculate the projection operators $P_{\pm}$ onto the eigenspaces 
associated with the eigenvalues $\lambda^{\pm}(\boldsymbol{k})$. 
A straightforward calculation provides 
\begin{align} 
   P_+ &=\frac{1}{2\lambda^+_{}}\begin{pmatrix} 
       (\lambda^+_{} +\mu)\1 & -i\boldsymbol{k}\cdot \boldsymbol{\sigma} \\
       i\boldsymbol{k}\cdot \boldsymbol{\sigma}  & (\lambda^+_{} -\mu)\1 \end{pmatrix} 
       = \frac{1}{2\lambda^+_{}}(\lambda^+_{} I +K_{\mu}), 
      \label{P+}   \\
    P_- & =\frac{1}{2\lambda^-_{}}\begin{pmatrix} 
       (\lambda^-_{} +\mu)\1 & -i\boldsymbol{k}\cdot \boldsymbol{\sigma} \\
       i\boldsymbol{k}\cdot \boldsymbol{\sigma}  & (\lambda^-_{} -\mu)\1 \end{pmatrix} 
       = \frac{1}{2\lambda^-}(\lambda^-_{} I + K_{\mu}), 
       \label{P-}
\end{align} 
respectively, where $I$ denotes the $4\times 4$ identity matrix.  

\subsection{Eigen-vector bundles}
In this subsection, we denote the eigenvectors by $|u^+_{\rm up}(\boldsymbol{k},\mu)_a\rangle$ {\it etc}, 
in order to stress their dependence on $\boldsymbol{k}$ and $\mu$. 
We denote by $V^{\pm}(\boldsymbol{k},\mu)$ the eigenspaces    
associated with the eigenvalues $\lambda^{\pm}(\boldsymbol{k},\mu)$, where 
$\dim_{\mathbb{C}}V^{\pm}(\boldsymbol{k},\mu)=2$ for $\mu\neq 0$. 
Then, the totality of the eigenspaces, 
$V^{\pm}(\mu)=\bigsqcup_{\boldsymbol{k}\in \mathbb{R}^3}V^{\pm}(\boldsymbol{k},\mu)$,  
form respective vector bundles  over $\mathbb{R}^3$ for $\mu\neq 0$, 
which we call the eigen-vector bundles associated with the eigenvalues. 
From the table \eqref{change in exc pt}, we see that
the ``up" eigenvectors $|u^+_{\rm up}(\boldsymbol{k},\mu)_a\rangle$ and 
$|u^-_{\rm up}(\boldsymbol{k},\mu)_a\rangle$ have no exceptional point at $\boldsymbol{k}=0$ 
for $\mu<0$ and for $\mu>0$,  respectively.  
This implies that $V^+(\mu)$ and $V^-(\mu)$ are trivial vector bundles for $\mu<0$ and for $\mu>0$, 
respectively.  
Furthermore, the ``down" eigenvectors $|u^-_{\rm down}(\boldsymbol{k},\mu)_a\rangle$ and 
$|u^+_{\rm down}(\boldsymbol{k},\mu)_a\rangle$ have no exceptional point for 
$\mu<0$ and for $\mu>0$, respectively, so that the eigen-vector bundles $V^-(\mu)$ and 
$V^+(\mu)$ are trivial for $\mu<0$ and for $\mu>0$. 
It then seems that the bundle structure $V^+(\mu)\oplus V^-(\mu)$ receives 
no topological change when the parameter passes the critical point $\mu=0$ from the  
$\mu<0$ side to the $\mu>0$ side.  
However, topological change takes place, accompanying the variation of the parameter 
$\mu$ passing the critical value $\mu$, as will be shown in the succeeding subsections. 

A naive way to detect a change in term of topological quantity is to refer to Chern numbers 
assigned to the vector bundles $V^{\pm}(\mu)$. 
However, this way proves bad. 
Though one can calculate the connections $A^{\pm}$ and the curvatures $F^{\pm}$ for $V^{\pm}(\mu)$ 
with $\mu\neq 0$, the first Chern forms, which are defined to be ${\rm tr}F^{\pm}$ 
within a constant factor, vanishes, since they take values in $\mathfrak{su}(2)$ as a consequence of 
the fact that the structure groups are $SU(2)$ (see  \eqref{transf U} and \eqref{transf U^-}). 
Further, the Chern-Simon form ${\rm tr}(A \wedge dA +\frac23 A\wedge A \wedge A)$ also vanishes and gives 
no information of topological property.   

In spite of this fact, we are interested in looking for what change takes place when 
the parameter $\mu$ runs through $\mu=0$. 
To this end, we look into the decomposition $V^+(\boldsymbol{k},\mu)\oplus V^-(\boldsymbol{k},\mu)$. 
Though this decomposition fails at $\boldsymbol{k}=0$ when $\mu=0$, 
it may be viewed as surviving in a sense.  In fact, we can verify that 
\begin{subequations}
\label{eigenspaces joined} 
\begin{align}
    \lim_{\mu \uparrow 0} V^+(0,\mu) & ={\rm span}_{\mathbb{C}}\{|e_3\rangle, |e_4\rangle \} = 
    \lim_{\mu \downarrow 0} V^-(0,\mu) , \\ 
   \lim_{\mu \uparrow 0} V^-(0,\mu) & ={\rm span}_{\mathbb{C}}\{|e_1\rangle, |e_2\rangle \} = 
    \lim_{\mu \downarrow 0} V^+(0,\mu) ,
\end{align}
\end{subequations}
where $\mu \uparrow 0$ and $\mu \downarrow 0$ means that $\mu$ tends to zero, taking negative 
and positive values, respectively, and where $|e_k\rangle,k=1,\dots,4$, are the standard basis vectors with 
the $k$-th component one and the others zero.  
It then turns out that the degenerate positive (resp. negative) eigenvalue  
$\lambda^+(0,\mu)=|\mu|$ for $\mu<0$ (resp. $\lambda^-(0,\mu)=-|\mu|$ for $\mu<0$)  
goes down (resp. up) to degenerate negative (resp. positive) eigenvalue $\lambda^-(0,\mu)=-|\mu|$ 
for $\mu>0$ 
(resp. $\lambda^+(0,\mu)=|\mu|$ for $\mu>0$) through the totally degenerate zero eigenvalue 
at $\boldsymbol{k}=0$ as $\mu$ runs in the positive direction, passing $\mu=0$.  
Put another way, two energy levels cross each other at $\boldsymbol{k}=0$, when $\mu=0$. 
It is to be noted that the level crossing occurs at $\boldsymbol{k}=0$ only and no crossing at 
$\boldsymbol{k}\neq 0$.  
This phenomenon is a realization of the singularity alluded to in the last sentence of Sec.~\ref{R-dependence} 
and we may expect that the level crossing gives rise to a certain topological change. 

\subsection{Introducing  ``Q matrices"}
In order to seek for a suitable topological invariant to describe a topological change against the parameter 
in the semi-quantum model, 
we are reminded of the fact that for the semi-quantum model 
corresponding to the 2D Dirac equation the winding number (or a delta-Chern) 
associated with the $2\times 2$  semi-quantum Hamiltonian  
plays a key role \cite{IZ2015,IZ2016}.   
In an analogous manner, we are to seek for a suitable winding number (or mapping degree) 
associated with the Hamiltonian $K_{\mu}(\boldsymbol{k})$.    

According to a paper  \cite{SchFurLud}, a ``$Q$ matrix" is associated with a projection matrix. 
Let $P$ be a projection matrix of rank $m$ (different from the operator given in \eqref{parity-like}), 
of which the complementary projection operator 
is assumed to be of rank $n$. 
The $Q$ matrix is defined to be 
\begin{equation} 
    Q=2P-I, 
\end{equation}
where $I$ is the identity matrix of order $m+n$.  
Since $P^2=P,\,P^{\dag}=P, {\rm tr}\,P=m$, the $Q$ matrix satisfies 
\begin{equation}  
    Q^{\dag}=Q, \quad Q^2=I, \quad {\rm tr}Q=m-n. 
\end{equation} 
For the semi-quantum Hamiltonian $K_{\mu}$, 
the $Q$ matrices corresponding to the projection operators \eqref{P+} and \eqref{P-} 
are obtained as follows: 
\begin{equation}
\label{Q+-}
    Q_+=\frac{1}{\lambda^+}K_{\mu}, \quad Q_-=\frac{1}{\lambda^-}K_{\mu}, 
\end{equation}
respectively, where $m=n=2$.  

As is well known,  Hamiltonians with chiral symmetry can take an off-diagonal block form \cite{SchFurLud}.  
We have treated the chiral operator of the form $\sigma_1\otimes \1$.  
However, we may choose to express the chiral operator as $\sigma_3\otimes \1$, 
which is known as a canonical expression of the chiral operator. 
In fact, since 
\begin{equation}
      h^{-1}\sigma_1h=\sigma_3, \quad h=\frac{1}{\sqrt{2}}\begin{pmatrix} 1 &  1 \\ 1 & -1 \end{pmatrix},  
\end{equation}
the matrix $\sigma_1\otimes \1$ is unitarily equivalent to $\sigma_3\otimes \1$; 
\begin{equation}
   (h^{-1}\otimes \1)(\sigma_1\otimes \1)(h\otimes \1)= \sigma_3\otimes \1. 
\end{equation}
It is easy to verify that the unitary matrix $g=h\otimes \1$ brings 
the $Q$ matrices \eqref{Q+-} into an off-diagonal block form, 
\begin{align}
   g Q_{\pm} g^{-1}= & \frac{1}{\lambda^{\pm}} \begin{pmatrix} 0 & \mu \1+i\boldsymbol{k}\cdot \boldsymbol{\sigma} \\ 
         \mu \1-i\boldsymbol{k}\cdot \boldsymbol{\sigma} & 0 \end{pmatrix},  \quad 
    g = h\otimes \1 = \frac{1}{\sqrt{2}} \begin{pmatrix}  \1 & \1 \\ \1 & -\1 \end{pmatrix},  
   \label{off-diag Q+-}      
\end{align} 
respectively.  

\subsection{Mappings of $\mathbb{R}^3$ to $S^3$}
From \eqref{off-diag Q+-}, we may pick up the off-diagonal block matrices  
\begin{equation}
\label{block matrices}
  \frac{1}{\lambda^{\pm}}(\mu \1 \pm i\boldsymbol{k}\cdot \boldsymbol{\sigma}).  
\end{equation}
Since 
\begin{equation}
     \det\Bigl( \frac{1}{\lambda^{\pm}}(\mu \1 \pm i\boldsymbol{k}\cdot \boldsymbol{\sigma} )\Bigr)=
    \frac{1}{|\lambda^{\pm}|^2}\det (\mu \1 \pm i\boldsymbol{k}\cdot\boldsymbol{\sigma}) =1, 
\end{equation}
these matrices lie in  $SU(2)\cong S^3$. 
Since the multiplication of \eqref{block matrices} by $\pm i$ gives no topological change, 
we are allowed to take, independently of the choice of $Q_{\pm}$,  the mappings 
\begin{equation}
\label{mappings +-} 
    \boldsymbol{k}\mapsto  \frac{1}{|\lambda^{\pm}|}(\boldsymbol{k}\cdot \boldsymbol{\sigma} \pm i\mu \1)
   =\frac{1}{|\lambda^{\pm}_{}(\boldsymbol{k})|}\begin{pmatrix} 
      k_3\pm i\mu & k_1-ik_2 \\ k_1+ik_2 & k_3\pm i \mu \end{pmatrix}, 
\end{equation} 
and to regard this mapping as defining the mapping of $\mathbb{R}^3$ to $S^3$,   
\begin{equation}
\label{maps to S3}
    q_{\pm}:\; \boldsymbol{k}\longmapsto \frac{1}{\sqrt{k^2+\mu^2}}\begin{pmatrix} k_1+ik_2 \\ 
k_3\pm i\mu \end{pmatrix}
   \in S^3 \subset \mathbb{C}^2 \cong \mathbb{R}^4,  
\end{equation}
where the identification  $\mathbb{C}^2 \cong \mathbb{R}^4$ is defined through 
\begin{equation}
\label{identify}
    (x_1+ix_2, x_3+ix_4)\longmapsto (x_1,x_2,x_3,x_4). 
\end{equation}

\subsection{Mapping degrees or winding numbers}
We first study the mapping 
\begin{equation} 
\label{map to S3,+}
    q_+(\boldsymbol{k})= \frac{1}{\sqrt{k^2+\mu^2}}\begin{pmatrix} k_1+ik_2 \\ k_3 + i\mu \end{pmatrix}
   \in S^3 \subset \mathbb{C}^2 \cong \mathbb{R}^4. 
\end{equation} 
If we let $k=|\boldsymbol{k}|\to \infty$, the limit points of the mapping \eqref{map to S3,+} are expressed as 
\begin{equation}
\label{k infty} 
    q_+(\boldsymbol{k})=
    \frac{k}{\sqrt{k^2+\mu^2}}\begin{pmatrix} (k_1+ik_2)/k \\ (k_3 + i\mu)/k \end{pmatrix} \longrightarrow 
      \begin{pmatrix} n_1+in_2 \\ n_3 \end{pmatrix} 
   \in S^2 \subset S^3 , 
\end{equation}    
where $S^2$ denotes the equator of the $S^3$, to which the vector $\boldsymbol{n}=(n_k)$ belongs.  

\begin{figure}[h]
\begin{center}
\setlength{\unitlength}{1.0mm}
\begin{picture}(85,46)
\put(5,20){\line(1,0){75}}
\put(40,2){\vector(0,1){40}}
\put(40,20){\circle{31}}
\put(42,41){$\mu$}
\put(78,21){$\mathbb{R}^3\cong \mathbb{C}\times \mathbb{R}$}
\put(7,38){\line(1,0){73}}
\path(40,20)(73,38)
\put(81.5,35){$\mathbb{R}^3_{\mu}$\;{\rm with}\;$\mu>1$}
\put(66,39.6){$(\boldsymbol{k},\mu)$}
\put(56,23){$S^3$}
\put(7,6.5){\line(1,0){73}}
\path(40,20)(73,6.5)
\put(72,8){$(\boldsymbol{k},\mu)$}
\put(76,1){$\mathbb{R}^3_{\mu} \;{\rm with}\; -1<\mu<0$}
\end{picture}
\end{center}
\caption{A schematic description of $q_+(\boldsymbol{k})$
\label{map q+} } 
\end{figure}
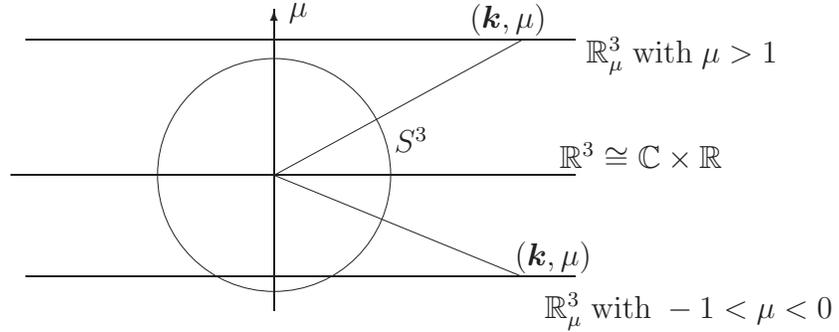

Let $\mathbb{R}^3_{\mu}$ denote the three-plane parallel  to the tangent plane to the unit sphere 
$S^3$ at the north or south pole, intersecting the fourth axis at $(0,\mu)\in \mathbb{R}^3\times \mathbb{R}$. 
The mapping $q_+(\boldsymbol{k})$ given in \eqref{map to S3,+} maps each point $(\boldsymbol{k},\mu)$ 
on the plane  $\mathbb{R}^3_{\mu}$ to the point at which 
the line joining $(\boldsymbol{k},\mu)$ and the origin crosses the unit sphere $S^3$ (see Fig.~\ref{map q+}).  
Equations \eqref{map to S3,+} and \eqref{k infty} imply that if $\mu>0$ the upper hemisphere is covered 
by $\mathbb{R}^3_{\mu}$ through the mapping $q_+$ and if $\mu<0$ the lower hemisphere is covered. 
It is to be noted that if $\mu<0$ the orientation of the lower hemisphere covered by  
the mapping is opposite to the naturally defined orientation of $S^3$.  

We proceed to discuss the mapping degree of \eqref{map to S3,+} in terms of integrals over $\mathbb{R}^3$. 
We have to start by finding the area element of the sphere $S^3 \subset \mathbb{R}^4$.  
Let $(x_k)$ be the Cartesian coordinates of $\mathbb{R}^4$. 
The area element of $S^3$ is obtained by contracting the canonical volume element 
$dx_1\wedge dx_2\wedge dx_3 \wedge dx_4$ by the radial vector field $\sum x_k\partial /\partial x_k$. 
In view of the orientation of the tangent plane to $S^3$ at the  north pole, 
we take the area element of $S^3$ as  
\begin{align}
\label{area form}
   \omega = & -x_1dx_2\wedge dx_3\wedge dx_4  + x_2dx_3 \wedge dx_4\wedge dx_1 \nonumber \\
       &  - x_3dx_4 \wedge dx_1\wedge dx_2  + x_4dx_1 \wedge dx_2\wedge dx_3 . 
\end{align}
The pull-back $q^*_+\omega$ of the area form $\omega$ is shown to take the form 
\begin{equation} 
   q^*_+\omega = \frac{\mu dk_1\wedge dk_2 \wedge dk_3}{(k^2+\mu^2)^2}. 
\end{equation}  
Introducing the spherical polar coordinates $(k,\theta,\phi)$ in the $\boldsymbol{k}$-space $\mathbb{R}^3$, 
we integrate $q^*_+\omega$ over $\mathbb{R}^3$ to obtain 
\begin{equation}
\label{volume integral} 
    \int_{\mathbb{R}^3}q^*_+\omega = \pi^2\frac{\mu}{|\mu|}.  
\end{equation} 
 
Since ${\rm vol}(S^3)=2\pi^2$, the mapping degree of $q_+:\,\mathbb{R}^3\to S^3$ is defined 
and evaluated as 
\begin{equation}
\label{winding no (q+)} 
     \nu[q_+]=\frac{1}{2\pi^2}\int_{\mathbb{R}^3}q^*_+\omega =\frac12 \frac{\mu}{|\mu|}.  
\end{equation} 
The factor $\frac12$ implies that half of the sphere $S^3$ is covered by the mapping $q_+$, and 
$\frac{\mu}{|\mu|}=\pm1$ means that the covering is positively- or negatively-oriented according as  
${\rm sgn}(\mu)$.  
This is completely in agreement with the schematic description of the mapping $q_+$, 
which is already shown in Fig.~\ref{map q+}.  

\begin{figure}[h]
\begin{center}
\setlength{\unitlength}{1.0mm}
\begin{picture}(85,46)
\put(5,20){\line(1,0){75}}
\put(40,0){\vector(0,1){40}}
\put(40,20){\circle{31}}
\put(42,41){$\mu$}
\put(78,21){$\mathbb{R}^3\cong \mathbb{C}\times \mathbb{R}$}
\put(7,38){\line(1,0){73}}
\path(73,38)(73,2)
\put(81.5,35){$\mathbb{R}^3_{\mu}$\;{\rm with}\;$\mu>1$}
\put(66,39.6){$(\boldsymbol{k},\mu)$}
\put(56,23){$S^3$}
\put(7,2){\line(1,0){73}}
\path(40,20)(73,2)
\put(74,3.5){$(\boldsymbol{k},-\mu)$}
\end{picture}
\end{center}
\caption{A schematic description of $q_-(\boldsymbol{k})$
\label{map q-} } 
\end{figure}
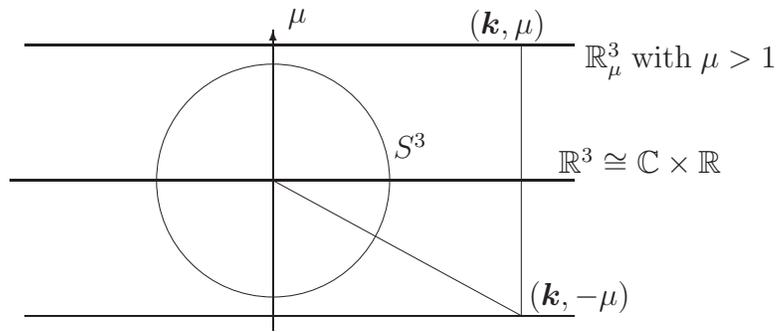

We turn to the mapping 
\begin{equation}
\label{map q_-} 
  q_-(\boldsymbol{k})=\frac{1}{\sqrt{k^2+\mu^2}}\begin{pmatrix} k_1+ ik_2 \\ k_3-i\mu \end{pmatrix}.  
\end{equation} 
Since the expression of $q_-(\boldsymbol{k})$ is obtained by replacing the parameter $\mu$  
in the definition of $q_+(\boldsymbol{k})$ by $-\mu$,  
the winding number of $q_-$ is obtained by replacing $\mu$ by $-\mu$ in \eqref{winding no (q+)};    
\begin{equation} 
   \nu[q_-]=-\frac12 \frac{\mu}{|\mu|}. 
\end{equation} 
Through the mapping $q_-$, a point $(\boldsymbol{k},\mu)$ on the plane $\mathbb{R}^3_{\mu}$ 
(see Fig.~\ref{map q-}) 
is mapped to the point $(\boldsymbol{k},-\mu)$ once, 
and then mapped to the point at which the line joining the origin and 
the $(\boldsymbol{k},-\mu)$ crosses the unit sphere.    

In place of $q_{\pm}$, we are allowed to consider the mappings 
\begin{equation} 
\label{h+-}
     h_{\pm}(\boldsymbol{k})=\frac{1}{\sqrt{k^2+\mu^2}}(\boldsymbol{k}\cdot \boldsymbol{\sigma}\pm i\mu \1), 
\end{equation}
which also come from \eqref{mappings +-} and are viewed as  mappings $\mathbb{R}^3\to SU(2)$. 
In what follows, we mainly treat $h_+$. 

The winding number (or mapping degree) of $h_+$ is defined to be 
\begin{equation} 
      \nu[h_+]=\frac{1}{24\pi^2}\int_{SU(2)}   {\rm tr}(h_+^{\dag}dh_+ \wedge h_+^{\dag}dh_+ \wedge h_+^{\dag}dh_+), 
\end{equation} 
where $24\pi^2$ is the volume of $SU(2)$ 
with respect to the volume form determined by the left- or the right-invariant one-forms. 
This winding number for a mapping $\mathbb{R}^3\to SU(2)$ is usually adopted in literature \cite{SchFurLud}.  
A straightforward but rather lengthy calculation with  \eqref{h+-}  provides   
\begin{equation} 
    {\rm tr}(h_+^{\dag}dh_+ \wedge h_+^{\dag}dh_+ \wedge h_+^{\dag}dh_+)=
    \frac{12\mu}{(k^2+\mu^2)^2}dk_1\wedge dk_2 \wedge dk_3 . 
\end{equation} 
Then, we obtain 
\begin{equation} 
    \nu[h_+]=\frac{1}{24\pi^2}\int_{\mathbb{R}^3} \frac{12\mu dk_1dk_2dk_3}{(k^2+\mu^2)^2} 
   = \frac12 \frac{\mu}{|\mu|},
\end{equation}
which is exactly the same as $\nu[q_+]$.  

\subsection{Jump in the mapping degree against the parameter}
\label{Jump in mapping degree} 
Though the mapping degrees $\nu[q_+]$ and $\nu[q_-]$ are half-integer valued, the change 
accompanying the variation in $\mu$ is integer valued; 
\begin{equation}
\label{change map deg} 
        \nu[q_{\pm}]|_{\mu>0} - \nu[q_{\pm}]|_{\mu<0} = \pm1. 
\end{equation} 
We show that the jump \eqref{change map deg} has a topological meaning. 
Since $q_+^*\omega$ is a three-form on $\mathbb{R}^3$, and 
since $\mathbb{R}^3$ is simply connected,  $q_+^*\omega$ must be an exact form. 
We express $q_+^*\omega$, in terms of the polar spherical coordinates $(k,\theta,\phi)$, as 
\begin{equation} 
    q_+^*\omega = \mu\frac{k^2dk}{(k^2+\mu^2)^2}\wedge \sin\theta\,d\theta \wedge d\phi. 
\end{equation} 
Denoting the canonical volume (or area) element on $S^2$ by 
\begin{equation} 
                  \varpi =\sin \theta\, d\theta \wedge d\phi, 
\end{equation} 
we can easily verify that $q_+^*\omega$ takes the form 
\begin{equation} 
     q_+^*\omega = \mu d(F\varpi), 
\end{equation} 
where 
\begin{equation} 
\label{primitive} 
   F(k) = -\frac{k}{2(k^2+\mu^2)} +\frac{1}{2\mu} \arctan \frac{k}{\mu}. 
\end{equation} 

Let $S_r$ denote the boundary of the ball $B_r$ of radius $r$ with the center at the origin.  
Then, the integral of $q_+^*\omega$ over $\mathbb{R}^3$ is evaluated as 
\begin{equation} 
 \int_{\mathbb{R}^3}q_+^*\omega  = \lim_{r\to \infty} \int_{B_r} q_+^*\omega 
   =  \lim_{r\to \infty} \mu F(r) \int_{S^2} \varpi, 
\end{equation} 
where $S^2$ denote the unit two-sphere, which may be viewed as the equator of $S^3$,  
the limit of $q_+^{}(S_r)$ as $r\to \infty$. 
In the limit as $r\to \infty$, one obtains from \eqref{primitive}
\begin{equation} 
     \lim_{r\to \infty} \mu F(r) = \frac{\pi}{4}{\rm sgn}(\mu), 
\end{equation}  
so that the mapping degree of $q_+$ is put in the form 
\begin{equation}
\label{wind no q+} 
    \nu[q_+] =\frac{1}{2\pi^2}\int_{\mathbb{R}^3} q_+^*\omega = \frac{1}{8\pi}{\rm sgn}(\mu)  \int_{S^2} \varpi. 
\end{equation} 

When the parameter $\mu$ passes the zero value in the positive direction, the present equation 
provides the accompanying jump in the mapping degree as  
\begin{equation} 
 \label{jump in q+}  
    \nu[q_+]|_{\mu>0} -\nu[q_+]|_{\mu<0} =\frac{1}{4\pi}  \int_{S^2} \varpi.
\end{equation} 
The right-hand side of the above equation is one, of course. 
We are allowed to interpret that the right-hand side is a topological quantity assigned to the equator 
$S^2$ as the limit of $q_+^{}(S_r)$ as $r\to \infty$. 

For the mapping $q_-(\boldsymbol{k})$ given in \eqref{map q_-}, 
we obtain, in place of  \eqref{wind no q+}, 
\begin{equation} 
    \nu[q_-] =\frac{1}{2\pi^2}\int_{\mathbb{R}^3} q_-^*\omega = \frac{-1}{8\pi}{\rm sgn}(\mu)  \int_{S^2} \varpi, 
\end{equation} 
and in place of \eqref{jump in q+}, 
\begin{equation} 
\label{jump in q-}
      \nu[q_-]|_{\mu>0} -\nu[q_-]|_{\mu<0} =\frac{-1}{4\pi}  \int_{S^2} \varpi.
\end{equation} 

\section{Correspondence between spectral flow and mapping degree}
\label{Correspondence}
We are now in a position to answer the question, raised in Introduction, as to the correspondence 
between band rearrangement and topological change or the bulk-edge correspondence. 
In order to characterize the band rearrangement resulting from the 3D Dirac equation with the APS 
and the chiral bag boundary conditions, we have introduced the notion of spectral flow and its extension 
in Sec.~\ref{Spectral flow}  
and found that the spectral flow in both cases of the APS and the chiral bag boundary 
conditions is $+1+(-1)=0$, where the spectral flow is taken as the ordinary or the extended one 
according as the APS or the chiral bag boundary condition is concerned.   
As we have already seen in Sec.~\ref{Spectral flow}, the zero value of the spectral flow 
does not mean a trivial phenomenon but implies that the redistribution of edge state eigenvalues 
takes place in opposite manners at the same time, independently of the choice of boundary conditions. 
We note in addition that the spectral flow is independent of the radius $R$ of the ball. 

A semi-quantum counterpart to the spectral flow of the full quantum model has been discussed in 
Sec.~\ref{Jump in mapping degree} in terms of the mapping degree $\nu[q_{\pm}]$.  
As we have already found, the change in the mapping degree for each of $\nu[q_{\pm}]$ 
is given in \eqref{change map deg} and interpreted as a winding number for the unit two-sphere 
(see \eqref{jump in q+} and \eqref{jump in q-}).   
The jumps $\nu[q_+]_{\mu>0}-\nu[q_+]_{\mu<0}=1$ and $\nu[q_-]_{\mu>0}-\nu[q_-]_{\mu<0}=-1$  
can be viewed as corresponding to the eigenvalue transfer for the state with $P<0$ 
and for the state with $P>0$, respectively (see Fig.~\ref{transient eigenvalue curve}). 
The sum of the jumps is zero, of course.  

It then turns out that the zero sum of  the jumps in winding numbers (or mapping degrees) 
for the semi-quantum model 
is in marked correspondence to the zero spectral flow for the full quantum model. 
Since the semi-quantum Hamiltonian $K_{\mu}$ is viewed as a bulk Hamiltonian (see 
Sec.~\ref{R-dependence}), the correspondence between spectral flow and  
a change in mapping degree can be looked upon as a bulk-edge correspondence. 

In order to get a better understanding of the correspondence between band rearrangement and 
topological change, we compare the respective correspondences for the 2D and the 3D models. 
The spectral flow for the 2D Dirac equation under both the APS and the chiral bag boundary conditions 
is $+1$ or $-1$ \cite{IZ2015,IZ2016} but the spectral flow for 3D Dirac equation is $+1-1=0$. 
Correspondingly, the delta-Chern characterized as a winding number for $S^1$ in the 2D 
semi-quantum model takes value $-1$ or $+1$ according as the eigenvalue is positive or negative \cite{IZ2016},
but the changes in winding numbers for $S^2$ in the 3D semi-quantum model take values $\pm1$ 
independently of the sign of the eigenvalue and the sum of those values is zero.  
Here, we notice that the $S^1$ and $S^2$ referred to in the 2D and the 3D semi-quantum models are 
viewed as the equator of $S^2$ and of $S^3$, respectively. 
Though the correspondence between band rearrangement and topological change hold true for the 
2D and the 3D models, the difference in appearance comes from 
the difference in the discrete symmetries that the Hamiltonians in respective models admit.

\section {Concluding remarks}
\label{conclusion}
In Sec.~\ref{setting-up}, we did not refer to the AZ classification \cite{AZ, HHZ} of Hermitian matrices 
for brevity.  
We here make remarks on the nomenclature such as A, AI, AII, AIII, {\it etc.}, used 
in the AZ classification.  
These symbols comes from the classification of symmetric spaces by Cartan (see \cite{Helg}).  
The classes A, AI, AII are called the Wigner-Dyson classes \cite{Wigner, Dyson}, 
and classified with respect to the time-reversal operator: 
Matrices belonging to the class A are merely Hermitian.  
If a Hermitian matrix admits the time-reversal symmetry, it belongs to the class AI or AII, 
according to whether the squared time-reversal operator is equal to the identity or to minus 
the identity.     
The Hamiltonian $\mathcal{H}$ given in \eqref{TRS Ham} belongs to the class AII, 
since the squared time-reversal operator is equal to minus the identity; 
$((\1 \otimes i\sigma_2)K)^2=-{\rm id}_{\mathbb{C}^2\otimes \mathbb{C}^2}$. 
If we further require $\mathcal{H}$ to admit the $SU(2)$ symmetry, 
in the form different from \eqref{semi-quantum SU(2) action on Ham}, by imposing that 
\begin{equation}
\label{SU(2)-inv} 
      (\1 \otimes g) \mathcal{H}(\1\otimes g)^{-1}=\mathcal{H}, \quad g\in SU(2), 
\end{equation} 
the Hamiltonian becomes a real symmetric matrix,  
\begin{equation}
\label{real symmetric}
   \mathcal{H}=\begin{pmatrix} a_{11} & a_{12} \\ a_{12} & a_{22} \end{pmatrix} \otimes \1, \quad 
      a_{jk}\in \mathbb{R},  
\end{equation}
which naturally reduces to a $2\times 2$ real symmetric matrix. 
Hamiltonians of this form belong to the class AI, since the squared time-reversal operator is the identity, 
where the relevant time-reversal operator is merely the complex conjugation. 

In addition, since the semi-quantum Hamiltonian $K_{\mu}(\boldsymbol{k})$ 
admits the chiral symmetry, it belongs to the class AIII, 
where Hamiltonians having the chiral symmetry are classified into AIII without reference 
to the time-reversal or the particle-hole symmetry.    
Further, on account of the particle-hole symmetry with 
$((\sigma_1\otimes i\sigma_2)K)^2=-{\rm id}_{\mathbb{C}^2\otimes \mathbb{C}^2}$, 
it belongs to the class CII, where Hamiltonians of class CII are required to have both 
the chiral symmetry with 
$((\sigma_1\otimes i\sigma_2)K)^2=-{\rm id}_{\mathbb{C}^2\otimes \mathbb{C}^2}$ 
and the time-reversal symmetry with 
$((\1 \otimes i\sigma_2)K)^2=-{\rm id}_{\mathbb{C}^2\otimes \mathbb{C}^2}$.   

We have to compare the conditions \eqref{semi-quantum SU(2) action on Ham} and 
\eqref{SU(2)-inv} concerning $SU(2)$ symmetry. 
Since distinction among parameters of a Hamiltonian is not required in the AZ classification, 
Eq.~\eqref{SU(2)-inv} is adopted as a symmetry condition. However,   
parameters of a semi-quantum Hamiltonian in this article have to be broken up into control parameters 
and dynamical variables,  so that \eqref{semi-quantum SU(2) action on Ham} has been adopted as 
a symmetry condition, while we have called \eqref{semi-quantum SU(2) action on Ham} 
$SU(2)$-equivariant.  

We here recall  that we have imposed the chiral symmetry in addition to the 
time-reversal symmetry in order to characterize the semi-quantum Hamiltonian 
$K_{\mu}(\boldsymbol{k})$ given in \eqref{SQ Dirac Hamiltonian}. 
In contrast to this, if we do not impose the chiral symmetry but merely require 
the Hermitian matrix \eqref{TRS Ham} to be traceless, the Hamiltonian admitting the time-reversal 
symmetry proves to take the form 
\begin{equation}
\label{4D Ham} 
   \tilde{K}_{\mu}(\boldsymbol{k},k_4)= \begin{pmatrix} 
     \mu \1 & k_4 \1-i\boldsymbol{k}\cdot \boldsymbol{\sigma} \\
      k_4 \1+i\boldsymbol{k}\cdot \boldsymbol{\sigma} & -\mu \1 \end{pmatrix},  
    \quad (\boldsymbol{k},k_4)\in \mathbb{R}^4, 
\end{equation} 
where the parameters in \eqref{TRS Ham} have been replaced according to 
$a=k_4-ik_3, b=-k_2-ik_1, c=\mu$ and $d=-\mu$. 
A comparison of \eqref{4D Ham} with \eqref{SQ Dirac Hamiltonian} or \eqref{3D Ham, TR PH} 
shows that the chiral symmetry makes the semi-quantum Hamiltonian 
$\tilde{K}_{\mu}(\boldsymbol{k},k_4)$ defined on $\mathbb{R}^4$ reduce to the $K_{\mu}(\boldsymbol{k})$ 
defined on $\mathbb{R}^3$. Furthermore, 
if $k_j, \,j=1,\dots,4$, are replaced by $-i\frac{\partial}{\partial x_j}$, 
the $\tilde{K}_{\mu}$ brings about the 4D Dirac Hamiltonian defined on $\mathbb{R}^4$. 
A question as to the correspondence between band rearrangement and topological change 
for the 4D Dirac Hamiltonian is reserved for a future study. 
 
An additional question is raised as for the correspondence between band rearrangement 
in a full quantum model and topological change in the corresponding semi-quantum model. 
The conclusion described in Sec.~\ref{Correspondence} was obtained through the free Dirac equation 
defined on the bounded domain, irrespective of the choice of the APS 
and the chiral bag boundary conditions.  
The additional question is as to whether a similar correspondence holds true or not if a full quantum model 
is defined on an unbounded domain, say $\mathbb{R}^2$ or $\mathbb{R}^3$, in the presence of 
an electric or magnetic field.  

Last but not least, we remark that classification schemes for vector bundles 
have been studied according to the AZ classification \cite{DNG, DNG2015}. 

\appendix
\section{Spinor harmonics and the Hamiltonian in the polar spherical coordinates}
\label{SU(2) representation}
We make a brief review of the representation spaces for the orbital angular momentums. 
As is well known, the spherical harmonics on $S^2$ are given by  
\begin{equation} 
    Y^n_{\ell}(\theta,\phi)=(-1)^n \sqrt{\frac{2\ell+1}{4\pi}}\sqrt{\frac{(\ell-n)!}{(\ell+n)!}}
                                   e^{in\phi}P^{n}_{\ell}(\cos\theta), 
\end{equation}
where $P^n_{\ell}(x)$ are the associated Legendre functions defined to be  
\begin{equation}
   P^n_{\ell}(x)=\frac{1}{2^{\ell}\ell !} (1-x^2)^{n/2}\frac{d^{\ell+n}}{dx^{\ell+n}}(x^2-1)^{\ell}, 
    \quad |n|\leq \ell. 
\end{equation}
Further, the associated Legendre functions are known to satisfy the recursion formulas   
\begin{subequations} 
 \begin{align} 
    (\nu+\mu+1)xP^{\mu}_{\nu}(x)-\sqrt{1-x^2}P^{\mu+1}_{\nu}(x) & = (\nu-\mu+1)P^{\mu}_{\nu+1}(x), \\
    xP^{\mu}_{\nu}(x)+(\nu+\mu)\sqrt{1-x^2}P^{\mu-1}_{\nu}(x) & =P^{\mu}_{\nu+1}(x), \\
   xP^{\mu}_{\nu}(x)-(\nu-\mu+1)\sqrt{1-x^2}P^{\mu-1}_{\nu}(x) & = P^{\mu}_{\nu-1}(x), \\
   (\nu-\mu)xP^{\mu}_{\nu}(x)+\sqrt{1-x^2}P^{\mu+1}_{\nu}(x) & =(\nu+\mu)P^{\mu}_{\nu-1}(x). 
 \end{align}
\end{subequations} 
We recall also that the orbital angular momentum operators $L_k$ acts on $Y^n_{\ell}$ according to 
\begin{subequations}
\label{up-down-still}
 \begin{align} 
     (L_1+iL_2)Y^n_{\ell}=& \sqrt{(\ell-n)(\ell+n+1)}Y^{n+1}_{\ell}, \\
     (L_1-iL_2)Y^n_{\ell}=& \sqrt{(\ell+n)(\ell-n+1)}Y^{n-1}_{\ell},  \\ 
     L_3Y^n_{\ell} = & nY^n_{\ell}. 
   \end{align}
\end{subequations}
On account of the formula $P^n_{\ell}(-x)=(-1)^{n+\ell}P^n_{\ell}(x)$ and $e^{in(\phi+\pi)}=(-1)^n e^{in\phi}$, 
the spherical harmonics have the parity 
\begin{equation}    
     Y^n_{\ell}(-\boldsymbol{r})=(-1)^{\ell}Y^n_{\ell}(\boldsymbol{r}), \quad |\boldsymbol{r}|=1. 
\end{equation} 

We now compose the basis spinors which are simultaneous eigenstates of $J_3$ and $\boldsymbol{J}^2$.     
For $j=\ell+\frac12$, the basis state of $J_3$ 
takes the form $(c_1 Y^{m-\frac12}_{\ell}, c_2 Y^{m+\frac12}_{\ell})^T$.  In fact, we obtain 
\begin{equation} 
  \begin{pmatrix}  L_3 +\frac12 & 0 \\ 0 & L_3-\frac12 \end{pmatrix} 
    \begin{pmatrix} c_1 Y^{m-\frac12}_{\ell} \\ c_2 Y^{m+\frac12}_{\ell} \end{pmatrix} = 
   m \begin{pmatrix} c_1 Y^{m-\frac12}_{\ell} \\ c_2 Y^{m+\frac12}_{\ell} \end{pmatrix}, \quad 
   |m|\leq j=\ell+\frac12, 
\end{equation} 
where $m$ take values of half-integers.   
The constants $c_1,c_2$ are to be determined by the condition that  
$(c_1 Y^{m-\frac12}_{\ell}, c_2 Y^{m+\frac12}_{\ell})^T$ is also an eigenstate of $\boldsymbol{J}^2$ 
together with the normalization condition $|c_1|^2+|c_2|^2=1$.  
As is easily seen, the square $\boldsymbol{J}^2$ is written out as 
\begin{align} 
     \boldsymbol{J}^2= & 
     = \begin{pmatrix}  \boldsymbol{L}^2+L_3 +\frac34 & L_1-iL_2 \\
          L_1+iL_2 &  \boldsymbol{L}^2-L_3 +\frac34 \end{pmatrix}. 
\end{align}
Operating $\Phi=(c_1 Y^{m-\frac12}_{\ell}, c_2 Y^{m+\frac12}_{\ell})^T$ with $\boldsymbol{J}^2$, 
we obtain the condition for $\Phi$ to be an eigenstate associated with the eigenvalue $j(j+1)$ 
of $\boldsymbol{J}^2$, which results in 
\begin{equation} 
    \frac{c_1}{\sqrt{\ell+m+\frac12}}=\frac{c_2}{\sqrt{\ell-m+\frac12}}. 
\end{equation}
This equation and the normalization condition for $c_1,c_2$ are put together to give 
the basis \eqref{basis l+1/2,l}.   
For $j=(\ell+1)-\frac12$, in a similar manner, we find that 
\begin{equation} 
  \frac{c_1}{\sqrt{\ell-m+\frac32}}=-\frac{c_2}{\sqrt{\ell+m+\frac32}}. 
\end{equation}
With the normalization condition for $c_1,c_2$, we obtain the basis \eqref{basis l+1/2,l+1}. 

In the rest of this appendix, we make a brief review of the Dirac Hamiltonian expressed 
in terms of the polar spherical coordinates.  
As is well known, the standard moving frame attached to the polar spherical coordinates is given by 
\begin{equation} 
\label{spherical moving frame}
 \mbox{\boldmath $e$}_r=
\begin{pmatrix} \sin\theta \cos\phi \\ \sin\theta \sin \phi \\ \cos \theta \end{pmatrix}, \quad 
 \boldsymbol{e}_{\theta}=
     \begin{pmatrix} \cos\theta \cos \phi \\ \cos \theta \sin \phi \\ -\sin \theta \end{pmatrix}, \quad 
 \boldsymbol{e}_{\phi}=\begin{pmatrix} -\sin \phi \\ \cos \phi \\ 0 \end{pmatrix} , 
\end{equation} 
where $\boldsymbol{e}_{\theta}$ and $\boldsymbol{e}_{\phi}$ are defined on the domain without the $z$-axis or 
with $\theta\neq 0,\pi$. 
The associated vector fields (or operators) are expressed, respectively, as 
\begin{equation} 
\label{spherical-gradients}
 \boldsymbol{e}_r\cdot \boldsymbol{\nabla} = \frac{\partial}{\partial r}, \quad  
 \boldsymbol{e}_{\theta} \cdot \boldsymbol{\nabla} =  \frac{1}{r}\frac{\partial}{\partial \theta}, \quad 
  \boldsymbol{e}_{\phi} \cdot \boldsymbol{\nabla} =  \frac{1}{r \sin\theta }\frac{\partial}{\partial \phi},   
\end{equation}
where $\boldsymbol{\nabla}$ denotes the ordinary gradient operator. 

By using the identity 
\begin{equation} 
     \boldsymbol{e}_r \boldsymbol{e}_r^T +  \boldsymbol{e}_{\theta} \boldsymbol{e}_{\theta}^T + 
      \boldsymbol{e}_{\phi} \boldsymbol{e}_{\phi}^T ={\rm id}_{\mathbb{R}^3},
\end{equation}
where the superscript ${}^T$ denotes the transposition, 
we can put the Dirac Hamiltonian \eqref{3D_Dirac_Ham} in the form 
\begin{align}
\label{polar form} 
   H_{\mu} & = -i \boldsymbol{\gamma} \cdot(\boldsymbol{e}_r \boldsymbol{e}_r^T +  
     \boldsymbol{e}_{\theta} \boldsymbol{e}_{\theta}^T + 
      \boldsymbol{e}_{\phi} \boldsymbol{e}_{\phi}^T) \boldsymbol{\nabla} + \mu \gamma_0^{} \nonumber \\
  & = -i\gamma_r \partial_r -\frac{i}{r}(\gamma_{\theta}^{}X_{\theta}+\gamma_{\phi}X_{\phi}) +\mu\gamma_0^{}. 
\end{align}
where 
\begin{subequations} 
\begin{align}
\label{gammas}
  \gamma_a & =\boldsymbol{\gamma}\cdot \boldsymbol{e}_a, \quad a\in \{r,\theta,\phi\}, \\
 \frac{1}{r}X_b & =\boldsymbol{e}_b \cdot \boldsymbol{\nabla}, \quad b\in \{\theta,\phi\}, 
  \label{X_b}
\end{align}
\end{subequations} 
and where the operators $X_b$ are determined through \eqref{spherical-gradients} and \eqref{X_b}. 
We here remark that the products among $\sigma_a$ and among $\gamma_a$, $a\in\{r,\theta,\phi\}$   
are given by 
\begin{equation}
\label{products of sigmas} 
   \sigma_r\sigma_{\theta}=i\sigma_{\phi}, \quad   \sigma_{\theta}\sigma_{\phi}=i\sigma_{r}, \quad 
    \sigma_{\phi}\sigma_{r}=i\sigma_{\theta}, \quad \sigma_r^2=\sigma_{\theta}^2=\sigma_{\phi}^2=\1, 
\end{equation}
\begin{equation} 
    \gamma_r^{}\gamma_{\theta}^{}=i\1 \otimes \sigma_{\phi}, \quad 
    \gamma_{\theta}^{}\gamma_{\phi}=i\1 \otimes \sigma_{r}, \quad 
    \gamma_{\phi}\gamma_{r}^{}=i\1 \otimes \sigma_{\theta}, \quad 
    \gamma_r^2=\gamma_{\theta}^2=\gamma_{\phi}^2=\1 \otimes \1, 
\end{equation} 
respectively, and they satisfy 
\begin{equation} 
     \sigma_a\sigma_b +\sigma_b \sigma_a =2\delta_{ab}\1, \quad a,b\in\{\small{r,\theta,\phi}\}, 
\end{equation}
\begin{subequations}
 \begin{align} 
   \gamma_a\gamma_b+\gamma_b\gamma_a & =2\delta_{ab}\1 \otimes \1, 
   \quad a,b\in \{r,\theta,\phi\}, \\
   \gamma_a\gamma_0^{} +\gamma_0^{} \gamma_a & =0. 
 \end{align}
\end{subequations}
Further calculation with the above formula provides 
\begin{align}
  -\frac{i}{r}\gamma_r^{}(\gamma_{\theta}^{} X_{\theta}+\gamma_{\phi}^{}X_{\phi})+
       \mu \gamma_r^{}\gamma_0^{} 
 & = \frac{i}{r} \begin{pmatrix} -i\sigma_{\phi} X_{\theta}+i\sigma_{\theta} X_{\phi} & \mu r \sigma_r \\
           \mu r \sigma_r & -i\sigma_{\phi} X_{\theta}+i\sigma_{\theta} X_{\phi} \end{pmatrix} 
 \nonumber \\
 & = \frac{i}{r} \begin{pmatrix} 
  \boldsymbol{\sigma}\cdot \boldsymbol{L} & \boldsymbol{\sigma}\cdot \boldsymbol{r} \\
    \boldsymbol{\sigma}\cdot \boldsymbol{r} &  \boldsymbol{\sigma}\cdot \boldsymbol{L}
  \end{pmatrix} , 
\label{angular form}
\end{align} 
where use has been made of 
\begin{subequations}
 \begin{align} 
     L_1+iL_2 & =e^{i\phi}\Bigl(\frac{\partial}{\partial \theta}+i\cot\theta \frac{\partial}{\partial \phi}\Bigr), \\
      L_1-iL_2 & =-e^{-i\phi}\Bigl(\frac{\partial}{\partial \theta}-i\cot\theta \frac{\partial}{\partial \phi}\Bigr), \\
     L_3 & = -i\frac{\partial}{\partial \phi}. 
 \end{align} 
\end{subequations}  
and 
\begin{align} 
\label{sigma_X}
   i(\sigma_{\phi}X_{\theta}-\sigma_{\theta}X_{\phi}) 
   & =  -\begin{pmatrix}  L_3 & L_1-iL_2 \\ L_1+iL_2 & -L_3 \end{pmatrix} 
      =-\boldsymbol{\sigma}\cdot \boldsymbol{L}. 
\end{align}
Eqs.~\eqref{polar form} and \eqref{angular form} together with $\gamma_r^2=I$ 
are combined to provide \eqref{polar decomposition}.   

\medskip
\par\noindent
{\bf Acknowledgements}.    
The authors would like to thank Dr. G. Dhont for drawing graphs of  bulk and edge sate eigenvalues  
as functions of the control parameter and as those of the angular momentum eigenvalue as well.   
Part of this work was supported by a Grant-in Aid for Scientific Research 18K03297 (T.I) from JSPS.


\begin{thebibliography}{99}
\bibitem{AbBe}
A.A. Abrikosov and S.D. Beneslavskii, 
Possible existence of substances intermediate between metals and dielectrics, 
Soviet Phys. JETP, {\bf 32}, 699-708 (1971). 

\bibitem{AkBe}
A.R. Akhmerov and C.W. J. Beenakker, 
Boundary conditions for Dirac fermions on a terminated honeycomb lattice, 
Phys, Rev. {\bf B77}, 085423 (2008). 
 
\bibitem{AZ}
A. Altland and M.R. Zirnbauer, 
Nonstandard symmetry classes in mesoscopic normal-superconducting hybrid structures,  
Phys. Rev. {\bf B55}, 1142-1161 (1977). 

\bibitem{ABPP}
M. Asorey, A.P. Balachandran, and J.M. P{\'e}rez-Pardo, 
Edge states: topological insulators, superconductors and QCD chiral bags, JHEP (12)073 (2013).  

\bibitem{APS}  
M.F. Atiyah, V.K. Patodi, and I.M. Singer,   
Spectral asymmetry and Riemannian geometry. I, II, III,    
Math. Proc. Cambridge Phil. Soc. {\bf 77},  43-69 (1975), {\bf 78}, 405-432 (1975),   
{\bf 79}, 71-99 (1976).

\bibitem{AOS}
J.E. Avron, D. Osadchy, and R. Seiler,  
A topological look at the quantum Hall effect, 
Phys. Today 38-42 (2003). 

\bibitem{Arnoldmodes}
V.I. Arnold, 
Modes and quasimodes, 
Funct. Anal. and its Appl. {\bf 6}, 94-101 (1972). 

\bibitem{ArnoldChern}
V.I. Arnold, 
Remarks on eigenvalues and eigenvectors of Hermitian matrices. Berry
phase, adiabatic connections and quantum Hall effect, 
Selecta Mathematica {\bf 1}, 1-19 (1995).

\bibitem{BHZ}
B.A. Bernevig, T.L. Hughes, and S.-C. Zhang, 
Quantum spin Hall effect and topological phase transition in HgTe quantum wells, 
Science, {\bf 314}, 1757-1761 (2006).  

\bibitem{TopInsTopSup}
B.A. Bernevig with T.L. Hughes, 
{\it Topological Insulators and Topological Superconductors}, 
Princeton Univ. Press, Princeton, New Jersey (2013).  

\bibitem{Berry}
M.V. Berry,  
Quantal phase factors accompanying adiabatic change, 
Proc. R. Lond. {\bf A392}, 45-57 (1984). 

\bibitem{ChrJam}
D. Chru{\'s}ci{\'n}ski and A. Jami{\l}kowski, 
{\it Geometric phases in classical and quantum mechanics}, 
Birkh{\"a}user, Boston (2004).  

\bibitem{DNG}
G. De Nittis and K. Gomi, 
Classification of ``Real" Bloch-bundles: Topological quantum systems of type AI, 
J.  Geom. Phys. {\bf 86},  303–338 (2014).  

\bibitem{DNG2015}
G. De Nittis and K. Gomi, 
Classification of ``Quaternionic" Bloch-Bundles: 
Topological Quantum systems of type AII, 
Commun. Math. Phys. {\bf 339}, 1-55 (2015). 

\bibitem{Dirac}
P.A.M. Dirac, 
{\it The principles of quantum mechanics} (fourth edition), 
Oxford University Press, Oxford (1958). 

\bibitem{DIZ}
G. Dhont, T. Iwai, and B. Zhilinskii, 
Topological phase transition in a molecular Hamiltonian with symmetry and pseudo-symmetry, 
Studied through quantum, semi-quantum and classical models, 
SIGMA, {\bf 13}, 054, 34 pages (2017). 

\bibitem{Dyson}
F.J. Dyson, 
Statistical theory of the energy levels of complex systems, 
J. Math. Phys., {\bf 3}, 140-156 (1962). 

\bibitem{FaurePRL} 
F. Faure and B. I. Zhilinskii,
Topological Chern indices in molecular spectra, 
Phys. Rev. Lett., {\bf 85}, 960-963 (2000).

\bibitem{FaureLMP}
F. Faure and B. I. Zhilinskii,
Topological properties of the Born-Oppenheimer approximation and 
implications for the exact spectrum, 
Lett. Math. Phys. {\bf 55}, 239-247 (2001). 

\bibitem{FaureAAM}
F. Faure and B. Zhilinskii,
Qualitative features of intra-molecular dynamics. What can be learned from symmetry and topology, 
Acta Appl. Math. {\bf 70}, 265-282 (2002) 

\bibitem{FSFF}
T. Fukui, K. Shiozaki, T. Fujiwara, and S. Fujimoto, 
Bulk-edge correspondence for Chern topological phases: A viewpoint from 
a generalized index theorem, 
J. Phys. Soc. Jpn. {\bf 81} 114602 (2012).  

\bibitem{Graf-Porta}
G.M. Graf and M. Porta,  
Bulk-edge correspondence for two-dimensional topological insulators, 
Commun. Math. Phys., {\bf 324}, 851-895 (2013). 

\bibitem{GrVo}
P.G. Grinevich and G.E. Volovik, 
Topology of gap nodes in superfluid ${}^3$He: $\pi_4$ homotopy group for ${}^3$He-B Disclination, 
J. Low Temp Phys., {\bf 72}, 371-380 (1988). 

\bibitem{Guichardet}
A. Guichardet, 
On rotation and vibration motions of molecules, 
Ann. Inst. Henri Poincar{\'e}, {\bf 40}, 329-342 (1984)

\bibitem{HasKan}
M.Z. Hassen and C.L. Kane, 
Colloquium: Topological insulators, 
Rev. Mod. Phys. {\bf 82}, 3045-3067 (2010). 

\bibitem{HHZ}
P. Heinzner, A. Huckleberry, and M.R. Zirnbauer, 
Symmetry classes of disordered fermions, 
Comm. Math. Phys., {\bf 257}, 725-771 (2005).  

\bibitem{Helg}
S. Helgason, 
{\it Differential geometry, Lie groups, and symmetric spaces}, 
New York, Academic Press, 1978.  

\bibitem{Herz}
G. Herzberg, 
{\it Molecular spectra and molecular structure}, vol.2, Krieger (1989).  

\bibitem{HLH}
G. Herzberg, H.C. Longuet-Higgins, 
Intersection of potential energy surfaces in polyatomic molecules, 
Disc. Faraday Soc. {\bf 35}, 77-82 (1963). 

\bibitem{IPP}
A. Ibort and J.M. P{\'e}rez-Pardo, 
On the theory of self-adjoint extensions of symmetric operators and its applications to 
quantum physics, 
Internat. J. Geom. Methods Mod. Phys. {\bf 12}, 1560005 (2015). 

\bibitem{IwaiClass}
T. Iwai, 
A geometric setting for classical molecular dynamics, 
Ann. Inst. Henri Poincar\'e, {\bf 47}. 199-219 (1987). 

\bibitem{IwaiHolonomy} 
T. Iwai, A. Tachibana,
The geometry and mechanics of multi-particle systems, 
Ann. Inst. Henri Poincar\'e, {\bf 70}. 525-559 (1999). 

\bibitem{IwaiYama05}
T. Iwai and H. Yamaoka, 
Stratified dynamical systems and their boundary behaviour for three bodies in space, 
with insight into small vibrations, 
J. Phys., {\bf A38} 5709-5730 (2005). 

\bibitem{IwaiYama}
T. Iwai and H. Yamaoka, 
Rotational-vibrational energy spectra of triatomic molecules near relative equilibria, 
J. Math Phys., {\bf 49}, 043505 (2008).  

\bibitem{IwaiAnnPhys}
T. Iwai and  B. Zhilinskii, 
Energy bands: Chern numbers and symmetry, 
Ann. Phys. (NY), {\bf  326}, 3013-3066  (2011). 

\bibitem{Iwai}
T. Iwai and  B. Zhilinskii, Rearrangement of energy bands: Chern numbers in
the presence of cubic symmetry, Acta Appl. Math,  {\bf 120}, 153-175 (2012) 

\bibitem{IZ2013}
T. Iwai and  B. Zhilinskii, Qualitative feature of the rearrangement of molecular energy spectra 
from a ``wall-crossing" perspective, 
Phys. Lett. {\bf A 377}, 2481-2486 (2013)

\bibitem{IZChemAcc}
T. Iwai and B. Zhilinskii, 
Topological phase transition in the vibration-rotation dynamics of an isolated molecule, 
Theor. Chem. Acc. {\bf 133}, 1501 (13 pages) (2014).  

\bibitem{IZ2015}
T. Iwai and B. Zhilinskii, 
Local description of band rearrangements, Comparison of semi-quantum and full-quantum approach,   
Acta Appl. Math. {\bf 137}, 97-121 (2015).  

\bibitem{IZ2016}
T. Iwai and B. Zhilinskii,  
Band rearrangement through the 2D-Dirac equation: Comparing the APS and the chiral bag 
boundary conditions, Indag. Math. {\bf 27}, 1081-1106 (2016).  

\bibitem{IZ2017}
T. Iwai and B. Zhilinskii, 
Chern number modification in crossing the boundary between different band structures: 
Three-band models with cubic symmetry, 
Rev. Math. Phys. {\bf 29}, 1750004 (91 pages) (2017). 

\bibitem{KaneMele}
C.L. Kane and E.J. Mele, 
Quantum spin Hall effect in graphene, 
Phys. Rev. Lett. {\bf 95}, 146802 (2005).  

\bibitem{Kitaev}
A. Kitaev, Periodic table for topological insulators and superconductors, 
AIP Conf. Proc., 1134, 22-30 (2009).

\bibitem{Kohmoto}
M. Kohmoto, 
Topological invariant and the quantization of the Hall conductance, 
Ann. Phys. {\bf 160}, 343-354 (1985)

\bibitem{KS}
H. Koizumi and S. Sugano, 
Geometric phase in two Kramers doublet molecular systems, 
J. Chem. Phys. {\bf 102}, 4472-4481 (1995).

\bibitem{McC-F} 
E. McCann and V.I. Fal'ko, 
Symmetry of boundary operators of the Dirac equation for electrons in carbon nanotubes, 
J. Phys.: Condens. Matter {\bf 16}, 2371-2379 (2004). 

\bibitem{Mead}
C.A. Mead, 
The ``noncrossing" rule for electronic potential energy surface: The role of time-reversal invariance, 
J. Chem. Phys. {\bf 70}, 2276-2283 (1979).  

\bibitem{MS}
R.S.K. Mong and V. Shivamoggi, 
Edge states and the bulk-boundary correspondence in Dirac Hamiltonians, 
Phys. Rev. {\bf B}83, 125109 (2011).

\bibitem{VPVdp}
V. B. Pavlov-Verevkin, D. A. Sadovskii and B. I. Zhilinskii, 
On the dynamical meaning of diabolic points, Europhys. Lett. {\bf 6}, 573-78 (1988). 

\bibitem{PSB}
E. Prodan and H. Schulz-Baldes, 
{\it Bulk and Boundary Invariants for Complex Topological Insulators. From K-Theory to Physics}, 
Mathematical Physics Studies, Springer, Berlin (2016). 

\bibitem{Prokh}  
M. Prokhorova,   
The spectral flow for Dirac operators on compact planar domains with local boundary   
conditions,   
Comm. Math. Phys. {\bf 322}, 385-414 (2013).    

\bibitem{QHRZ}
X.-L. Qi, T.L. Hughes, S. Raghu, and S.-C. Zhang, 
Time-reversal-invariant topological superconductors and superfluids in two- and three-dimensions, 
Phys. Rev. Lett. {\bf 102}, 187001 (2009).  

\bibitem{RSFL}
S. Ryu, A.P. Schnyder, A. Furusaki, and A.W.W. Ludwig, 
Topological insulators and superconductors: tenfold way and dimensional hierarchy, 
New J. Phys., {\bf 12}, 065010 (2010).  

\bibitem{MolPhys88}
D. A. Sadovskii and B. I. Zhilinskii, 
Qualitative analysis of vibration-rotation Hamiltonians for 
spherical top molecules, Molec. Phys. {\bf 65}, 109-28 (1988). 

\bibitem{SadZhilMonodr}
D. A. Sadovskii and B. I. Zhilinskii,
Monodromy, diabolic points, and angular momentum coupling,
Phys. Lett. A {\bf 256}, 235-44 (1999). 

\bibitem{SchFurLud}
A.P. Schnyder, S. Ryu, A. Furusaki, and A.W.W. Ludwig, 
Classification of topological insulator and superconductors in three spatial dimensions, 
Phys.  Rev. B{\bf 78}, 195125 (2008).

\bibitem{TachIwai}
A. Tachibana and T. Iwai, 
Complete molecular Hamiltonian based on the Born-Oppenheimer adiabatic approximation, 
Phys. Rev. {\bf A33}, 2262-2269 (1986).  

\bibitem{TKNN}
D.J. Touless, M. Kohmoto, M.P. Nightingale, and M. den Nijs, 
Quantized Hall Conductance in a Two-Dimensional Periodic Potential, 
Phys. Rev. Lett. {\bf 49}, 405 (1982).

\bibitem{TSN}
Y. Tanaka, M. Sato, and N. Nagaosa, 
Symmetry and topology in superconductors --Odd-frequency paring and edge states--, 
J Phys. Soc. Jpn, {\bf 81}, 011013 (2012). 

\bibitem{WW}
E.T. Whittaker and G.N. Watson, {\it A course of modern analysis, fourth ed.}, 
Cam. Univ. Press, London, 1927.  

\bibitem{Wigner}
E.P. Wigner, 
On the statistical distribution of the width and spacings of nuclear resonance levels, 
Proc. Cambridge Philos. Soc. {\bf 47}, 790-798 (1951).  

\bibitem{WS}
F. Wilczek and A. Shapere (Eds.), {\it Geometric phases in Physics}, 
World Scientific, 1989.  

\bibitem{PhysRep2}
B. Zhilinskii, Symmetry, invariants and topology in molecular models, 
Phys. Rep. {\bf 341}. 85-172 (2001). 

\bibitem{Zirn}
M.R. Zirnbauer, Riemannian symmetric superspaces and their origin in random matrix theory, 
J. Math. Phys., {\bf 37}, 4986-5018 (1996). 

\end{thebibliography}
\end{document}